\documentclass[journal]{IEEEtran}
\usepackage{cite,mathtools}
\usepackage{graphicx}
\usepackage{subfigure}
\usepackage{amssymb}
\usepackage{xcolor}
\usepackage{url,breakurl}
\usepackage{float}
\ifCLASSINFOpdf
\else
\fi
\begin{document}

\title{Multi-Frequency Multi-Scenario Millimeter Wave MIMO Channel Measurements and Modeling for B5G Wireless Communication Systems}
%
%
%

\author{Jie~Huang,
        Cheng-Xiang~Wang,~\IEEEmembership{Fellow,~IEEE,}
        Hengtai~Chang,
        Jian Sun,~\IEEEmembership{Member,~IEEE,}
        and Xiqi Gao,~\IEEEmembership{Fellow,~IEEE}
\thanks{The authors would like to acknowledge the support from the National Key R\&D Program of China (No. 2018YFB1801101), the National Natural Science Foundation of China (No. 61901109, 61960206006, 61771293), the National Postdoctoral Program for Innovative Talents (No. BX20180062), the High Level Innovation and Entrepreneurial Talent Introduction Program in Jiangsu, the Research Fund of National Mobile Communications Research Laboratory, Southeast University (No. 2020B01), the Fundamental Research Funds for the Central Universities (No. 2242019R30001), Taishan Scholar Program of Shandong Province, EU H2020 RISE TESTBED2 project (No. 872172), and the Huawei Cooperation Project. The authors would also like to thank Yubei He, Yiling Huang, Qingshan Chen, Yunzeng Li, Cuiping Zhang, Hao Zhang, Zhigang Song, Lianjie Li, and Jianhao Wang from Shandong University for their help of the channel measurements.}
\thanks{J. Huang, C.-X. Wang (corresponding author), and X. Q. Gao are with the National Mobile Communications Research Laboratory, School of Information Science and Engineering, Southeast University, Nanjing, 210096, China, and also with Purple Mountain Laboratories, Nanjing, 211111, China (e-mail: \{j\_huang, chxwang, xqgao\}@seu.edu.cn).}
\thanks{H. Chang and J. Sun are with Shandong Provincial Key Lab of Wireless Communication Technologies, School of Information Science and Engineering, Shandong University, Qingdao, 266237, China (e-mail: hunter\_chang@126.com, sunjian@sdu.edu.cn).}
}

%
%

\markboth{IEEE Journal on Selected Areas in Communications}%
{Shell \MakeLowercase{\textit{et al.}}: Bare Demo of IEEEtran.cls for Journals}
%



\maketitle

\begin{abstract}
Millimeter wave (mmWave) bands have been utilized for the fifth generation (5G) communication systems and will no doubt continue to be deployed for beyond 5G (B5G). However, the underlying channels are not fully investigated at multi-frequency bands and in multi-scenarios by using the same channel sounder, especially for the outdoor, multiple-input multiple-output (MIMO), and vehicle-to-vehicle (V2V) conditions. In this paper, we conduct multi-frequency multi-scenario mmWave MIMO channel measurements with 4$\times$4 antennas at 28, 32, and 39 GHz bands for three cases, i.e., the human body and vehicle blockage measurements, outdoor path loss measurements, and V2V measurements. The channel characteristics, including blockage effect, path loss and coverage range, and non-stationarity and spatial consistency, are thoroughly studied. The blockage model, path loss model, and time-varying channel model are proposed for mmWave MIMO channels. The channel measurement and modeling results will be of great importance for further mmWave communication system deployments in indoor hotspot, outdoor, and vehicular network scenarios for B5G.
\end{abstract}


%
\IEEEpeerreviewmaketitle

\section{Introduction}
%
%
%
%
\IEEEPARstart{T}{he} fifth generation (5G) communication systems are being deployed worldwide and will be operated at several allocated millimeter wave (mmWave) bands. International telecommunication union (ITU) allocated eleven bands in the range of 24.25-86 GHz for wireless communications, including 26/28, 32, 38/39, and 60~GHz bands. 

Though there have been various mmWave channel measurement campaigns, most of them are limited in the measurement bandwidth (500 MHz typically), distance range (less than 300 m typically), scenario (indoor/outdoor static), antenna configuration (single antenna), etc. \cite{Huang17, Huang19, COMST19}. Meanwhile, as the channel sounder, measurement setup, measurement scenario, and data post-processing method are not the same for the measurements conducted by different research groups, it is difficult to have a fair comparison of the propagation characteristics at different frequency bands \cite{Huang17}. The underlying channels are still not fully investigated at multi-frequency bands and in multi-scenarios by using the same channel sounder, especially for the outdoor, multiple-input multiple-output (MIMO), and vehicle-to-vehicle (V2V) conditions, which are challenging for beyond 5G (B5G) communications. 

The early mmWave channel measurements conducted by T. S. Rappaport et al. proved the ability of mmWave communications for 5G \cite{Rap13}. 
In \cite{Ben11, Rap13_2, Sam13, Aza13, Mac14, Sam15, Sun16}, channel measurements were conducted at 28, 38, 60, and 73 GHz bands for several outdoor environments with rotated horn antennas. The path loss, root mean square (RMS) delay spread, angle of arrival (AoA), and outage probability were obtained. The NYU Simulator (NYUSIM) \cite{NYUSIM}, an open source software for mmWave channel modeling, was then proposed based on these channel measurement results. Its latest version includes three important channel characteristics, i.e., spatial consistency, human blockage shadowing loss, and outdoor-to-indoor (O2I) penetration loss. 

In ITU-R P.2406-0 \cite{ITU2406}, several outdoor channel measurements were conducted in the range of 0.8-73 GHz for urban, suburban, and residential environments. The different measurement results from NTT DOCOMO (Japan), Radio Research Agency (RRA), Electronics and Telecommunications Research Institute (ETRI), and Samsung (Korea), Intel (America), and Durham University (UK) were combined to produce the path loss and shadowing fading models as well as the statistics of RMS delay spread. 

In \cite{Wei14}, the outdoor urban environment was measured at 60~GHz. The results showed that the environment was highly time-variant. The human blockage was also investigated. In \cite{Zhao17}, 32 GHz channel measurements were conducted in outdoor microcells with rotated horn antennas. The QuaDRiGa open source model was extended and applied to the mmWave channel modeling. In \cite{Zhou17}, 28 GHz channel measurements were conducted in outdoor courtyard scenario with rotated horn antennas to obtain the path loss model. In \cite{Wang19_4, Zhang19}, channel measurements were conducted at 28 GHz and 39~GHz in outdoor urban macro (UMa) and urban micro (UMi) environments. The path loss, RMS delay spread, and angular spread were investigated.  

Most of the above-mentioned mmWave channel measurements are conducted with single antenna at both transmitter (Tx) and receiver (Rx) sides with several fixed locations. In general, horn antennas are used at both sides to compensate for the high path loss and sometimes the horn antenna is rotated at one side or both sides to obtain the angular information. For such channel measurement configurations, the environment needs to be quasi-static as it is time consuming for the data recording. It is not able to measure highly dynamic environments because the channel coherence time will be much smaller than the data recording time. Thus, the MIMO and V2V characteristics are not investigated. 

In the literature, only a few mmWave channel sounders have the ability to measure MIMO channel in time-variant environments, such as the channel sounders from Durham University \cite{Sana16}, National Institute of Standards and Technology (NIST) \cite{Pap16, Cau19}, and University of South California \cite{Wang18, Bas19, Wang19}. In the following, related mmWave outdoor channel measurements with MIMO or V2V characteristics are reviewed.

In \cite{Blu17_2, Blu18, Rah19, Cha19, Zoc19}, mmWave vehicular channel measurements were conducted at 60 GHz band. Several conditions of vehicular channels were measured, such as the in-vehicle, infrastructure-to-infrastructure (I2I), vehicle-to-infrastructure (V2I), and V2V channels. In \cite{Blu17_2}, the spatial consistency of the 60 GHz in-vehicle channel was investigated. The measured and calculated results indicated that a strong reverberation inside the vehicle produces similar power delay profiles (PDPs) within the range of approximately 10 wavelengths. In \cite{Blu18}, the 60 GHz V2I channel was measured in an urban highway environment. The channel was found to show clustering behaviour with a typical number of 4-5 reflected multipath components (MPCs). In \cite{Rah19}, the Doppler spread for 60 GHz I2I channel was studied. In \cite{Cha19}, the 60 GHz in-vehicle channel was measured and a tapped-delay-line (TDL) channel model was formulated from the proposed PDP model. In \cite{Zoc19}, the position-specific channel statistics for 60 GHz vehicular overtaking channel was investigated. In \cite{He19}, a survey of mmWave V2V channel propagation characteristics was given, in which the mmWave channel measurements were all with single antennas at both sides and only a few of them were in dynamic environments.

In \cite{Bas19, Wang17, Bas19_2}, outdoor channel measurements were conducted at 28 GHz with a phased-array channel sounder. The bandwidth is 400 MHz. At both Tx and Rx sides, the 8$\times$2 phased array was applied to measure the directional properties in dynamic environments. In \cite{Bas19}, channel measurements with moving scatterers and blocking objects were conducted to validate the performance of the channel sounder. In \cite{Wang17}, the stationary region was investigated at 28 GHz in a outdoor microcellular environment. In \cite{Bas19_2}, the O2I channel was measured. The path loss, delay spread, and angular spread for indoor and outdoor Rx locations for two different types of buildings were studied.

MmWave channel shows many new channel propagation characteristics, especially the space-time-frequency non-stationarity, cluster birth-death, spatial consistency, etc. The mmWave channel measurements are still insufficient for many potential applications, such as indoor hotspot, outdoor dynamic environments, MIMO and beamforming, and vehicular networks. MmWave communication systems will be deployed at multiple frequency bands and in multiple scenarios with multiple antennas. Thus, it is important to measure the real channels with such configurations for the evaluation of B5G communication systems. 
However, most of the above-mentioned mmWave channel measurements are conducted at single frequency band, in one scenario, or equipped with single antenna. To fill the above gaps, we conduct multi-frequency multi-scenario mmWave MIMO channel measurements and propose corresponding models in this paper. The major contributions and novelties include:

\begin{itemize}
\item[•] MmWave MIMO channel measurements are conducted at multi-frequency bands and in multi-scenarios by using the same channel sounder. Specifically, the human body and vehicle blockage measurements, outdoor path loss measurements, and V2V measurements are conducted with 4$\times$4 antennas at 28, 32, and 39 GHz bands.

\item[•] The channel characteristics, including blockage effect, path loss and coverage range, and non-stationarity and spatial consistency, are thoroughly studied. 

\item[•] The mmWave MIMO channel characteristics are compared among different frequency bands and scenarios. 

\item[•] The blockage model, path loss model, and time-varying channel model are proposed for mmWave MIMO channels. 

\end{itemize}	 	

The remainder of this paper is organized as follows. In Section II, a detailed description of the mmWave MIMO channel measurements are given, including blockage measurements, outdoor path loss measurements, and V2V measurements. MmWave channel models, including blockage model, outdoor path loss model, and time-varying channel model are introduced in Section III. In Section IV, the measurement and modeling results are thoroughly analyzed. Finally, conclusions and future works are given in Section V.

\section{MmWave MIMO Channel Measurements}
\subsection{MmWave MIMO channel sounder}
\begin{figure}[tb!]
\centering
\includegraphics[width=3.5in]{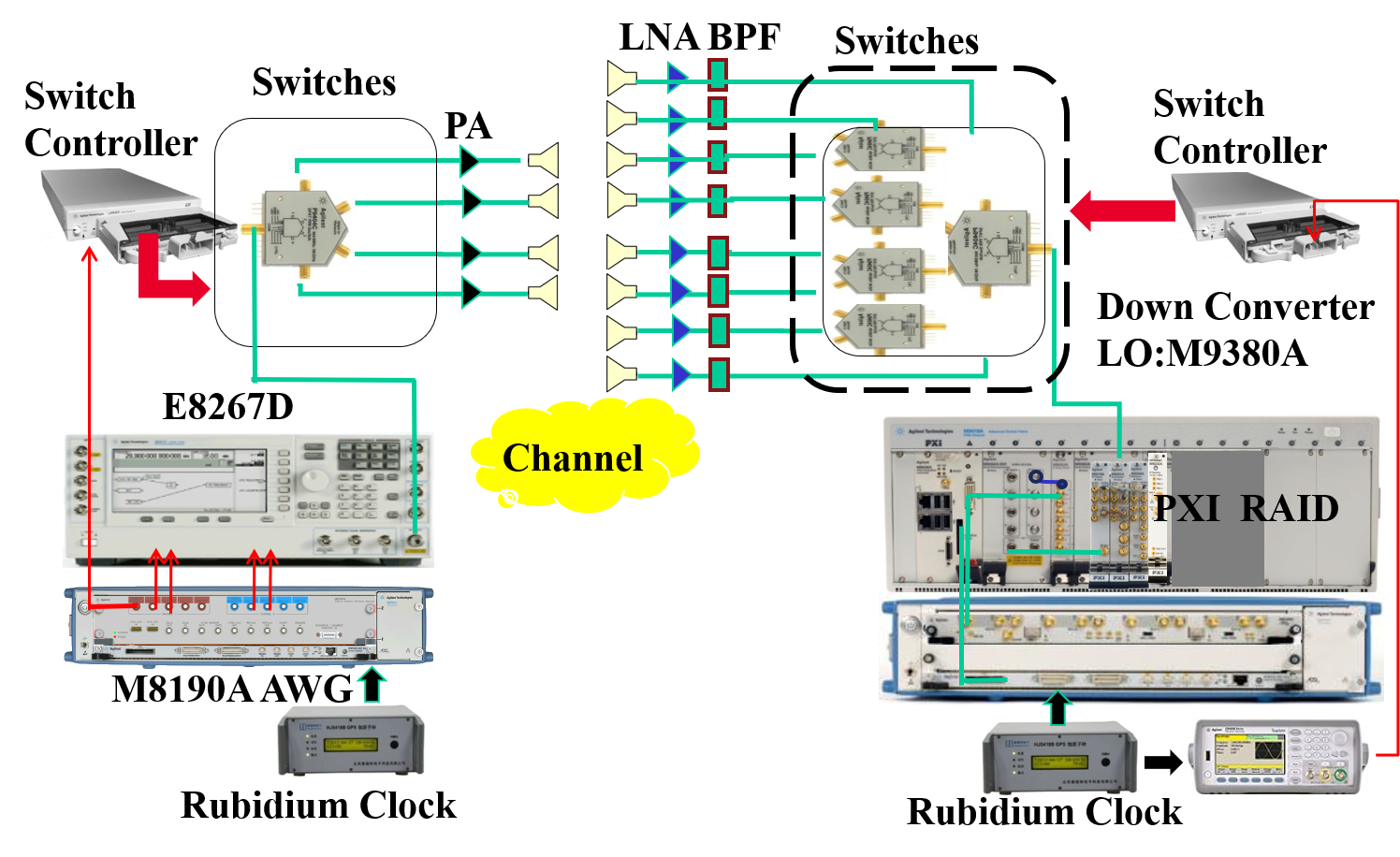}
\caption{The Keysight time domain channel sounder.}
\label{fig:sounder}
\end{figure}

As shown in Fig. \ref{fig:sounder}, the Keysight time domain channel sounder is used to conduct multi-frequency multi-scenario mmWave MIMO channel measurements. A summary of the detailed equipment and parameters is given in Table \ref{tab_equip}. The Tx side consists of a M8190A arbitrary waveform generator (AWG) with sampling rate of 12 GSa/s, an E8267D vector signal generator (VSG) with frequency range of 100 kHz to 44 GHz, an L4450A high speed four-way switch with switching time of less than 1 $\mu s$, a high-precision HJ5418 GPS Rubidium clock, Ainfo Tx antennas, and Connphy power amplifiers (PAs). The Tx side can switch in four channels in serial. The Rx side consists of a M9362A PXIe down converter, a M9352A PXI hybrid amplifier/attenuator, a M9300A PXIe frequency reference, a M9703B AXIe 12-bit digitizer, an E8257D analog signal generator, an L4450A high speed switch controller, a GPS Rubidium clock, Ainfo Rx antennas, and Connphy low noise amplifiers (LNAs). The Rx side can switch in four group of channels in serial and each group includes four channels in parallel, thus enables the 4$\times$16 MIMO channel measurements. The GPS Rubidium clocks at Tx and Rx sides are utilized to synchronize the 1 pulse per second (PPS) signal with the coordinated universal time (UTC). The latitude and longitude locations of Tx and Rx are recorded by the GPS. The channel sounder has wide frequency range, high bandwidth, high sampling speed, high dynamic range, and multiple antenna capabilities, thus fulfils the multi-frequency multi-scenario mmWave MIMO channel measurement requirements. 

\begin{table}[bt!]
\caption{A summary of channel sounder equipment and parameters.}
\setlength{\tabcolsep}{3pt}
\begin{tabular}{|p{120pt}|p{120pt}|}
\hline
Equipment& Parameter\\
\hline
M8190A AWG&	12 GSa/s sampling rate\\
\hline
E8267D VSG&	100 kHz-44 GHz band\\
\hline
L4450A four-way switch&	$<$1 $\mu s$ switching time\\
\hline
HJ5418 GPS Rubidium clock&	$<$1$\times$10$^{-12}$ precision\\
\hline
Connphy PA&	26.5-40 GHz band, 24 dBm output power, 30 dB gain\\
\hline
Connphy LNA&	26.5-40 GHz band, 30 dB gain\\
\hline
Ainfo horn antenna&	26.5-40 GHz band, 20/25 dBi gain\\
\hline
M9362A PXIe down converter&	10 MHz-50 GHz band\\
\hline
M9352A PXI hybrid amplifier/attenuator&	10 MHz-1 GHz band\\
\hline
M9300A PXIe frequency reference&	10 MHz and 100 MHz outputs\\
\hline
M9703B AXIe digitizer&	12-bit resolution, up to 3.2 GS/s, DC to 2 GHz input frequency range\\
\hline
E8257D analog signal generator&	100 kHz-67 GHz\\
\hline
\end{tabular}
\label{tab_equip}
\end{table}

Due to the limitations of the available number of Rx antennas and LNAs, only one group of the Rx channels is utilized to conduct 4$\times$4 mmWave MIMO channel measurements.

Directional horn antennas working in the range of 26.5-40~GHz are utilized. At the Tx side, the four antennas with 20 dBi gain and 18$^\circ$ half power beamwidth (HPBW) are connected to four PAs. The PA has a 30 dB gain and 24 dBm output P1dB power. The effective isotropic radiated power (EIRP) is 44~dBm. At the Rx side, the four antennas are connected to four LNAs and toward the same direction with separation distance about 7.5 cm to enable the multiplexing gain. The LNA has a 30 dB gain. Both 20 dBi and 25 dBi horn antennas are used at Rx side. The 25 dBi horn antenna has a HPBW of 10$^\circ$. The 25~dBi gain horizontal polarized, 25 dBi gain vertical polarized, 20 dBi gain vertical polarized, and 20 dBi gain horizontal polarized antennas are connected to Rx1 to Rx4, respectively. The polarization effects, i.e., cross-polarization ratio (XPR) can be investigated.

To enlarge the system dynamic range, a pseudo noise (PN) sequence with length of 2$^{12}$ = 4096 is used to obtain processing gain of $10log_{10}2^{12}$ = 36 dB. In addition, 104 and 800 zero points are added at the start and end of the waveform, respectively, to leave the time for antenna switching. Both the in-phase (I) and quadrature (Q) components are recorded. To further reduce the inter symbol interference (ISI), the transmitted waveform is interpolated with four times and filtered by a root raised cosine (RRC) pulse shaping filter with roll-off factor of 0.5. The received signal is then filtered by using the same matched RRC filter and down-sampled. The signal bandwidth is 1~GHz and the sampling rate is 1.28 GS/s. The length of the interpolated waveform is 20000 points with duration of 15.625 $\mu s$ for one snapshot. For the 4$\times$4 MIMO channel, it needs 62.5 $\mu s$ to complete one period. The waveform is then repeated several times to capture large number of snapshots for time-varying channels and also to improve the signal-to-noise ratio (SNR). The channel sounder also enables streaming mode for continuous data recording with a signal bandwidth of 320 MHz for several seconds. In streaming mode, the received signal is recorded by a digital down converter (DDC) with a ratio of 4 and a sampling rate of 1.6 GS/s. The signal is then saved to a redundant array of independent disk (RAID) with a rate up to 1 GB/s.  

\subsection{MmWave blockage measurements}
The 4$\times$4 mmWave MIMO blockage measurements are conducted in both indoor and outdoor scenarios at 28, 32, and 39 GHz bands with several configurations. 

\subsubsection{Indoor human blockage measurements} 
\begin{figure} [tb!]
\centering 
\subfigure[One person crosses the Tx-Rx path] { \label{fig:a} 
\includegraphics[width=3.5in,height=1.1in]{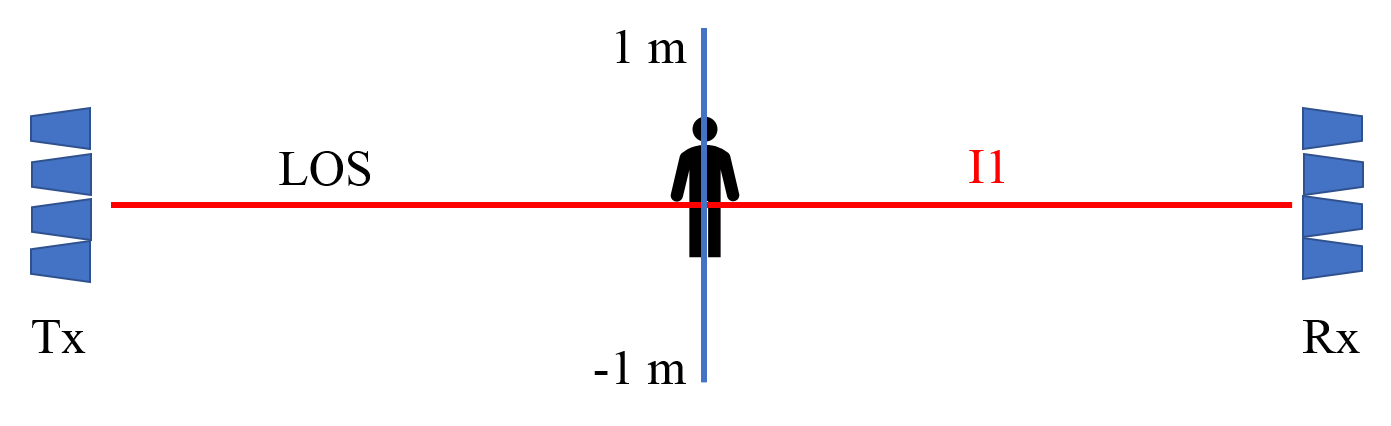} 
} 
\subfigure[Two persons cross the Tx-Rx path] { \label{fig:b} 
\includegraphics[width=3.5in,height=1.1in]{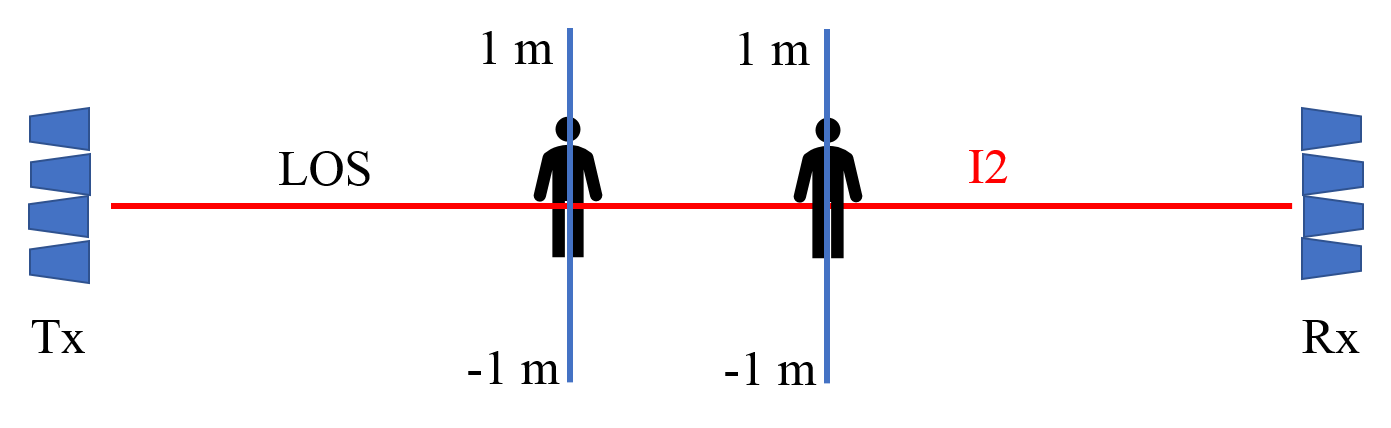} 
} 
\subfigure[One person travels along the Tx-Rx path] { \label{fig:c} 
\includegraphics[width=3.5in,height=0.9in]{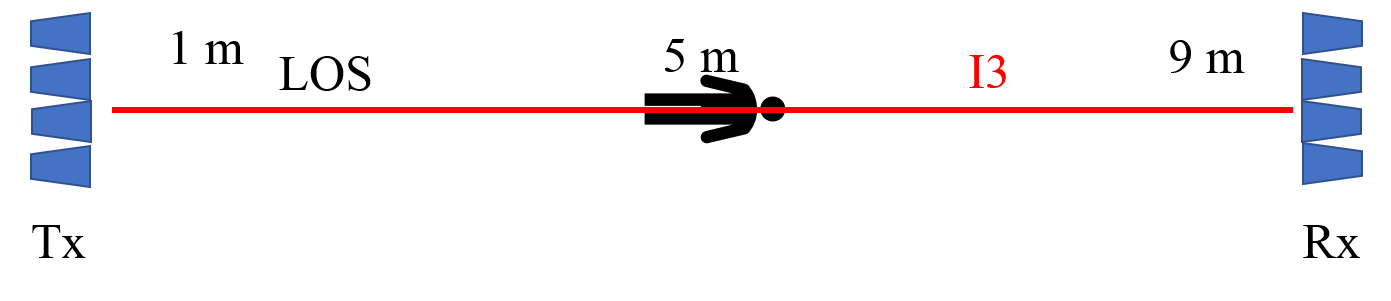} 
} 
\caption{An illustration of the indoor human blockage measurements.} 
\label{fig:block} 
\end{figure}

For the indoor scenario, Tx and Rx are separated with 10 m in a corridor environment. The four Tx antennas are toward the same direction. The measurements include three schemes: one person walks perpendicular to the line-of-sight (LOS) path at the center and from -1 m to 1 m with 0.1~m step size (I1), two persons walk perpendicular to the LOS path, both with 1 m to the center and from -1 m to 1 m with 0.1 m step size (I2), and one person walks along the LOS path from 1 m to 9 m with 0.2 m step size (I3). Channel measurements are conducted at 28, 32, and 39 GHz bands with the three schemes. An illustration of the measurements is shown in Fig. \ref{fig:block}. 


\begin{figure*}[bt!]
\centering
\begin{minipage}[t]{0.3\linewidth}
\centerline{\includegraphics[width=2.4in]{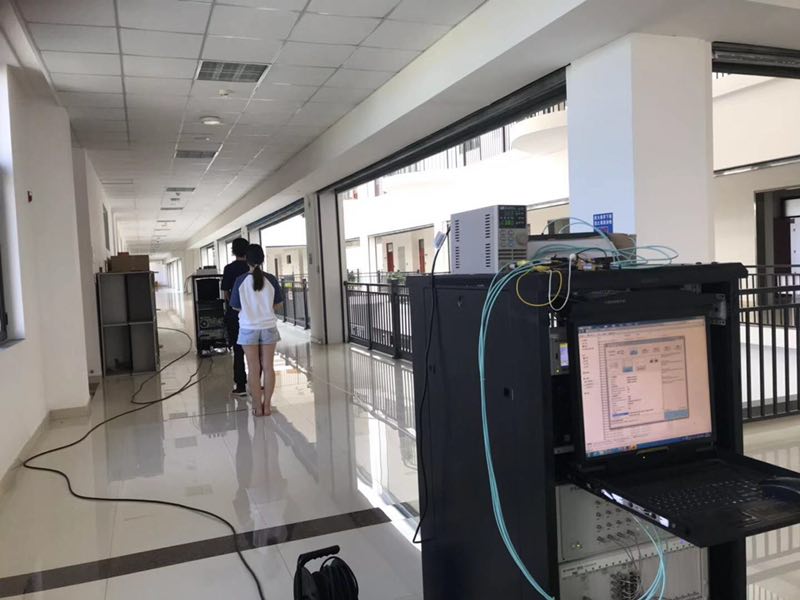}}
\footnotesize \centerline{(a) Indoor human blockage measurements}
\end{minipage}
\begin{minipage}[t]{0.3\linewidth}
\centerline{\includegraphics[width=2.4in]{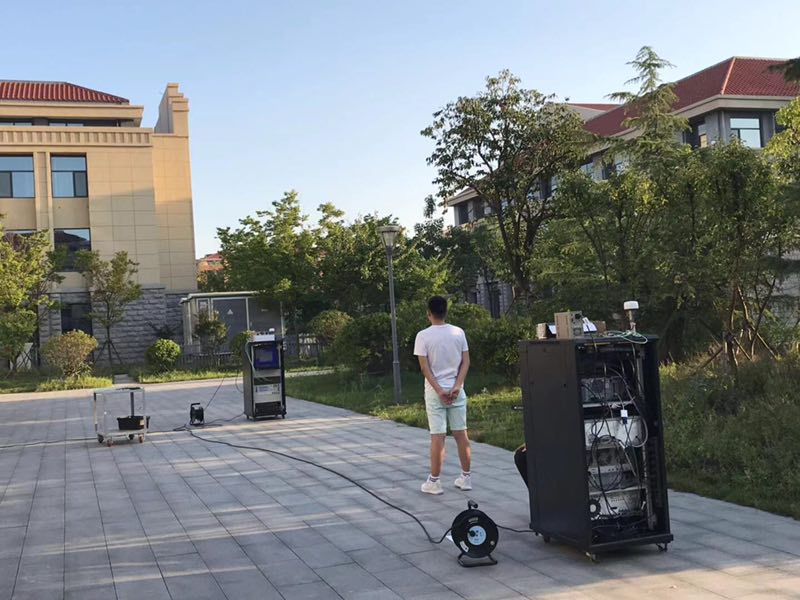}}
\footnotesize \centerline{(b) Outdoor human blockage measurements}
\end{minipage}
\begin{minipage}[t]{0.3\linewidth}
\centerline{\includegraphics[width=2.4in]{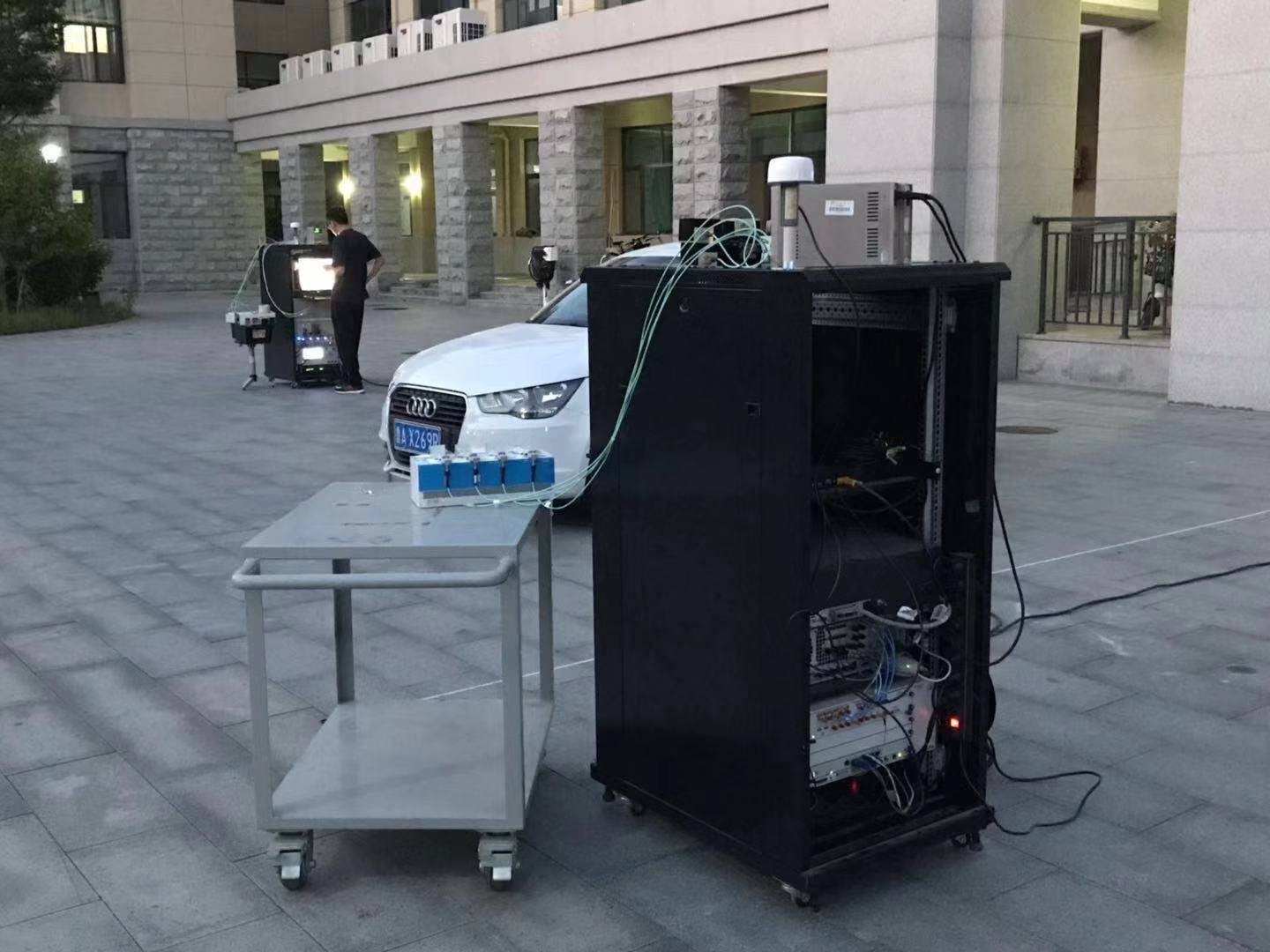}}
\footnotesize \centerline{(c) Outdoor vehicle blockage measurements}
\end{minipage}
\caption{The measurement scenarios for indoor human blockage, outdoor human blockage, and outdoor vehicle blockage measurements.}
\label{fig:blo}
\end{figure*}

\subsubsection{Outdoor human blockage measurements}
For the outdoor scenario, Tx and Rx are separated with 10 m in an open square environment which is surrounded by vegetation. The schemes are the same with indoor measurements. Channel measurements are conducted at 28 GHz and 32 GHz bands. As the human body is not only blocking the LOS component, but also may have relations with the reflected paths, the received signal with human blockage is an interaction with the human body and the environment. The human blockage between indoor and outdoor scenarios may have some differences, which are not studied in the literature and will be compared.

\subsubsection{Outdoor vehicle blockage measurements}
For outdoor scenarios, vehicles are very common objects, such as car, van, and truck. A car is relatively large in size compared to the wavelength of mmWave bands. It has a smooth surface and is composed of painted metals, glass windows, wheels, etc. Its interaction with mmWave is complex and may encounter different propagation mechanisms such as reflection, scattering, and diffraction \cite{Li19}. Here, channel measurements are conducted at 28, 32, and 39 GHz bands. A car is between the Tx and Rx. The Rx moves from the tail of the car to the head of the car with 0.2 m step size.

The measurement scenarios for indoor human blockage, outdoor human blockage, and outdoor vehicle blockage measurements are shown in Fig. \ref{fig:blo}.

\subsection{MmWave outdoor path loss measurements}

\begin{figure}[tb!]
\centering
\includegraphics[width=3.3in]{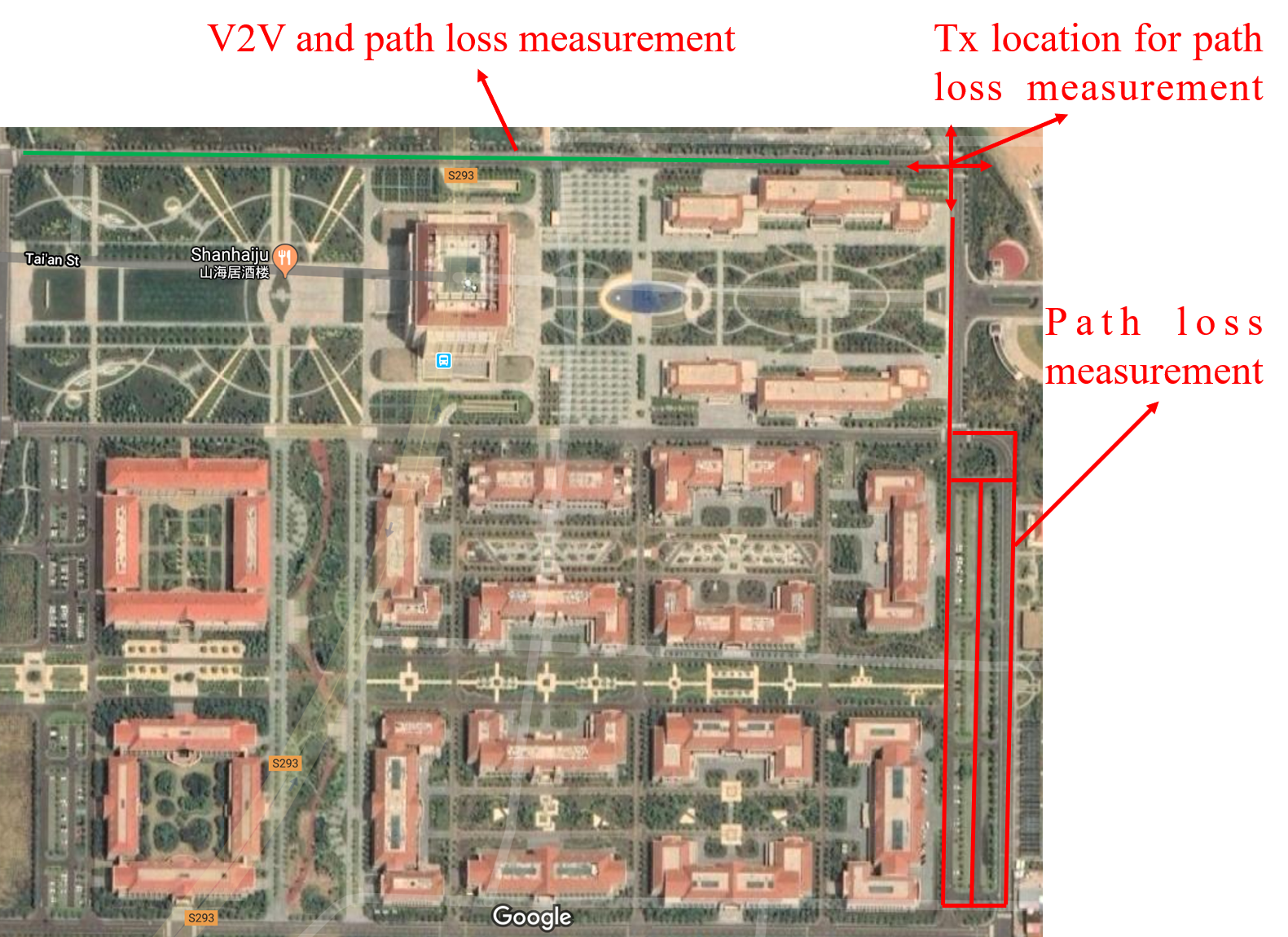}
\caption{Outdoor path loss and V2V channel measurement environment.}
\label{fig:map}
\end{figure}

The outdoor path loss measurements are conducted in Shandong University, Qingdao campus. It's a newly built modern campus with tall buildings and can be seen as an UMa scenario. The road is about 7 m wide with two lanes. As shown in Fig. \ref{fig:map}, the Tx is fixed around the crossroads with the four horn antennas toward four directions with 90$^\circ$ separation. The purpose is to mimic the base station (BS) with beamforming capability and study the coverage range for different directions. For the east-west road, buildings, open squares, and vegetation is distributed on one side, and an open space with vegetation is on the other side. For the south-north road, buildings, open squares, and vegetation is distributed on one side, and the university gate, an open square, and vegetation is distributed on the other side. The 1-4 Tx antennas are toward the west, north, east, and south, respectively. The Rx is on a pickup truck and moving on the two main roads back and forth. Channel measurements are conducted at 28, 32, and 39 GHz bands with 4$\times$4 MIMO antenna configurations. The maximum distance between Tx and Rx is about 800 m. In the measurement, the Rx antennas are toward the Tx location for most of the time and may be blocked by objects in the environment. The Rx first moves to the west and goes back to the Tx. Then, the Rx moves to the south and goes back to the Tx again. Signals are recorded along the movement route of Rx and the measurements are repeated for each frequency band. About 80 positions are recorded for each band.

\subsection{MmWave V2V measurements}

The V2V channel measurements are conducted with both Tx and Rx moving on the straight road with various relative velocities range from 0 km/h to 60 km/h. As the channel is highly dynamic, the streaming mode is used for continuous data collection for several seconds. For this measurement, only one Rx antenna is used to receive the signal and the four Tx antennas are used to transmit the signal in serial. Channel measurements are conducted at 28, 32, and 39 GHz bands for both same moving direction and opposite moving direction configurations. The purpose of the measurement is to capture the non-stationary and spatial consistency characteristics of mmWave V2V channels.

\section{Measurement Data Post-processing and MmWave MIMO Channel Modeling}
\subsection{Measurement data post-processing}
Denote the transmit signal as $x(t)$, the response of the measurement system as $g(t)$, and the channel impulse response (CIR) as $h(t)$. Note that the transmit signal is a time-invariant PN sequence, the response of the measurement system can be viewed as time-invariant, and the CIR for each snapshot is also assumed to be time-invariant as the data recording time is much less than the coherence time of the wireless channel. Thus, the wireless channel in each snapshot can be seen as a linear time-invariant (LTI) system, though the channel is time-invariant for a longer time duration, especially for mmWave high-mobility channel. The back-to-back calibration signal is

\begin{equation}
y_{th}(t)=x(t)*g(t)
\end{equation}
where * is the convolution operation. The received signal is 
\begin{equation}
y_{rx}(t)=x(t)*g(t)*h(t).
\end{equation}

The frequency domain response can be obtained from Fourier transform as
\begin{equation}
Y_{th}(f)=X(f)G(f)
\end{equation}

\begin{equation}
Y_{rx}(f)=X(f)G(f)H(f).
\end{equation}
The CIR is then obtained from inverse Fourier transform as
\begin{equation}
h(t)=IFFT(H(f))=IFFT(Y_{rx}(f)/Y_{th}(f)).
\end{equation}

The peak search algorithm is applied to extract MPCs in delay domain with a maximum number of 30. The maximum value between maximum power minus 25 dB and average noise floor plus 6 dB is used as the power threshold. The received power is given as 
\begin{equation}
P=\sum_{l=1}^{L}P_l
\end{equation}
where $L$ is the number of MPCs and $P_l$ is the power of the $l$-th path.

The path loss is 
\begin{equation}
PL (dB)=-P+G_t+G_r+P_{PA}-P_{t}+G_{LNA}-L_{c}
\end{equation}
where $G_t$ (20 dBi) and $G_r$ (20/25 dBi) are the gains of Tx and Rx antenna, respectively. $P_{PA}$ is the P1dB power of the PA (24 dBm), 
$P_t$ is the transmitted power for calibration (5 dBm), $G_{LNA}$ is the gain of LNA (30 dB), and $L_c$ is the additional loss of the cables in channel measurements (4 dB). 

The RMS delay spread is calculated as 
\begin{equation}
DS=\sqrt{\frac{\sum_{l=1}^{L}P_l\tau_{l}^{2}}{\sum_{l=1}^{L}P_l}-(\frac{\sum_{l=1}^{L}P_l\tau_{l}}{\sum_{l=1}^{L}P_l})^2}
\end{equation}
where $\tau_l$ is the delay of the $l$-th path.

\subsection{Blockage model}

In general, the human body is simulated as a rectangular absorbing screen or a perfectly conducting cylinder to study the blockage effects \cite{Qi17}. The knife-edge diffraction (KED) and geometrical theory of diffraction (GTD) models are applied and introduced.

\subsubsection{KED model}

The KED model assumes a human blocker to be represented as a rectangular screen which has a absorbing property. The METIS KED model \cite{KEDMETIS} and Kirchhoff KED model are two widely used methods.

The METIS KED model can calculate the loss caused by human blockage simply and accurately \cite{mmMAGIC}. A rectangular absorbing screen is used to simulate the human body. The screen is vertical to the ground and the screen orientation is parallel to the projection of the Tx--Rx connecting line in the top projection view. The edges on the left, right, top, and bottom of the screen are used to calculate the shadowing loss and denoted as $A_{l}$, $A_{r}$, $A_{t}$, and $A_{b}$, respectively, whose detailed calculations can be referred to \cite{Qi17}. The total loss can be obtained by

\begin{equation}
\label{EQ_2}
A (dB) =-20log_{10}(1-(A_{l}+A_{r})(A_{t}+A_{b})).
\end{equation}

One of the merits of the METIS KED model is that it can model the losses when the signal is blocked by several persons simultaneously. When the distribution of the persons is sparse, the total loss is summation of the multiple screens whose loss can be calculated as mentioned above.

In the Kirchhoff KED model \cite{Hri00, Fon07}, the Kirchhoff diffraction equation is used to simulate loss caused by human blockages. Assume $S\emph{}$ is a screen extended infinitely in the X--Y plane, $S_{0}$ refers to the aperture on the screen, $Q_{0}$ ($x_0$, $y_0$, $z_0$) is the intersection point of the Tx--Rx connecting line with the aperture, $d_{1}$ and $d_{2}$ denote the projection of Tx--$Q_0$ and Rx--$Q_0$ on Z axis, respectively.

The first Fresnel zone radius can be calculated as

\begin{equation}
\label{EQ_3}
R_{1}=\sqrt{\lambda\frac{d_{1}d_{2}}{d_{1}+d_{2}}}
\end{equation}
where $\lambda$ is the wavelength. The attenuation caused by the aperture can be calculated using the Kirchhoff diffraction equation, i.e.,

\begin{equation}
\label{EQ_5}
A=F_{d}(u,v)=\frac{j}{2}\iint_{S_{0}}^{}exp[-j\frac{\pi}{2}(u^{2}+v^{2})]dudv
\end{equation}
where $F_{d}(u,v)$ represents the Fresnel number. The parameters $u$ and $v$ can be written as

\begin{equation}
\label{EQ_6}
u=\sqrt{2}\frac{x-x_{0}}{R_{1}}
\end{equation}

\begin{equation}
\label{EQ_7}
v=\sqrt{2}\frac{y-y_{0}}{R_{1}}.
\end{equation}

\subsubsection{GTD model}
\label{GTD}

The GTD model simulates the human body as a perfectly conducting cylinder and calculates the attenuation as \cite{GTDbook}

\begin{equation}
\label{EQ_8}
\begin{split}
A=&\sum_{n=1}^{N}D_{n}^{e}\frac{exp(-jkS_{d})}{\sqrt{8jkS_{d}}}\\
&\times \{exp[-(jk+\Omega_{n}^{e})\gamma_{1}]+exp[-(jk+\Omega_{n}^{e})\gamma_{2}]\}
\end{split}
\end{equation}
where $N$ is the selected number of zero values of Airy function $Ai()$, $k$ is the wavenumber, $S_d$ is the distance between the Rx antenna and the point of tangency of the cylinder, $\Omega_{n}^{e}$ is the attenuation constant, $\gamma_{1}$ and $\gamma_{2}$ are the travel distances on the surface of the circle for the incident rays. The parameter $D_{n}^{e}$ is the amplitude weighting factor and can be calculated as

\begin{equation}
\label{EQ_15}
D_{n}^{e}=2M\{Ai'(-\alpha_{n})\}^{-2}e^{-\pi/6}
\end{equation}
where $Ai'()$ is the derivative function of the Airy function $Ai()$, $-\alpha_{n}$ is the zeros of the Airy function $Ai()$, and the parameter $M$ can be written as

\begin{equation}
\label{EQ_16}
M=(\frac{ka}{2})^{1/3}
\end{equation}
where $a$ is the radius of the cylinder. The attenuation constant $\Omega_{n}^{e}$ can be obtained as

\begin{equation}
\label{EQ_17}
\Omega_{n}^{e}=\frac{\alpha_{n}}{a}Me^{j\pi/6}.
\end{equation}

\subsection{Outdoor path loss model}

The alpha-beta-gamma (ABG) path loss model and close-in (CI) path loss model are two widely used models. The ABG model is
\begin{equation}
PL(d,f)[dB]=10\alpha log_{10}(d)+\beta +10\gamma log_{10}(f)+N(0,\sigma)
\end{equation}
where $d$ is the three-dimensional (3D) direct distance between Tx and Rx (m), $f$ is the operating frequency (GHz), $\alpha$ is the coefficient associated with the increase of the path loss with distance, $\beta$ is coefficient associated with the offset value of the path loss, $\gamma$ is coefficient associated with the increase of the path loss with frequency, $N(0,\sigma)$ is Gaussian distribution with standard deviation $\sigma$ (dB). The ABG model reverts to the floating-intercept (FI) model when using at a single frequency with $\gamma$ set to 0.

The CI model is
\begin{equation}
PL(d,f)[dB]=32.4+20log_{10}(f)+ 10n log_{10}(d)+N(0,\sigma)
\end{equation}
where $n$ is the path loss exponent (PLE). The free space path loss at 1 m is used as the reference. 

Though the ABG model is used in 3GPP and ITU models, the CI model has an identical mathematical form, while offering intuitive appeal, better model parameter stability, and better prediction performance with fewer parameters \cite{Sun16_2}. In this paper, both the CI model and FI model are used.

\subsection{Time-varying channel model}
The general 3D non-stationary 5G channel model can be used for the time-varying channel modeling \cite{Wang18_5G}. It's a general geometry based stochastic model (GBSM) and can capture the small-scale fading channel characteristics for several scenarios, such as massive MIMO \cite{Wu14, Wu15}, high-speed train \cite{Gha15, Gha17, Bian18, Liu19}, V2V \cite{Yuan14, Yuan15, Bian19}, and mmWave \cite{Wu17, Huang18, Huang17} scenarios. The channel model is based on the WINNER II and Saleh-Valenzuela (SV) channel models considering array-time cluster evolution.

The CIR of the time-varying channel equipped with $M_T$ Tx antennas and $M_R$ Rx antennas can be given as
\begin{equation}
\begin{split}
&h_{qp}(t,\tau)=\sqrt{\frac{K(t)}{K(t)+1}}h_{qp,0}(t)\delta (\tau-\tau_0(t))\\
&+\sqrt{\frac{1}{K(t)+1}}\sum_{n=1}^{N(t)}\sum_{m=1}^{M(t)}h_{qp,nm}(t)\delta (\tau-\tau_n(t)-\tau_{nm}(t))
\end{split}
\end{equation}
where $K(t)$ is the Ricean K factor, $h_{qp,0}(t)$ is the amplitude of the LOS component, $\tau_0(t)$ is the delay of the LOS component, $N(t)$ and $M(t)$ are the number of clusters and the number of rays in each cluster, respectively. $h_{qp,nm}(t)$ is the amplitude of the non-LOS (NLOS) component, $\tau_n(t)$ and $\tau_{nm}(t)$ are the delays of the cluster and the ray within the cluster, respectively.

The distance vectors and angles are calculated from the geometry relationships and parameter computation methods. The generation of cluster powers and delays are similar to the WINNER model. An array-time cluster evolution framework based on birth-death process is applied to capture the non-stationary property.

\begin{figure} [tb!]
\centering 
\subfigure[28 GHz, along the Tx-Rx path] { \label{fig:a} 
\includegraphics[width=3.5in,height=2.2in]{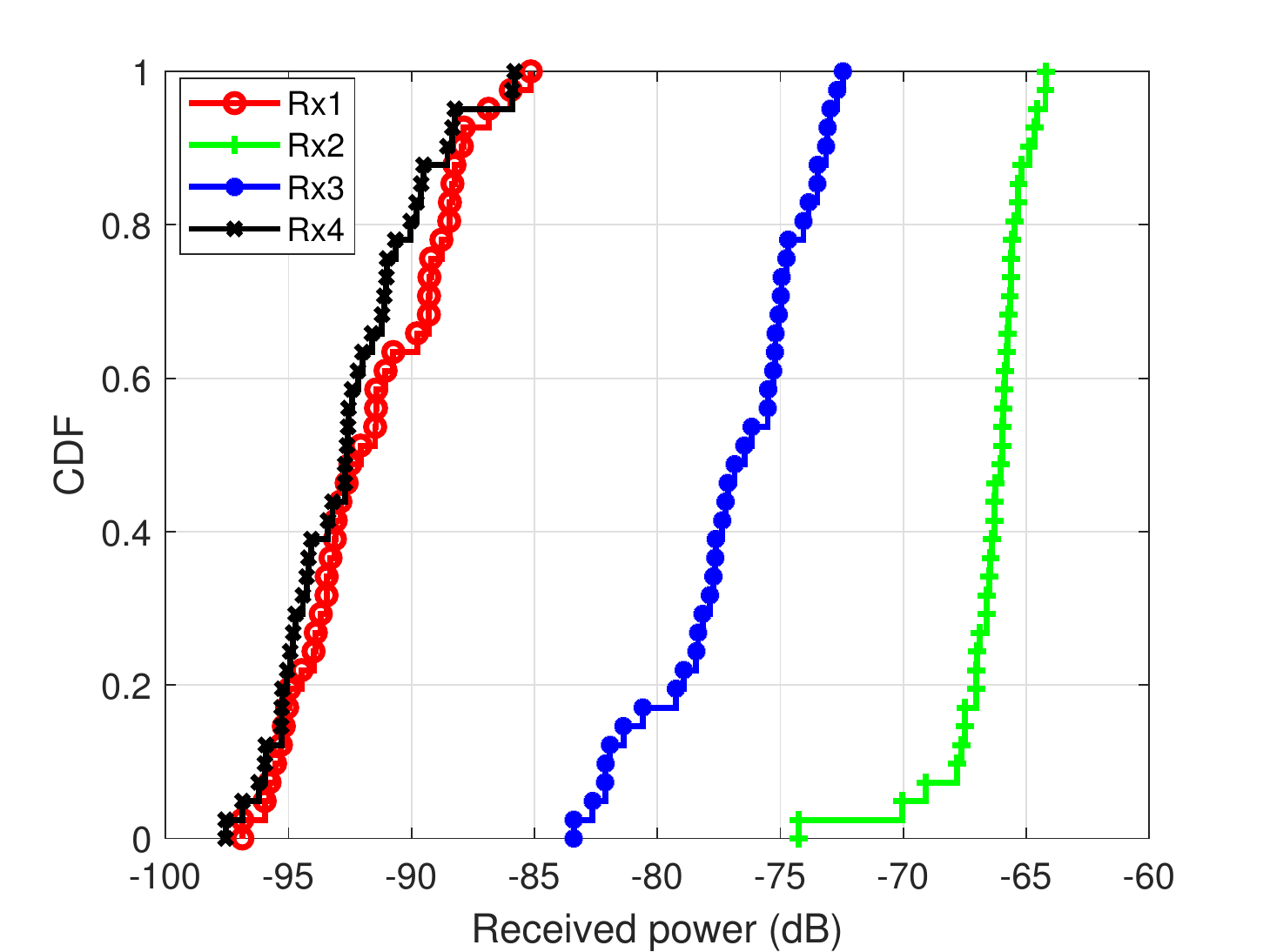} 
} 
\subfigure[32 GHz, along the Tx-Rx path] { \label{fig:b} 
\includegraphics[width=3.5in,height=2.2in]{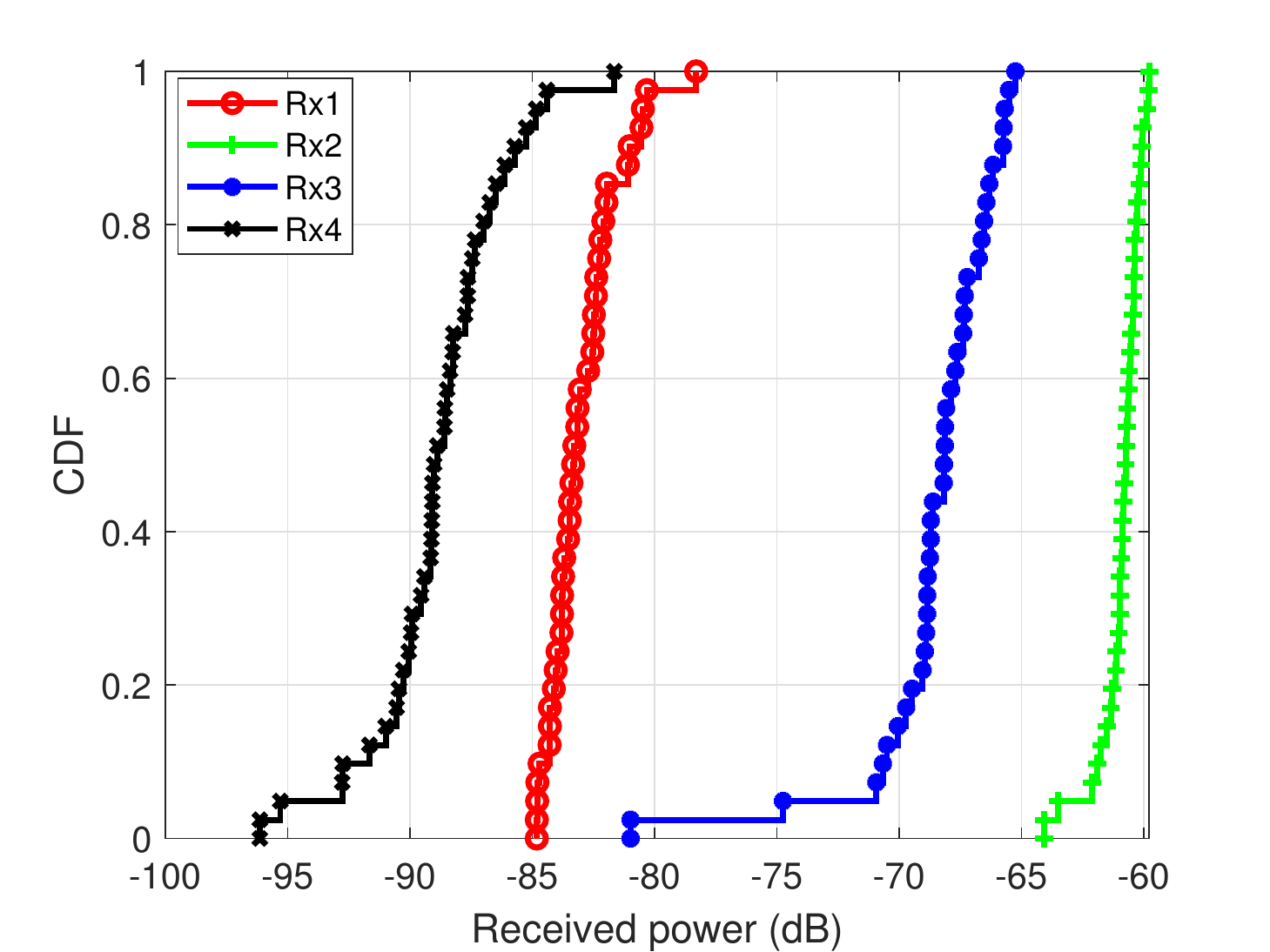} 
} 
\subfigure[39 GHz, along the Tx-Rx path] { \label{fig:c} 
\includegraphics[width=3.5in,height=2.2in]{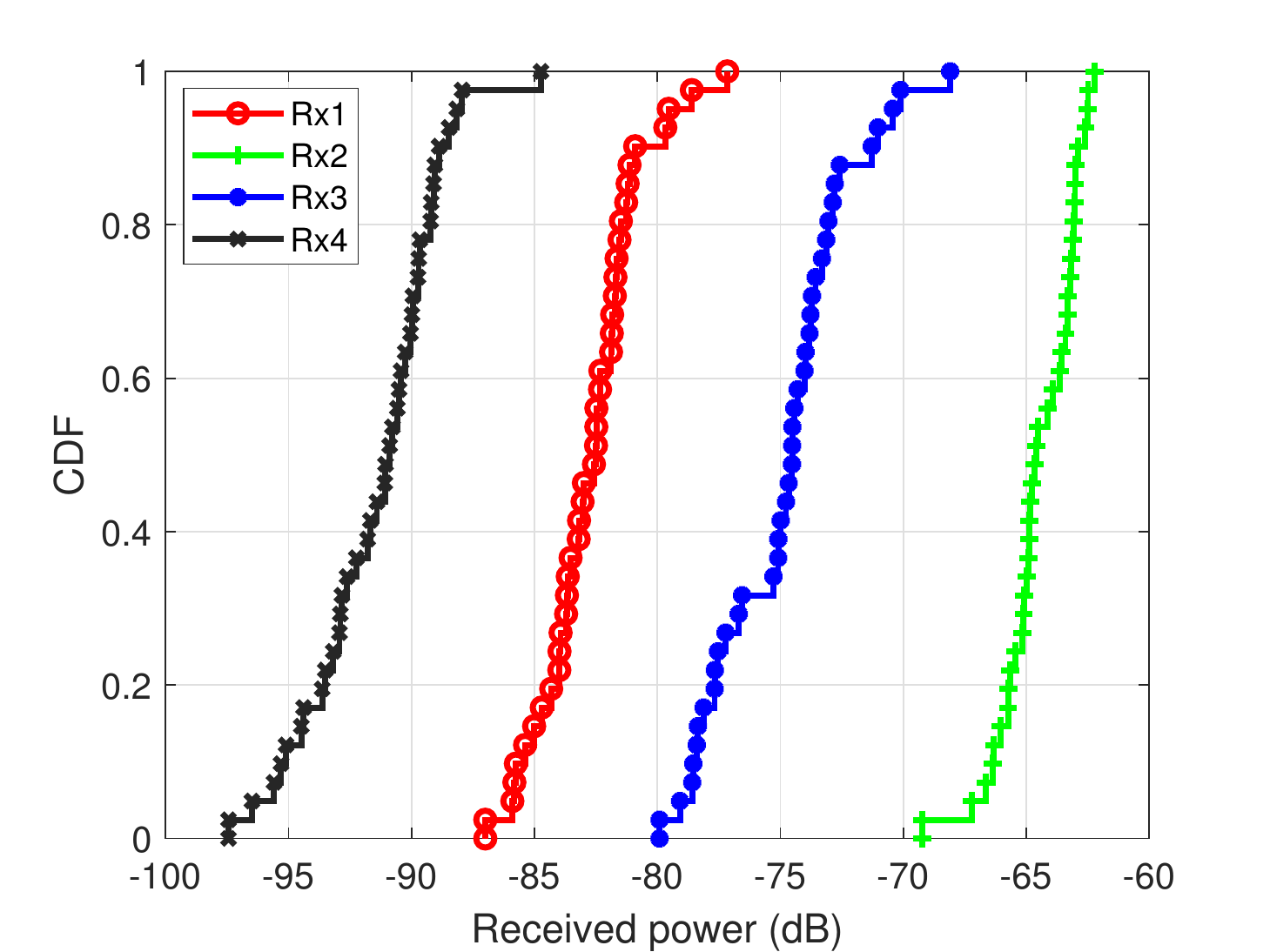} 
} 
\caption{Indoor human blockage measurement results along the Tx-Rx path at 28, 32, and 39 GHz bands.} 
\label{fig:ih_along} 
\end{figure}

\begin{figure} [tb!]
\centering 
\subfigure[Human blockage attenuations vs. distance to Tx along the Tx-Rx path for 28, 32, and 39 GHz bands using the GTD, METIS, and Kirchhoff models.] { \label{fig:a} 
\includegraphics[width=3.5in,height=2.2in]{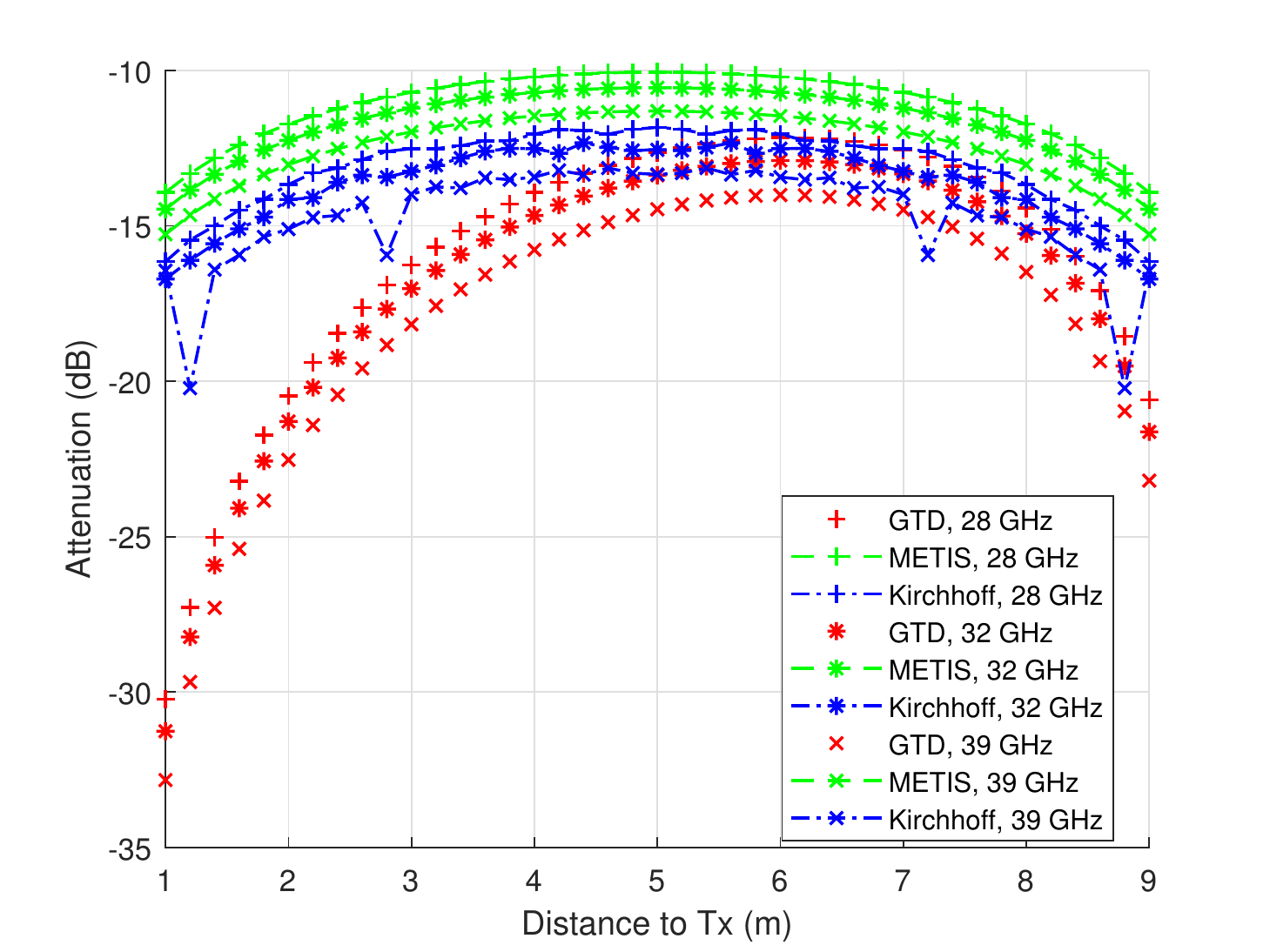} 
} 
\subfigure[Comparison of measured and modeled human blockage attenuations along the LOS path at 28 GHz band.] { \label{fig:b} 
\includegraphics[width=3.5in,height=2.2in]{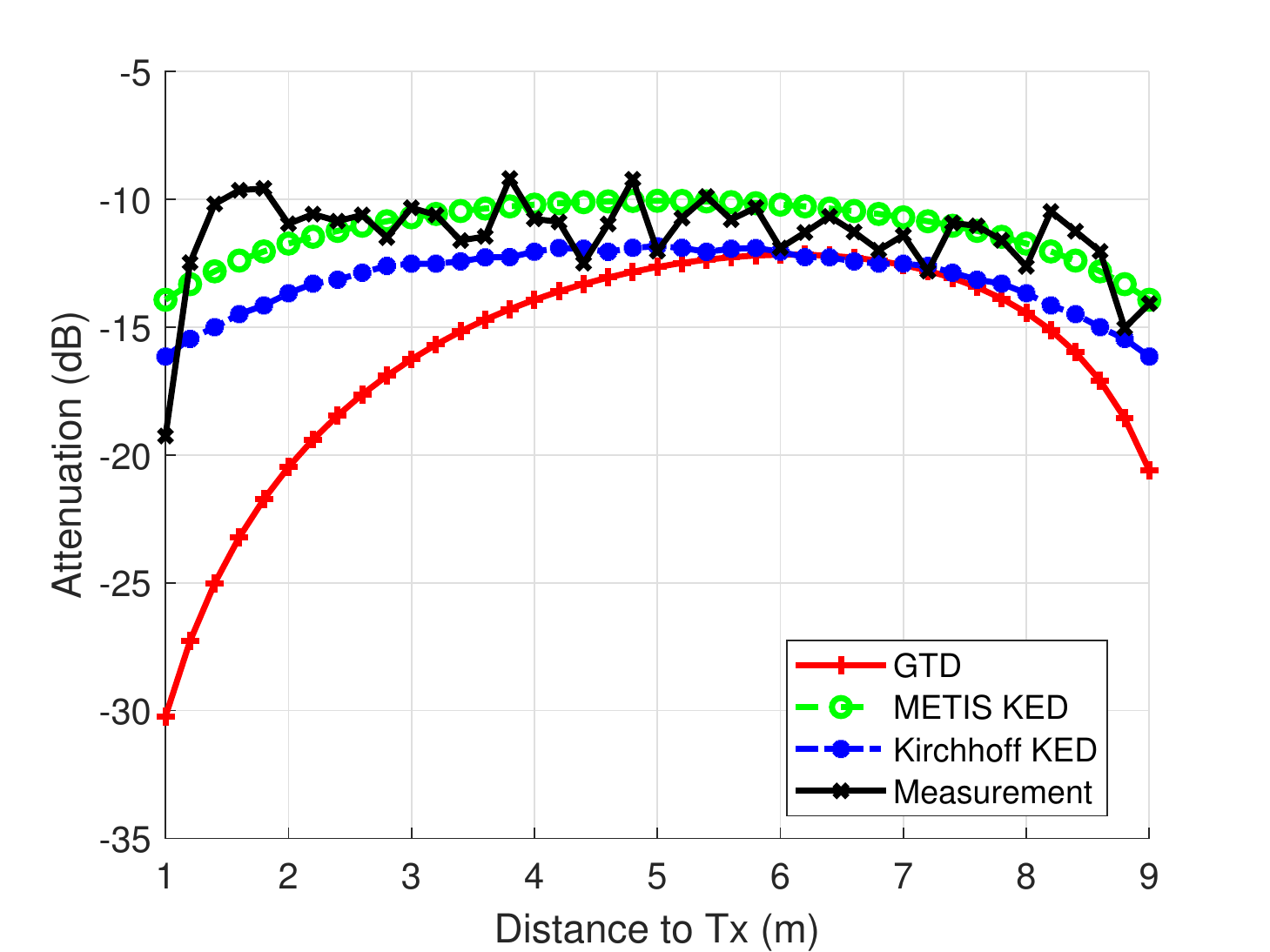} 
} 
\caption{Measured and modeled human blockage attenuations along the LOS path.} 
\label{fig:hb_along} 
\end{figure}

\section{Measurement and Modeling Results and Analysis}
\subsection{Blockage effect}
The indoor human blockage measurement results along the LOS path at 28, 32, and 39 GHz bands are shown in Fig.~\ref{fig:ih_along}. Specifically, the cumulative distribution functions (CDFs) of the received powers are presented. As the four Tx antennas are toward the same direction, the received signals are similar for different Tx antennas. Thus, only the received signals for Tx1 are illustrated. Note that the power is the calibrated power relative to 0 dBm. As Rx1 and Rx4 are with horizontal polarization, they have a lower signal level compared to Rx2 and Rx3 because vertical polarization has a higher signal level. The three frequency bands show similar human blockage attenuations. As the frequency increases, the attenuation also increases slightly. The attenuations along the LOS path can be seen as a constant value with some variations. In general, vertical polarization has smaller variations, which are about 5 dB, while it is about 10 dB for horizontal polarization. The reason may be that human body has a de-polarization effect on mmWave propagations. The XPR is about 15-25 dB for the three bands and similar for the 20 dBi and 25 dBi gain horn antennas.

Fig. \ref{fig:hb_along} shows the measured and modeled human blockage attenuations. The METIS model has the smallest attenuations, while the GTD model has the largest attenuations. As frequency increases from 28 GHz to 39 GHz, the modeled attenuation shows a slightly increasing trend. The attenuation difference between 28 GHz and 39 GHz is within 1 dB. The METIS KED model and Kirchhoff KED model show symmetry at Tx and Rx sides, while the GTD model shows larger attenuation near Tx side, which is about 10 dB larger than the Rx side. By comparing the measured and modeled human blockage attenuations, we can see that in the case of along the LOS path, the measured attenuation variates roughly between the METIS KED and Kirchhoff KED modeled attenuations, which means the METIS KED model and Kirchhoff KED model can be seen as the lower bound and upper bound of the attenuation, respectively.

\begin{figure} [tb!]
\centering 
\subfigure[28 GHz, one person crosses] { \label{fig:a} 
\includegraphics[width=0.45\columnwidth]{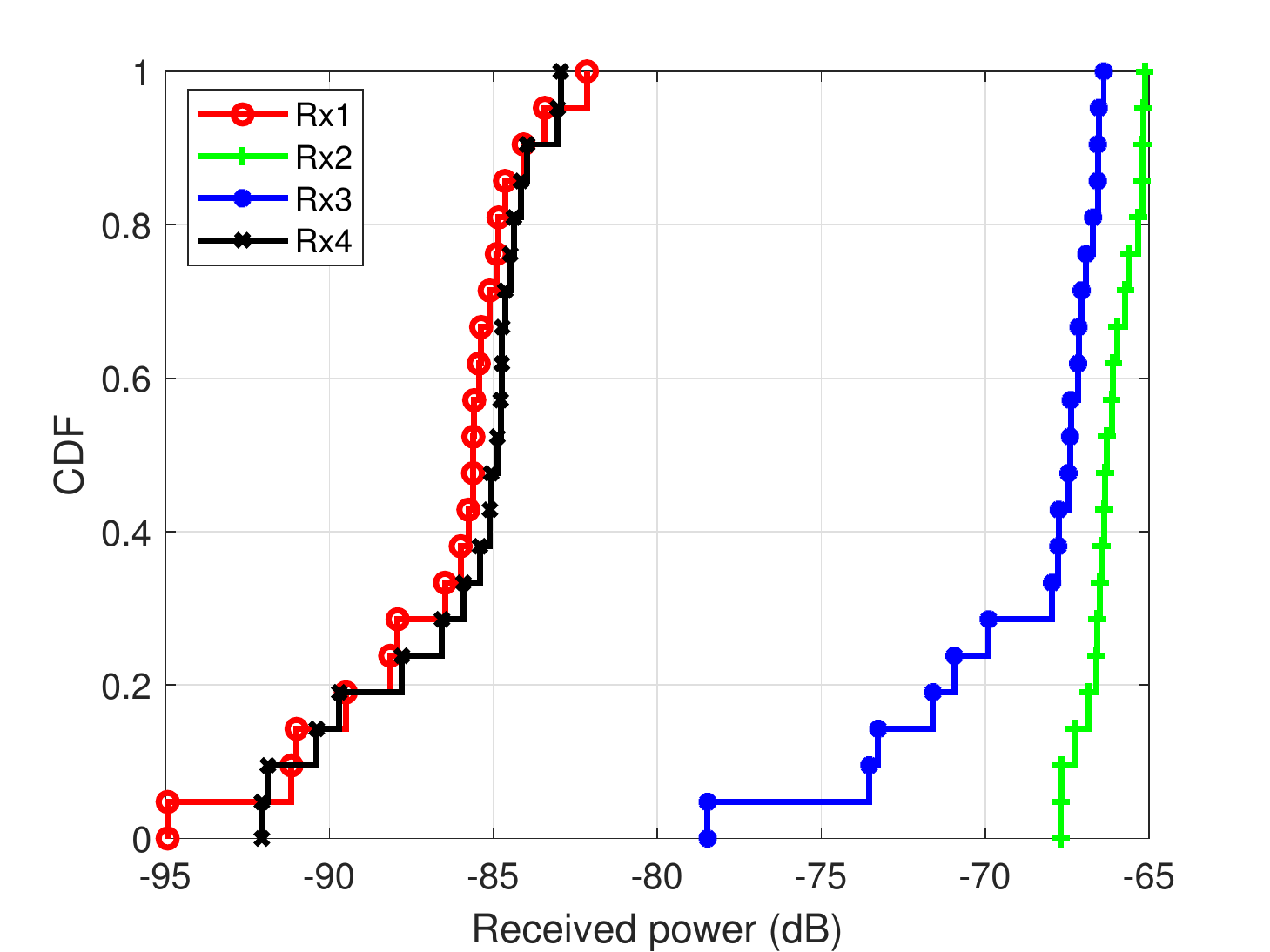} 
} 
\subfigure[28 GHz, two persons cross] { \label{fig:b} 
\includegraphics[width=0.45\columnwidth]{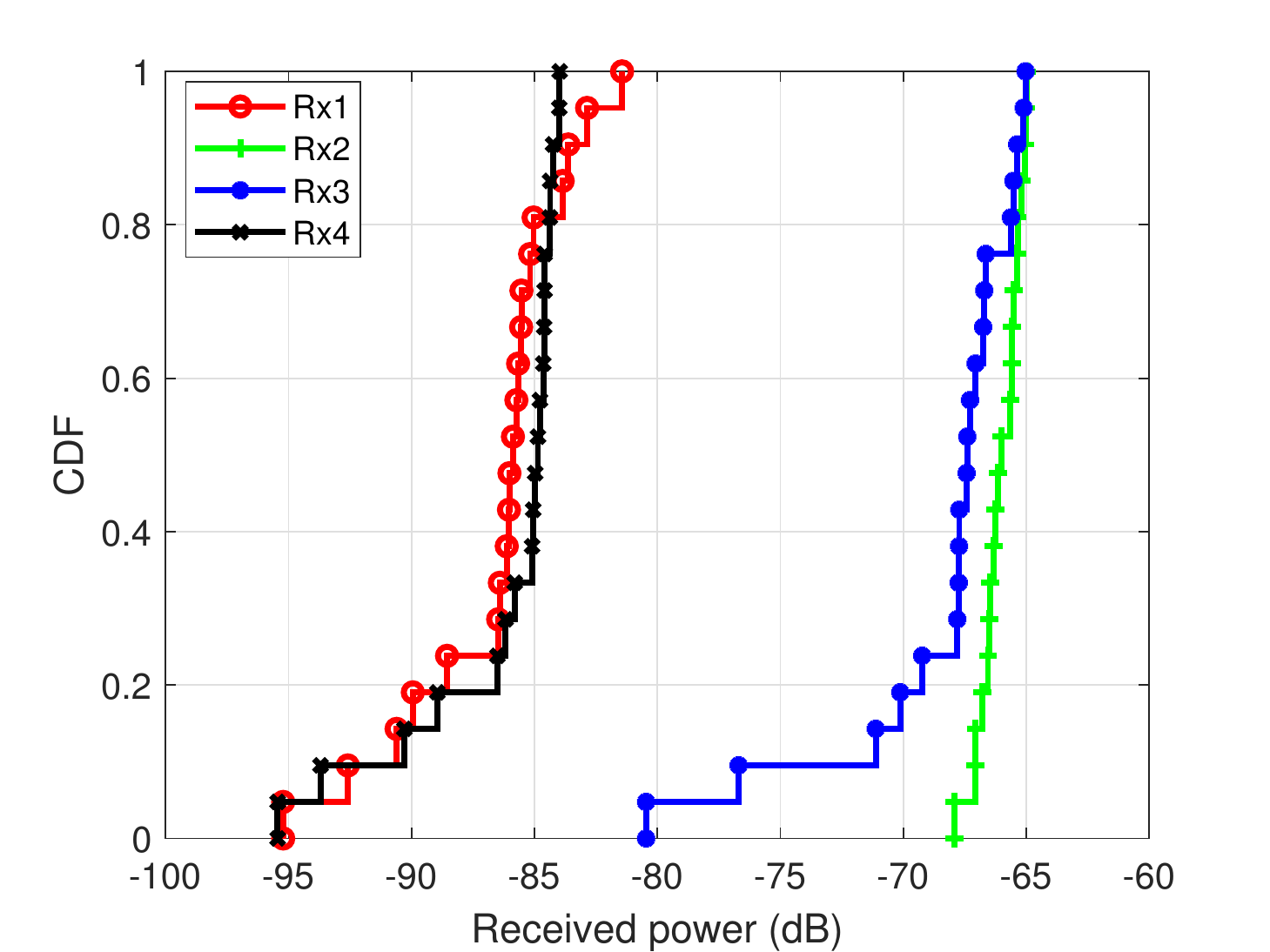} 
} 

\subfigure[32 GHz, one person crosses] { \label{fig:c} 
\includegraphics[width=0.45\columnwidth]{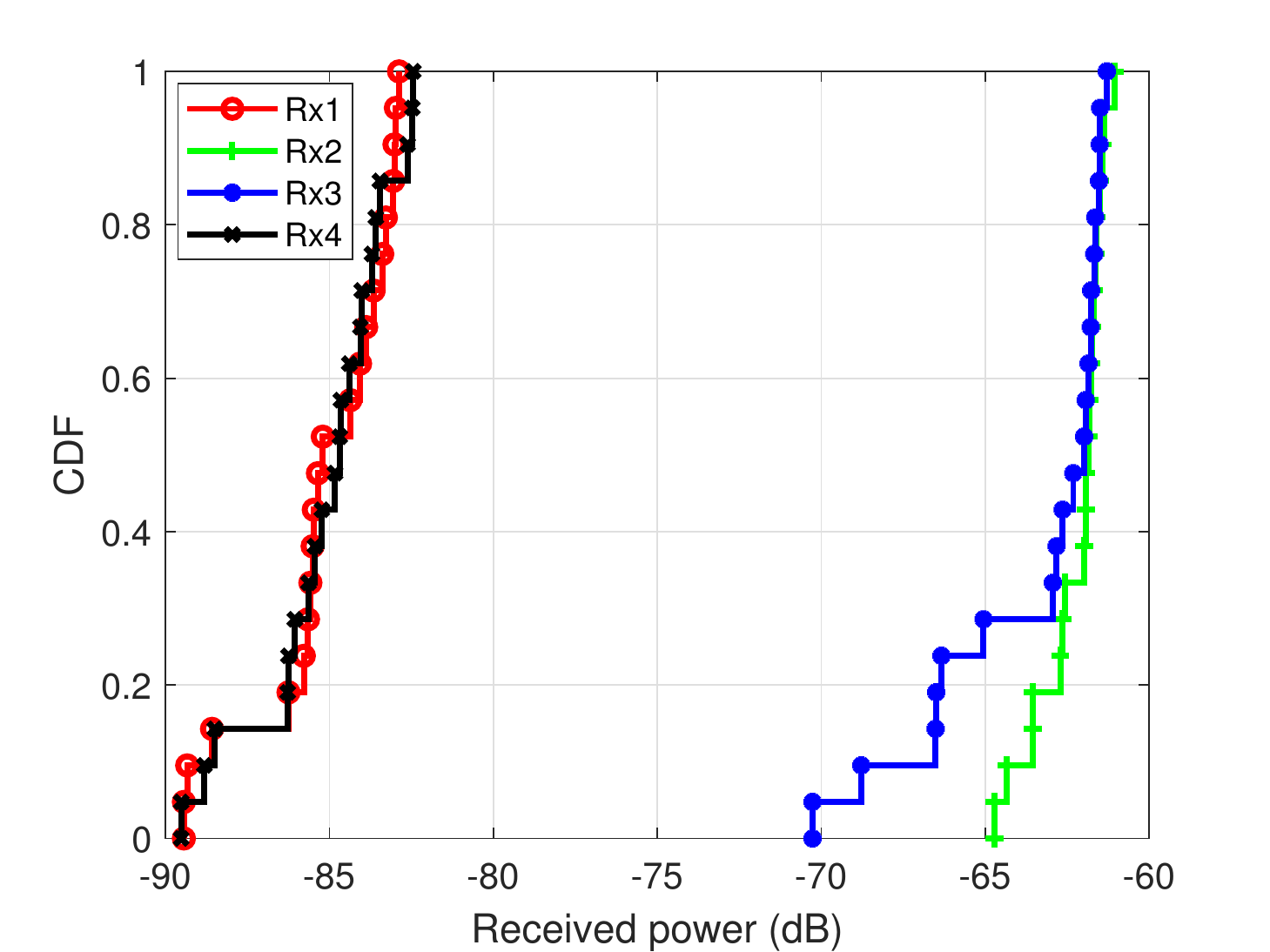} 
} 
\subfigure[32 GHz, two persons cross] { \label{fig:d} 
\includegraphics[width=0.45\columnwidth]{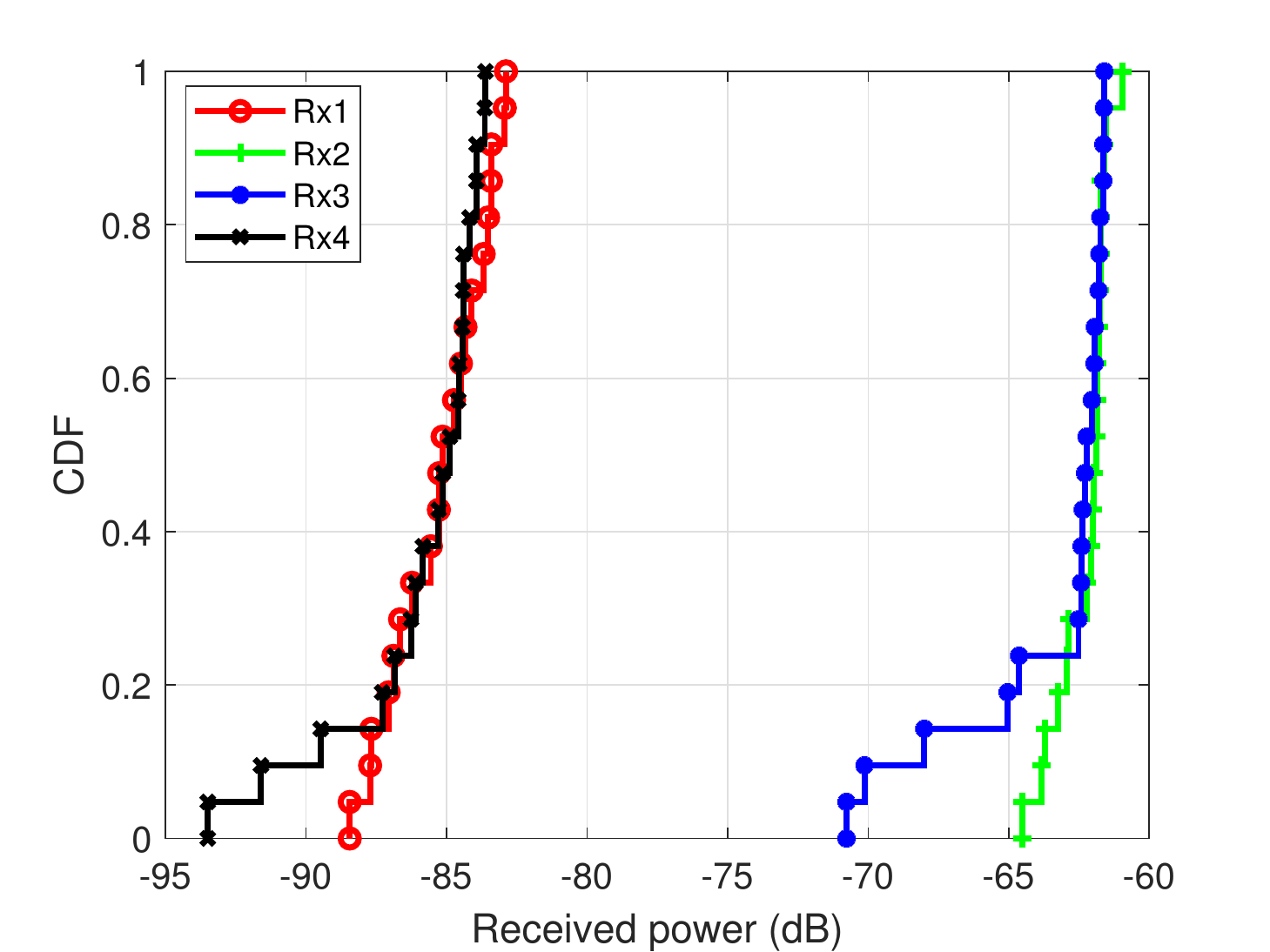} 
} 

\subfigure[39 GHz, one person crosses] { \label{fig:e} 
\includegraphics[width=0.45\columnwidth]{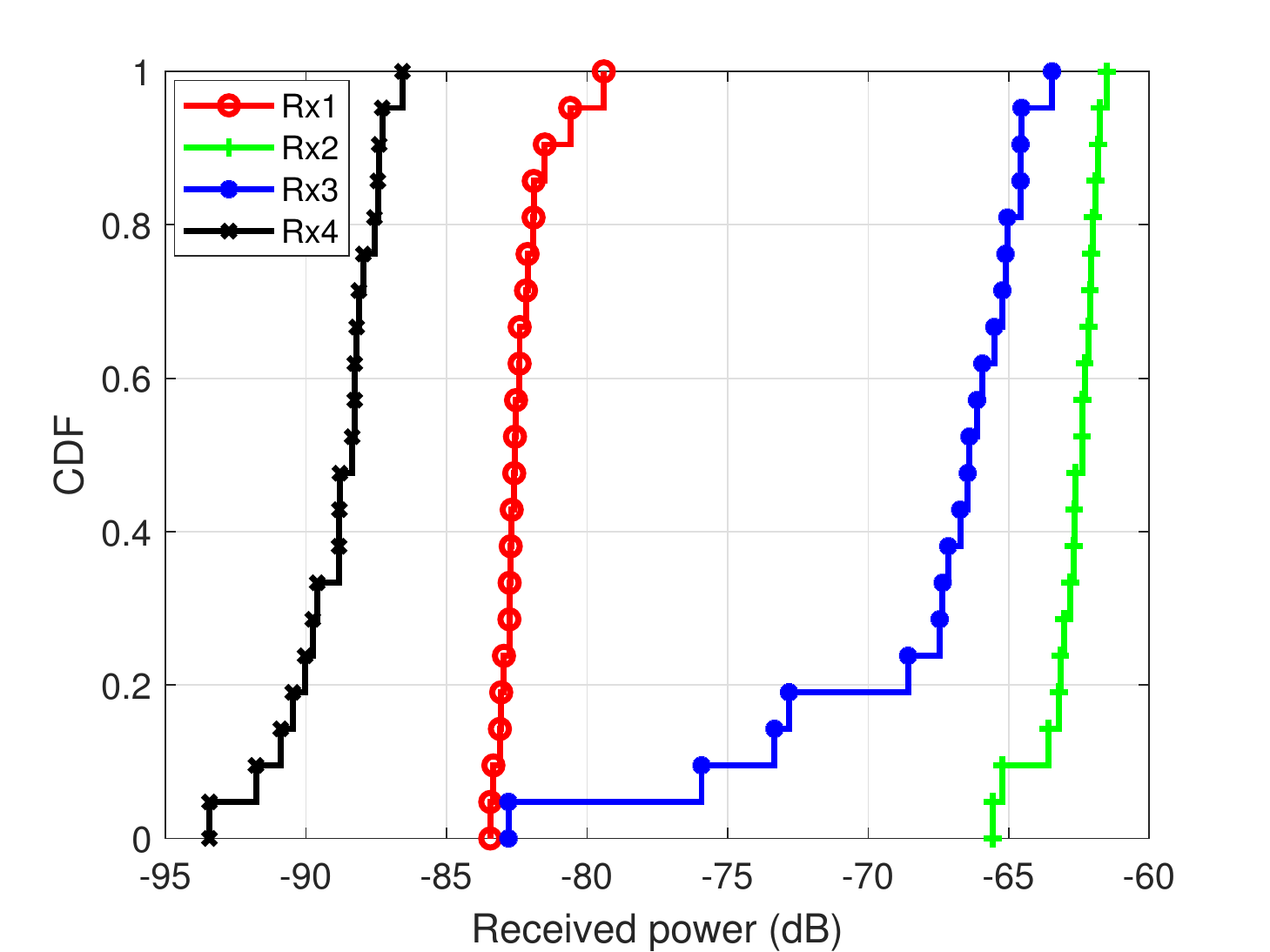} 
} 
\subfigure[39 GHz, two persons cross] { \label{fig:f} 
\includegraphics[width=0.45\columnwidth]{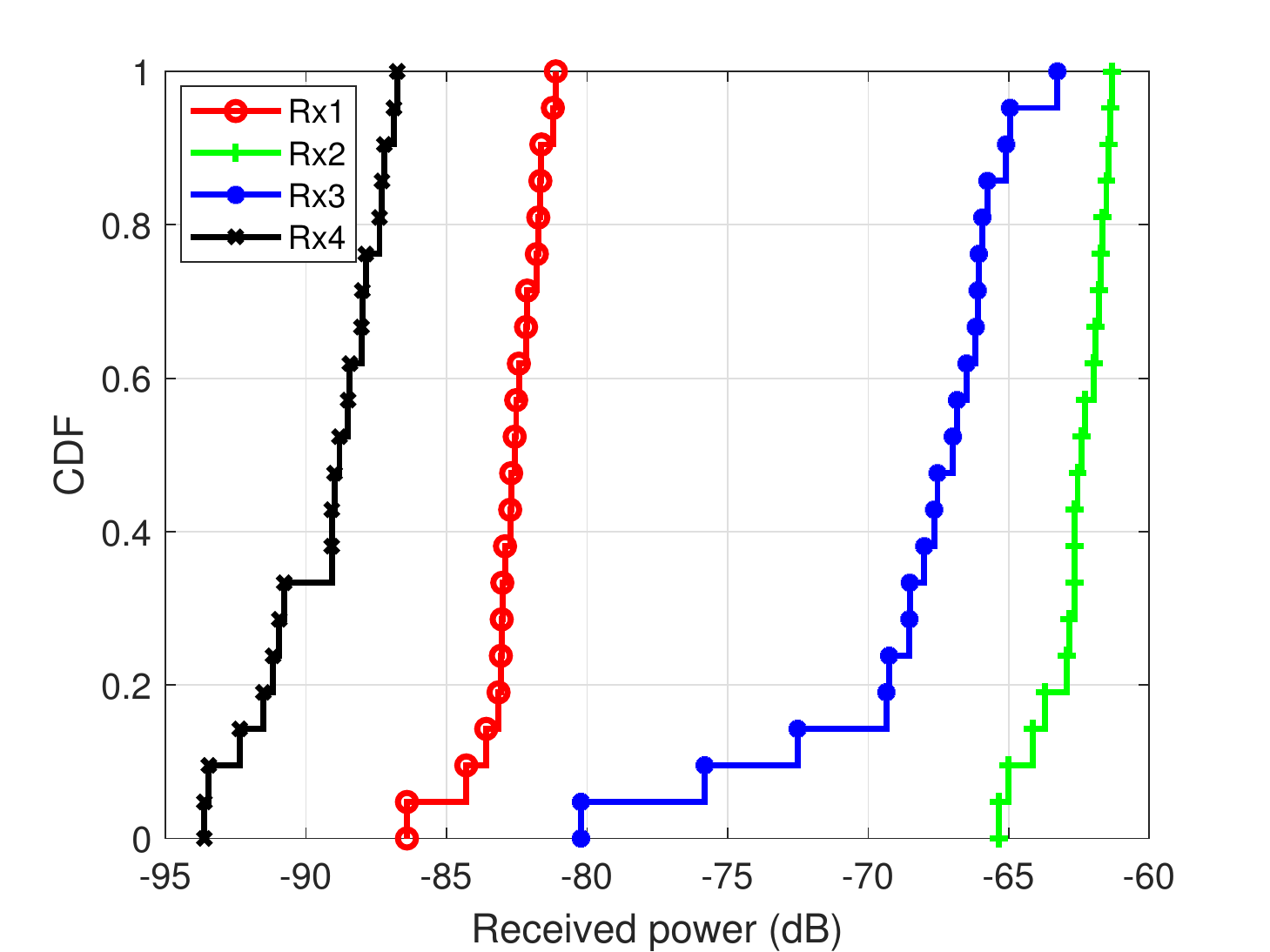} 
} 
\caption{Indoor human blockage measurement results with one person and two persons crossing the Tx-Rx path at 28, 32, and 39 GHz bands.} 
\label{fig:ih_cross} 
\end{figure}

The indoor human blockage measurement results with one person and two persons cross the LOS path are shown in Fig.~\ref{fig:ih_cross}. The Rx2 channel has little attenuation with either one person or two persons move across the LOS path. The reason may due to the saturation of the LNA in channel measurements. The other received signals show obvious human blockage attenuations when the person blocks the LOS path. The human blockage attenuation is about 10-15 dB. It shows no much difference either one person or two persons cross the LOS path. The vertical polarization has a larger attenuation than that of horizontal attenuation. The reason may due to the de-polarization effect.

\begin{figure} [tb!]
\centering 
\subfigure[Human blockage attenuations vs. distance to Tx-Rx path crossing the Tx-Rx path for 28, 32, and 39 GHz bands using the GTD, METIS, and Kirchhoff models.] { \label{fig:a} 
\includegraphics[width=3.5in,height=2.2in]{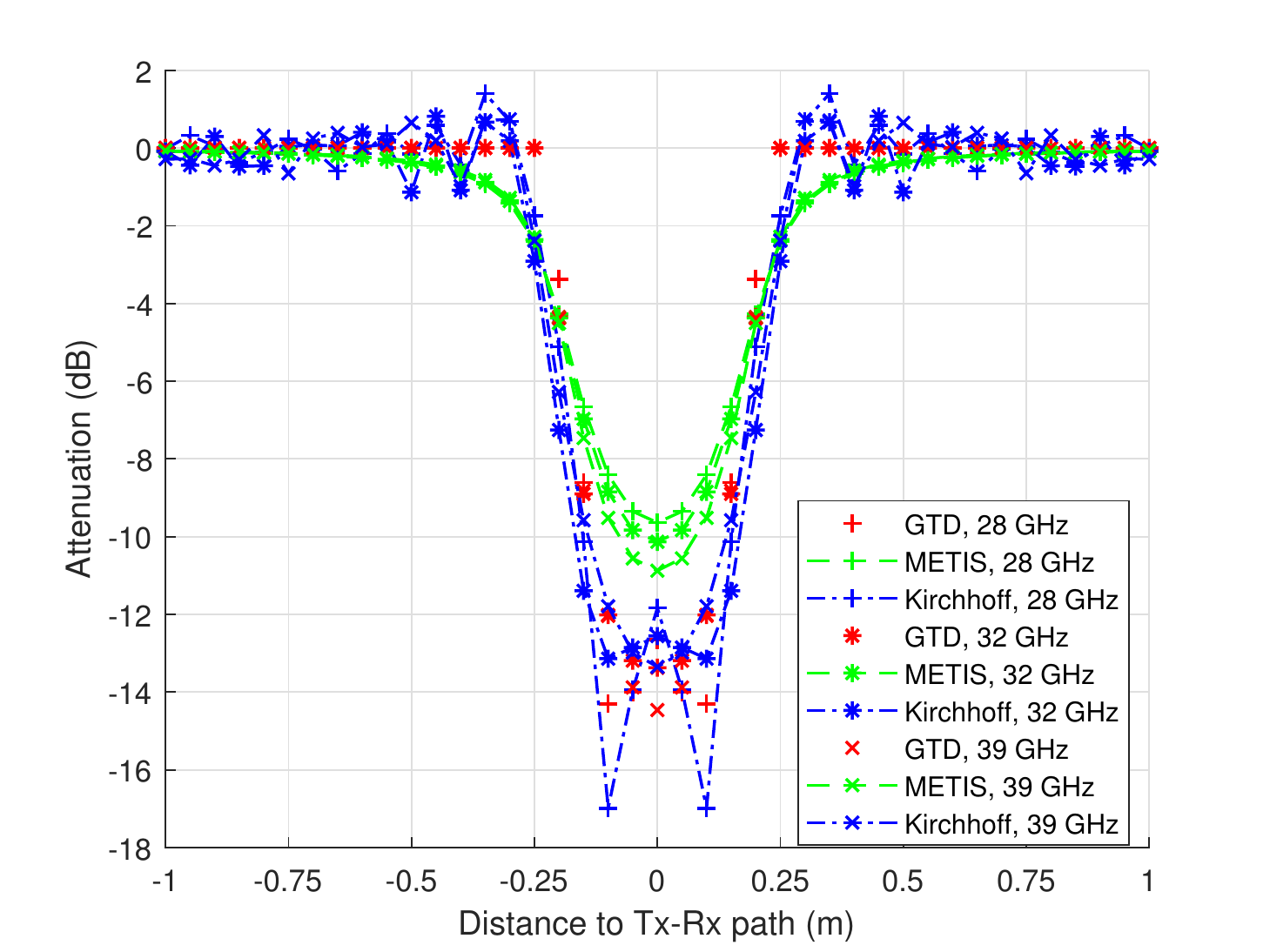} 
} 
\subfigure[Comparison of measured and modeled human blockage attenuations across the LOS path at 28 GHz band.] { \label{fig:b} 
\includegraphics[width=3.5in,height=2.2in]{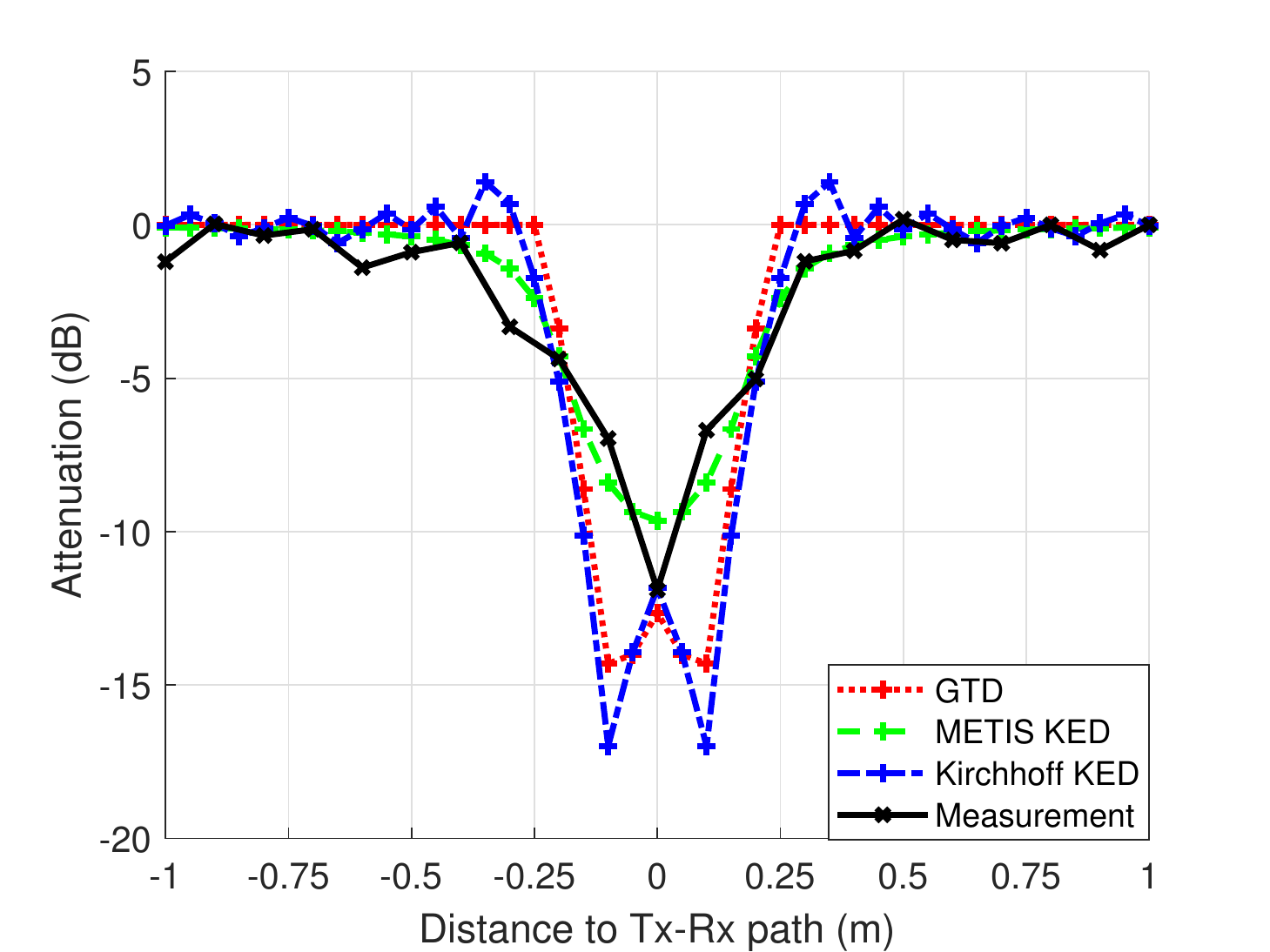} 
} 
\caption{Measured and modeled human blockage attenuations across the LOS path.} 
\label{fig:hb_cross} 
\end{figure}

Fig. \ref{fig:hb_cross} shows the measured and modeled human blockage attenuations with one person crossing the LOS path. The METIS model has an attenuation about 10 dB, while the Kirchhoff model and GTD model have an attenuation about 15 dB. As seen from the comparison of measured and modeled human blockage attenuation results, in the case of crossing the LOS path, the METIS KED model can act as the lower bound, while the Kirchhoff KED model and GTD model can be the upper bound.

\begin{figure} [tb!]
\centering 
\subfigure[28 GHz, along the path] { \label{fig:a} 
\includegraphics[width=0.45\columnwidth]{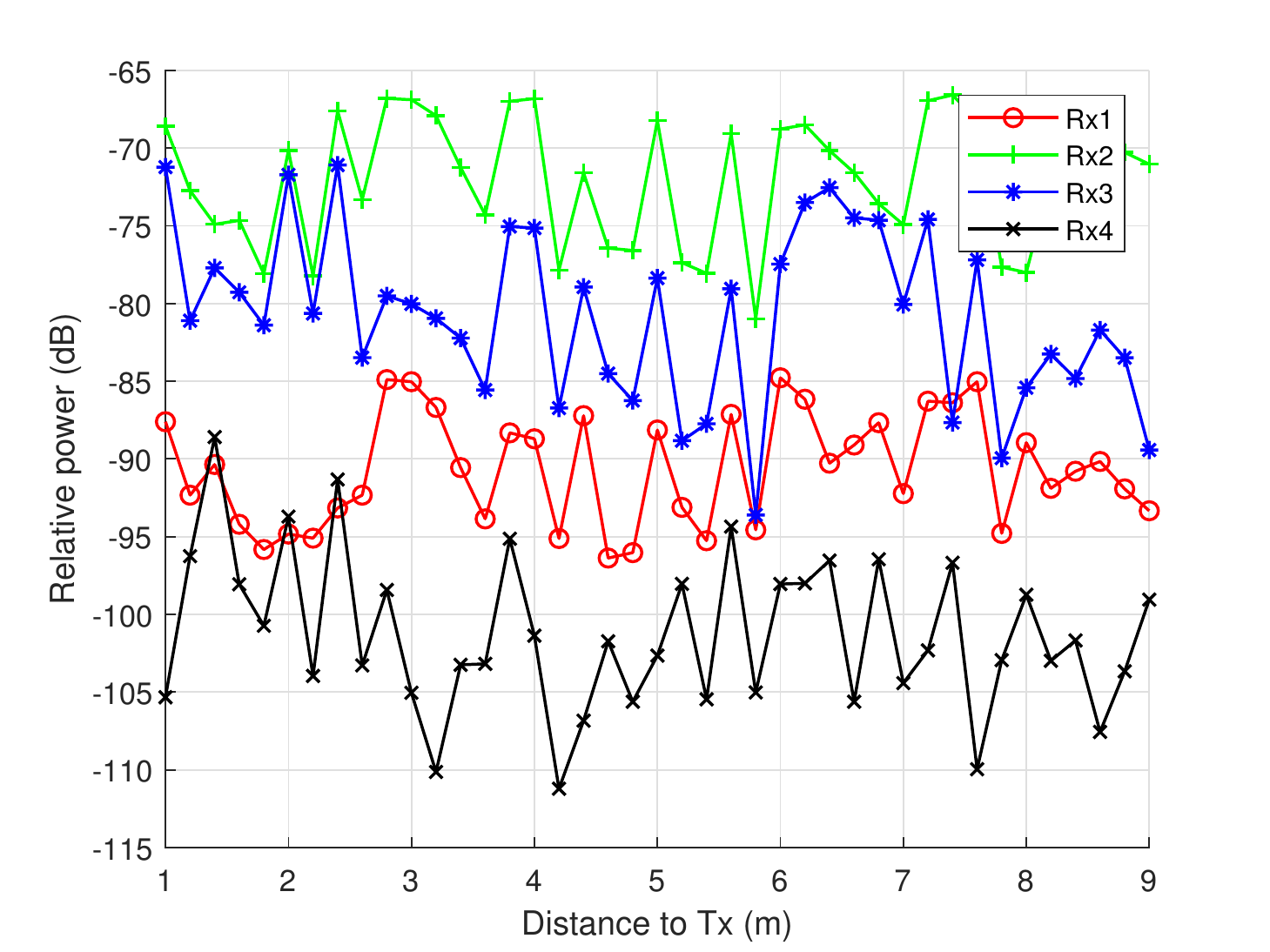} 
} 
\subfigure[32 GHz, along the path] { \label{fig:b} 
\includegraphics[width=0.45\columnwidth]{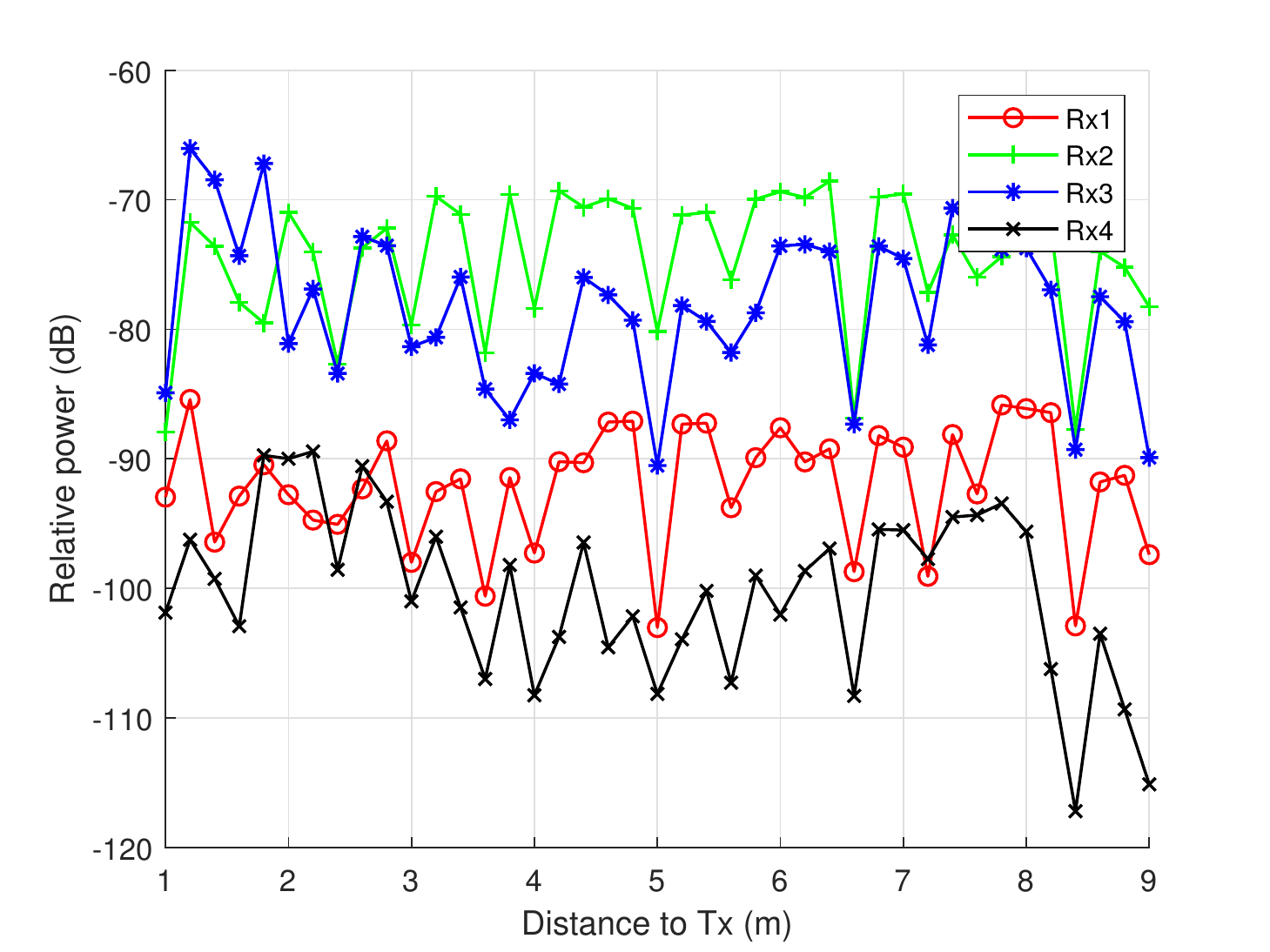} 
} 

\subfigure[28 GHz, one person crosses] { \label{fig:c} 
\includegraphics[width=0.45\columnwidth]{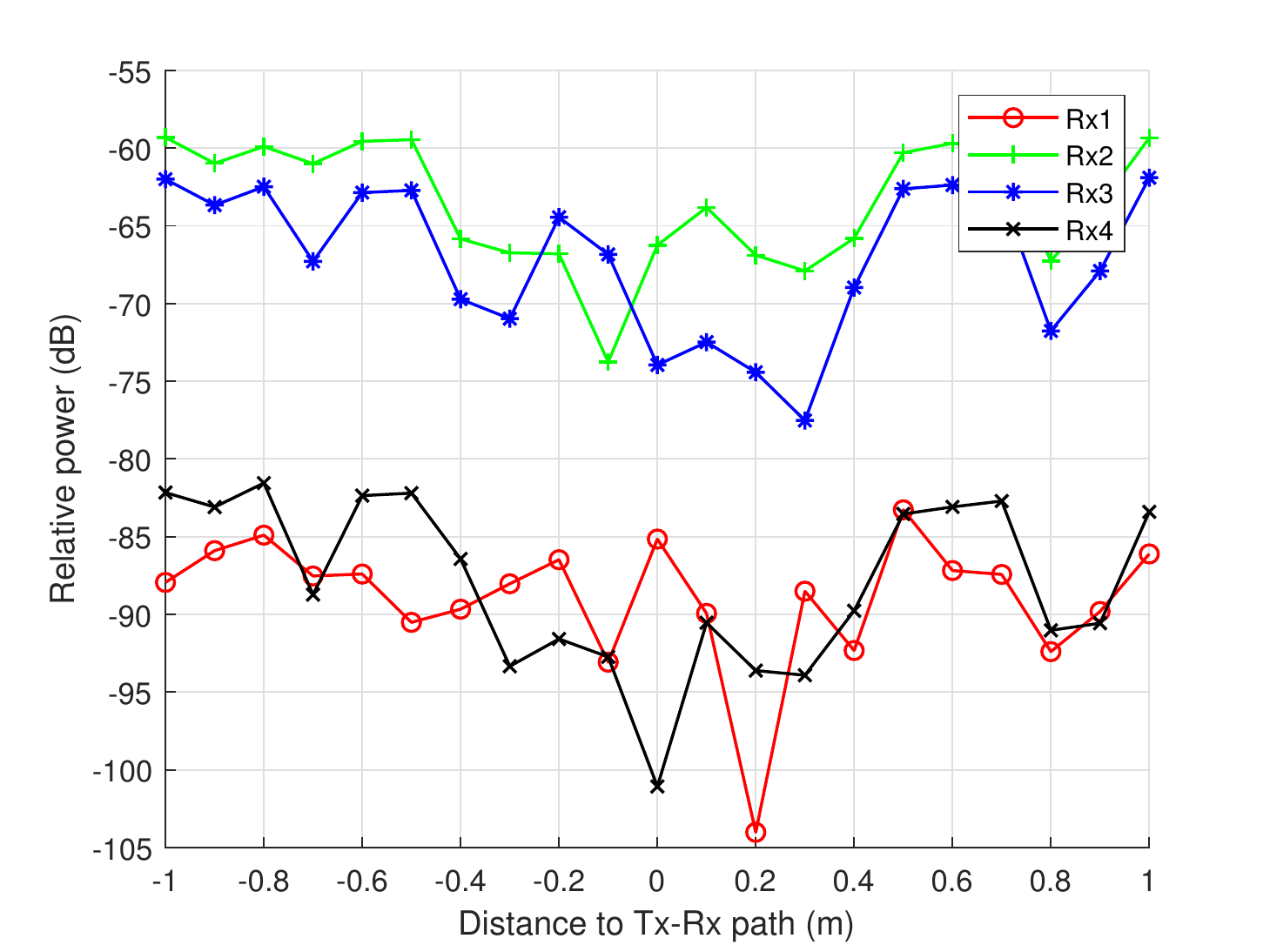} 
} 
\subfigure[32 GHz, one person crosses] { \label{fig:d} 
\includegraphics[width=0.45\columnwidth]{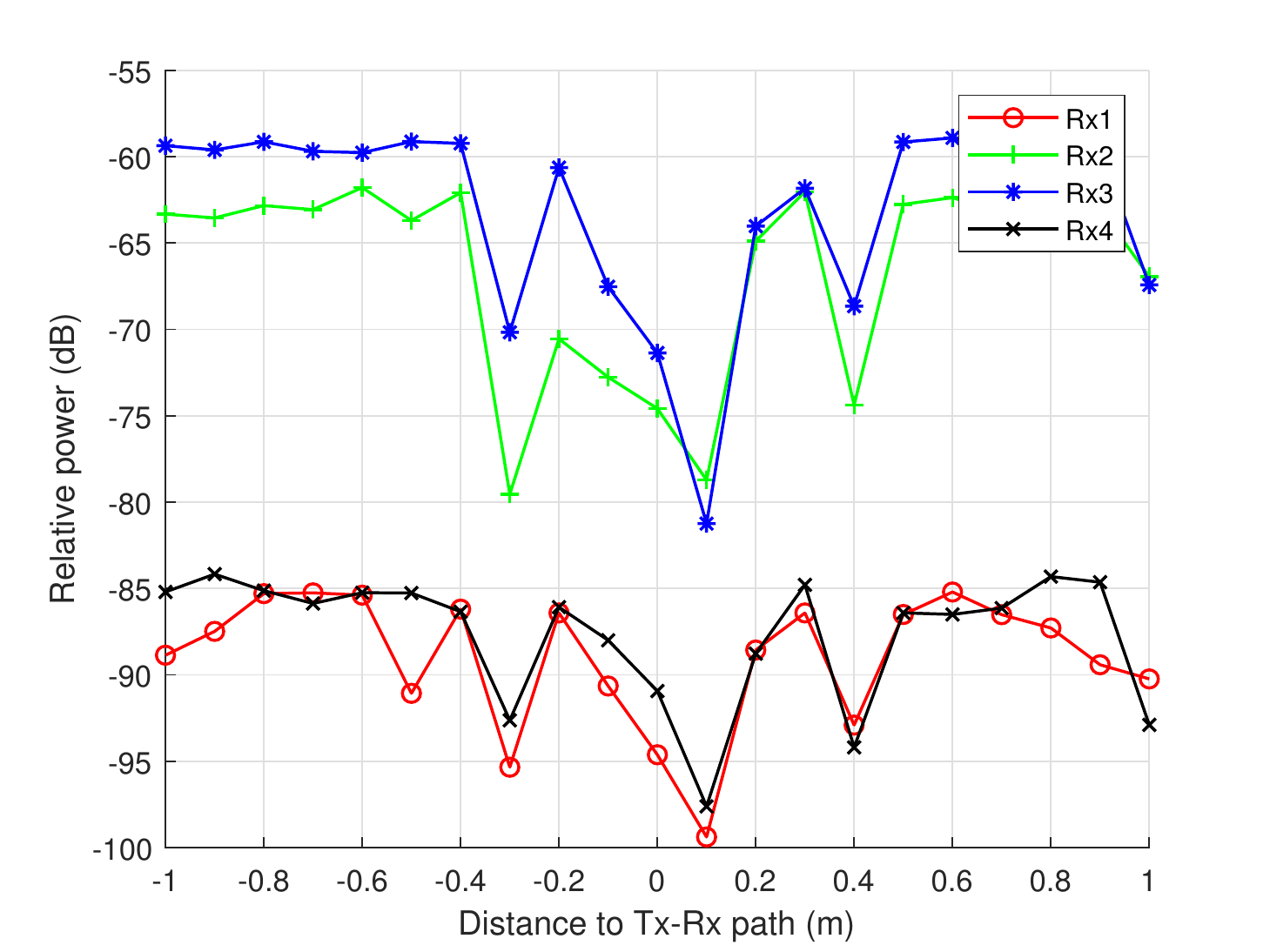} 
} 

\subfigure[28 GHz, two persons cross] { \label{fig:e} 
\includegraphics[width=0.45\columnwidth]{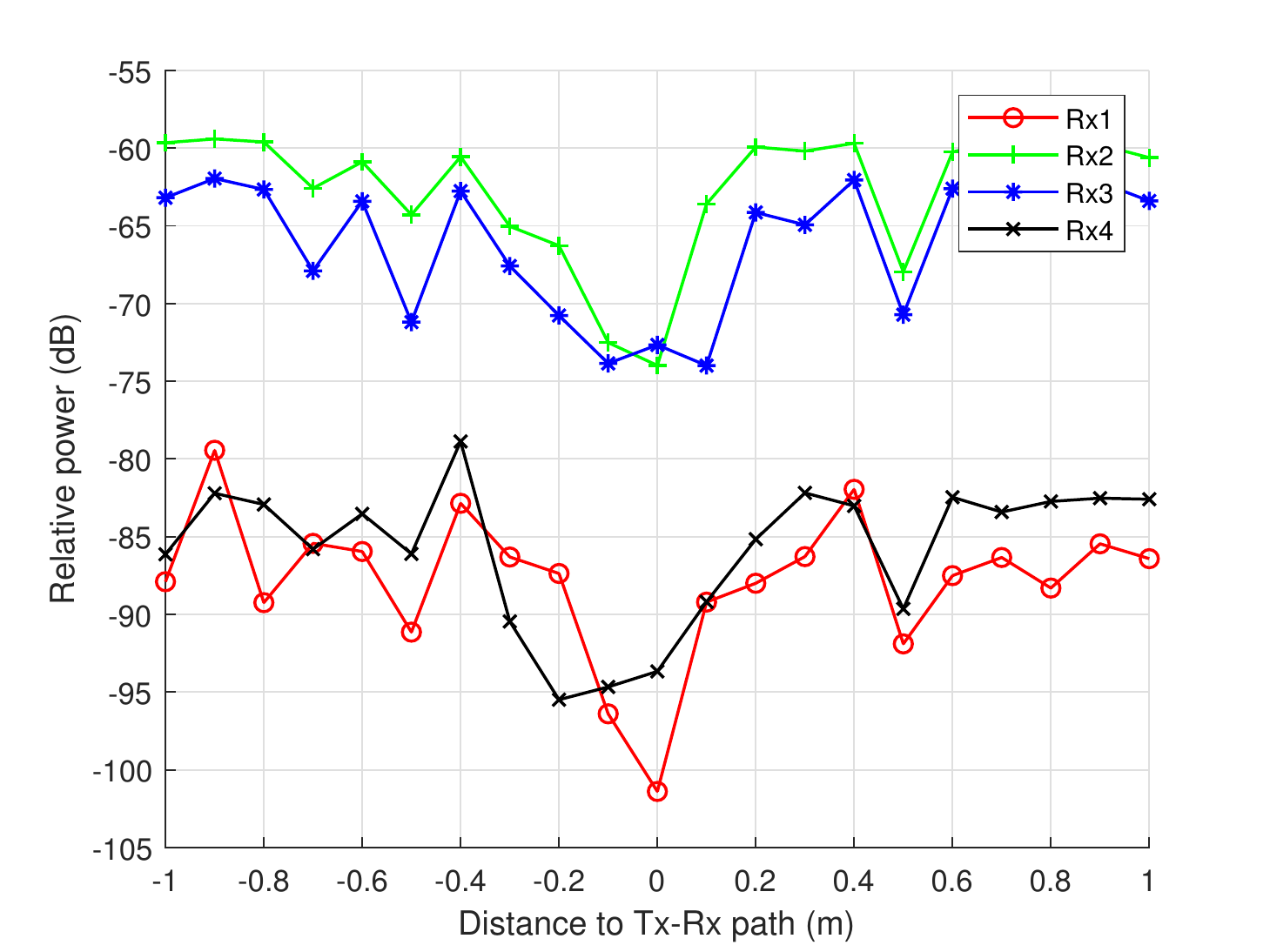} 
} 
\subfigure[32 GHz, two persons cross] { \label{fig:f} 
\includegraphics[width=0.45\columnwidth]{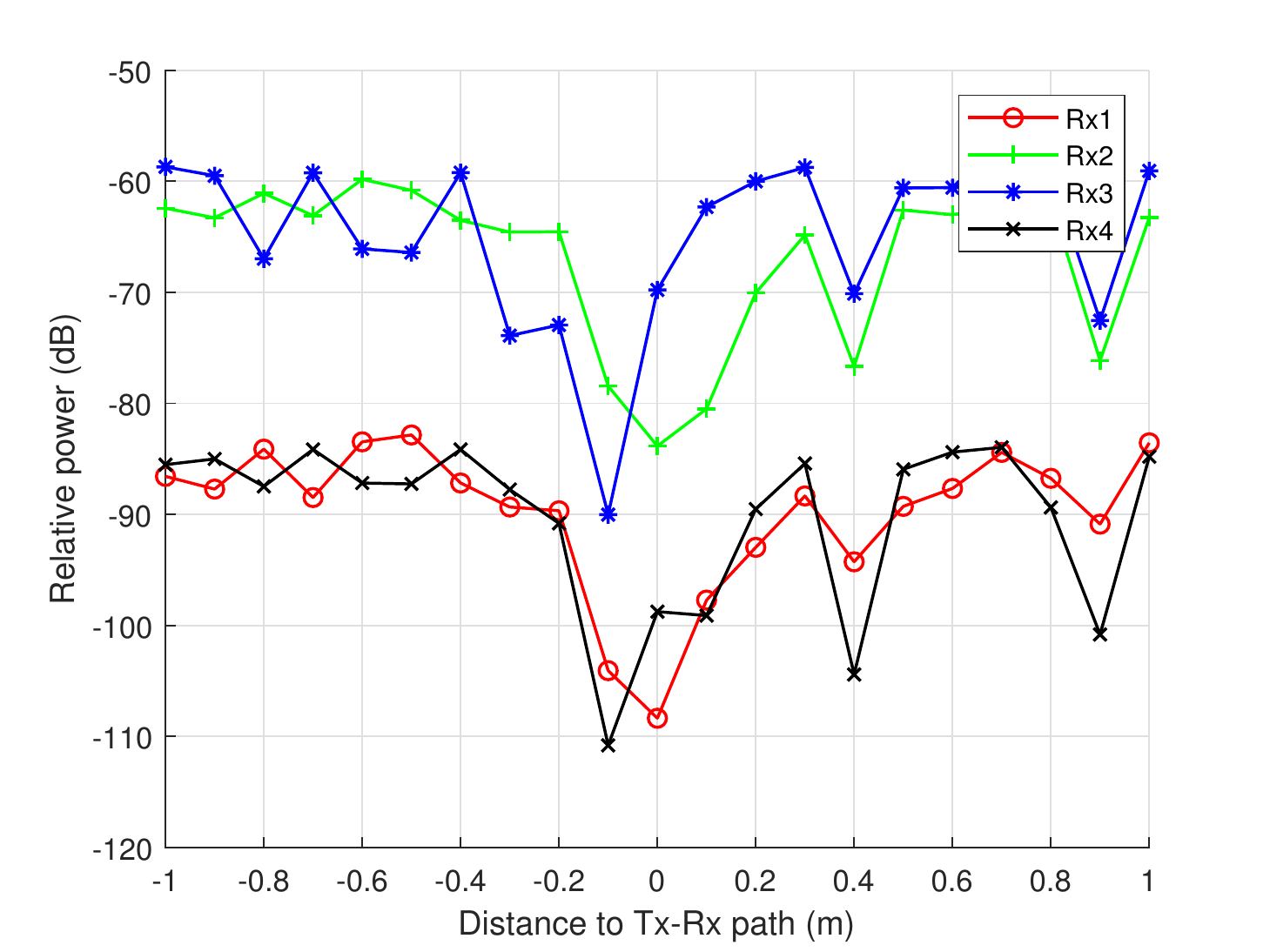} 
} 
\caption{Outdoor human blockage measurement results at 28 GHz and 32 GHz bands.} 
\label{fig:oh} 
\end{figure}

\begin{figure} [tb!]
\centering 
\subfigure[Indoor CIRs at 28 GHz with one person crossing the LOS path.] { \label{fig:a} 
\includegraphics[width=3.5in,height=2.2in]{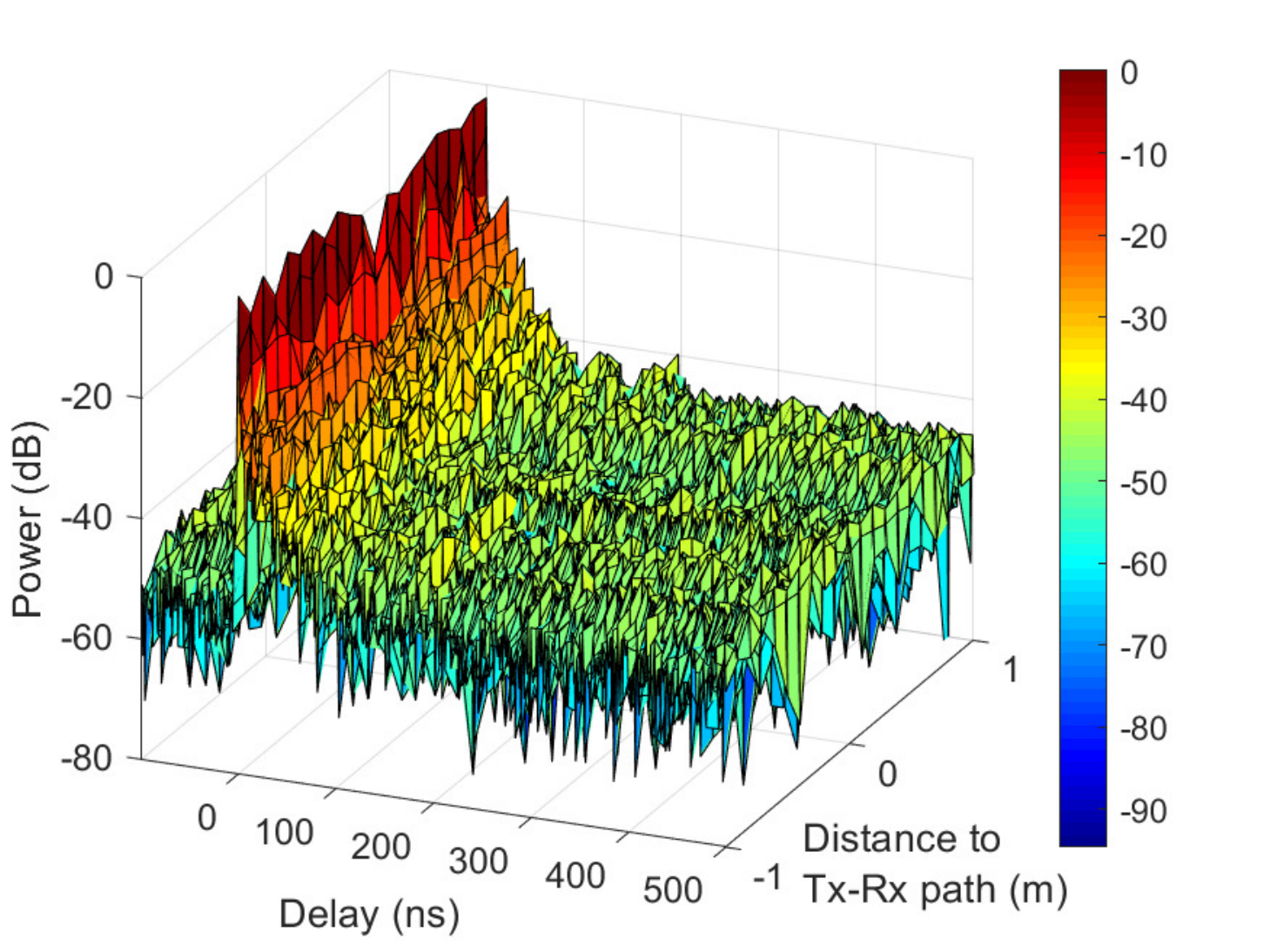} 
} 
\subfigure[Outdoor CIRs at 28 GHz with one person crossing the LOS path.] { \label{fig:b} 
\includegraphics[width=3.5in,height=2.2in]{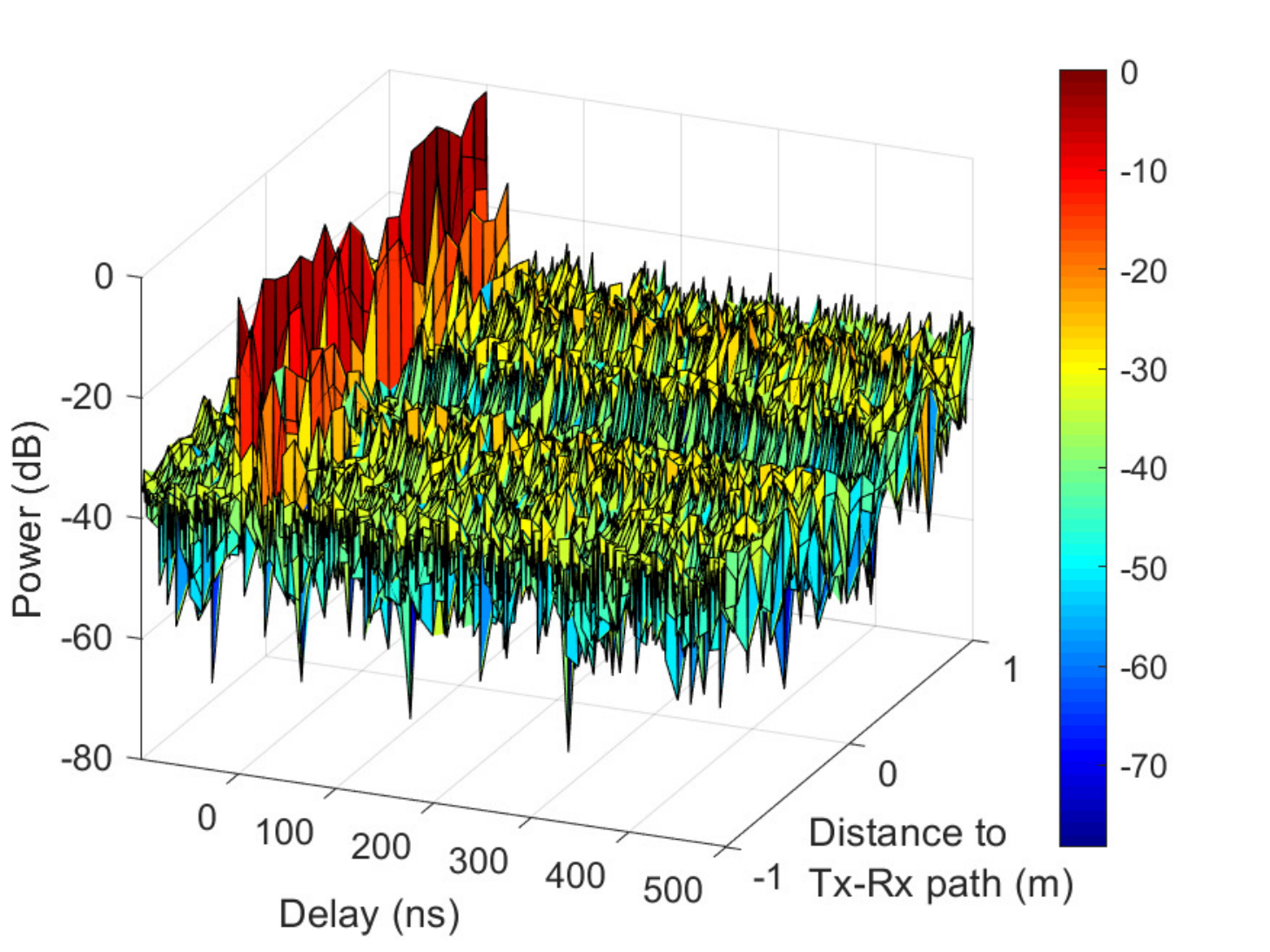} 
}  
\caption{Comparison of indoor and outdoor CIRs with one person crossing the LOS path at 28 GHz band.} 
\label{fig:in_out} 
\end{figure}

As a comparison, Fig. \ref{fig:oh} shows the outdoor human blockage measurement results at 28 GHz and 32 GHz bands with different schemes. 
For the case of along the LOS path, the outdoor human blockage has a larger attenuation, which is about 5-10 dB larger than that of indoor situations. Meanwhile, the attenuation along the path has a larger variation. For the case of cross the LOS path, the outdoor human blockage also has a larger attenuation than that of indoor situations. For both 28 GHz and 32 GHz bands, the outdoor human blockage attenuation is about 15-20 dB. The reason is that the outdoor environment has less MPCs than the indoor environment. An example of the indoor and outdoor CIRs at 28 GHz with one person crossing the LOS path is shown in Fig. \ref{fig:in_out} to validate it. Note that the maximum received power is normalized to 0 dB.

\begin{figure} [tb!]
\centering 
\subfigure[28 GHz] { \label{fig:a} 
\includegraphics[width=3.5in,height=2.2in]{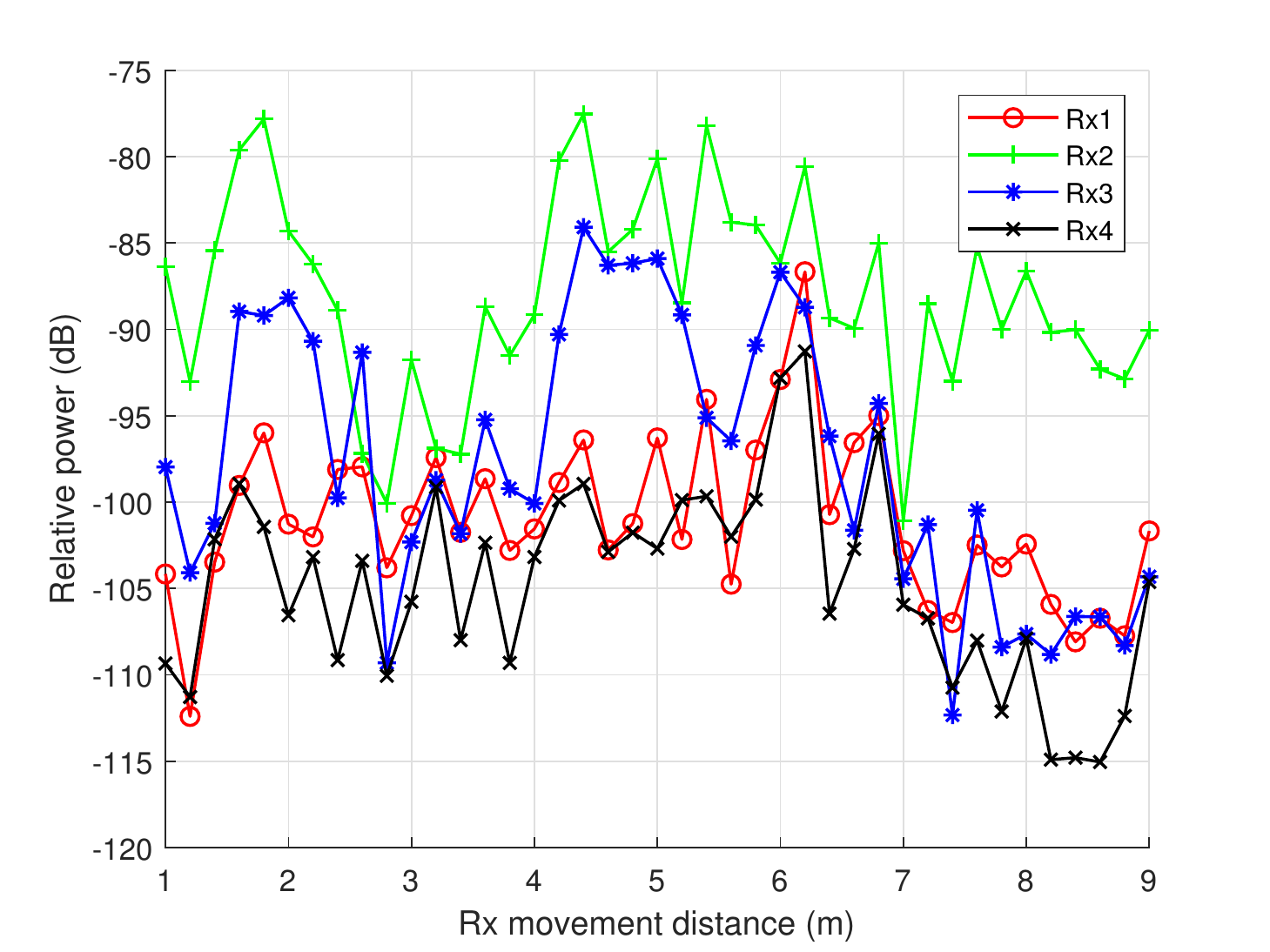} 
} 
\subfigure[32 GHz] { \label{fig:b} 
\includegraphics[width=3.5in,height=2.2in]{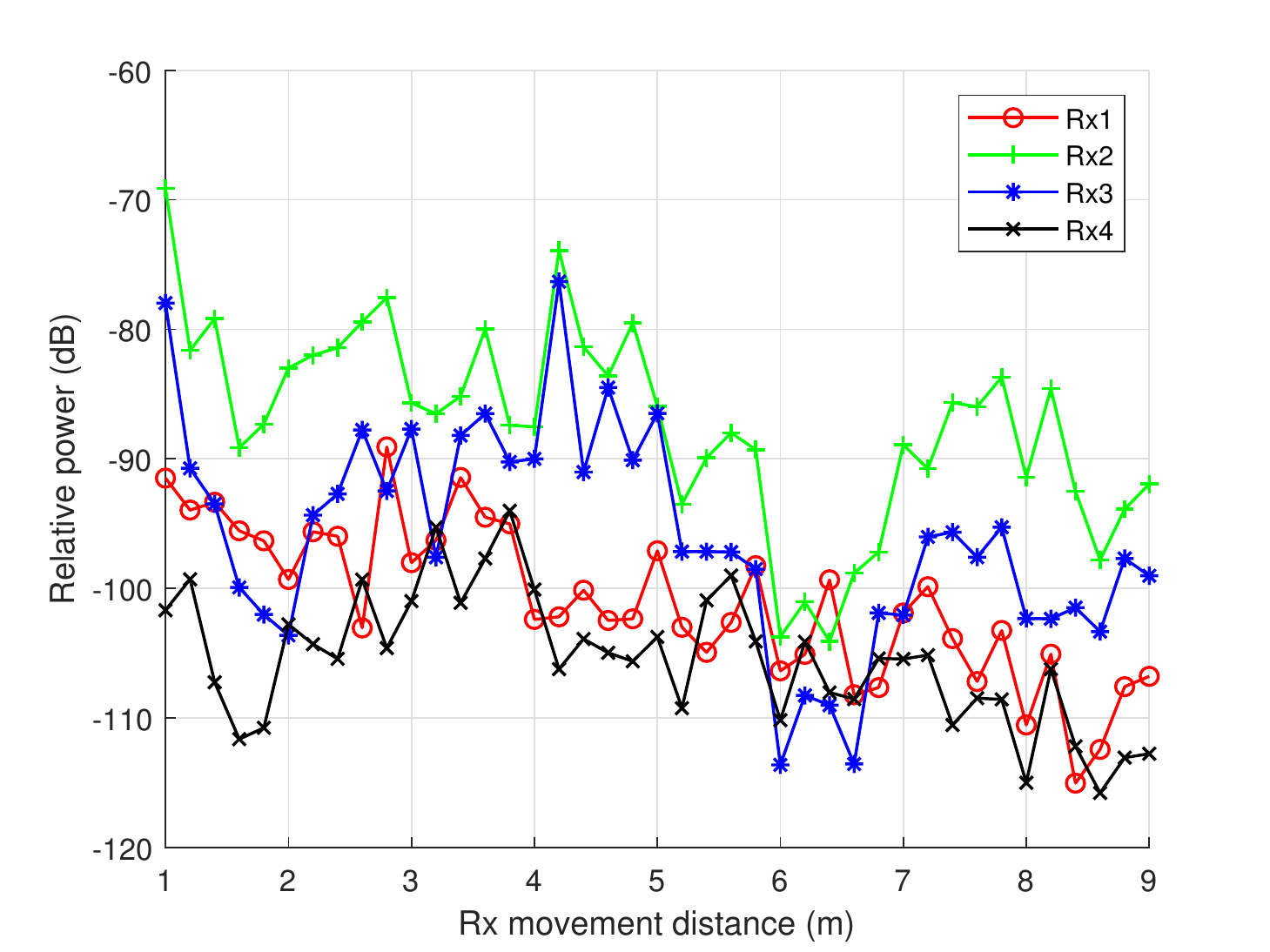} 
} 
\subfigure[39 GHz] { \label{fig:c} 
\includegraphics[width=3.5in,height=2.2in]{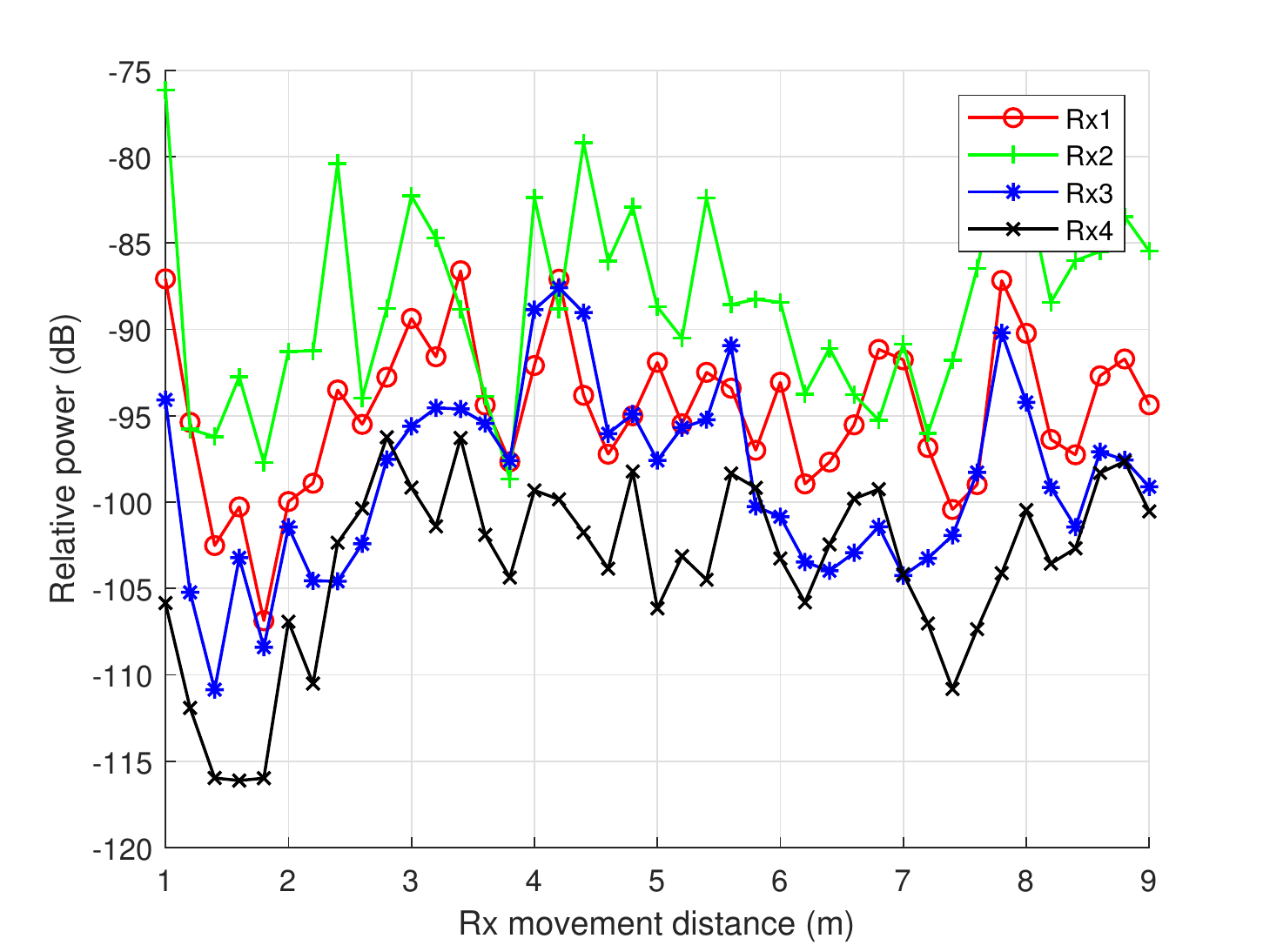} 
} 
\caption{Outdoor vehicle blockage measurement results at 28, 32, and 39~GHz bands.} 
\label{fig:oc} 
\end{figure}

Fig. \ref{fig:oc} shows the outdoor vehicle blockage measurement results. The signal level for the three frequency bands are similar for each Rx channel. It has a much larger attenuation than that of the human body blockage and shows large variations around different places of a car. Due to the complex propagation mechanisms, the XPR becomes smaller. The horizontal polarization can even have a higher signal level than the vertical polarization. The 25 dBi horizontal polarized Rx2 antenna can receive the signal at a similar level with 20 dBi vertical polarized Rx3 antenna. All these phenomenons indicate that an accurate blockage and scattering model for vehicles should be proposed and taken into account for vehicular channel modeling in the future.

\subsection{Outdoor path loss and coverage range}

\begin{figure} [tb!]
\centering 
\subfigure[28 GHz] { \label{fig:a} 
\includegraphics[width=3.5in,height=2.5in]{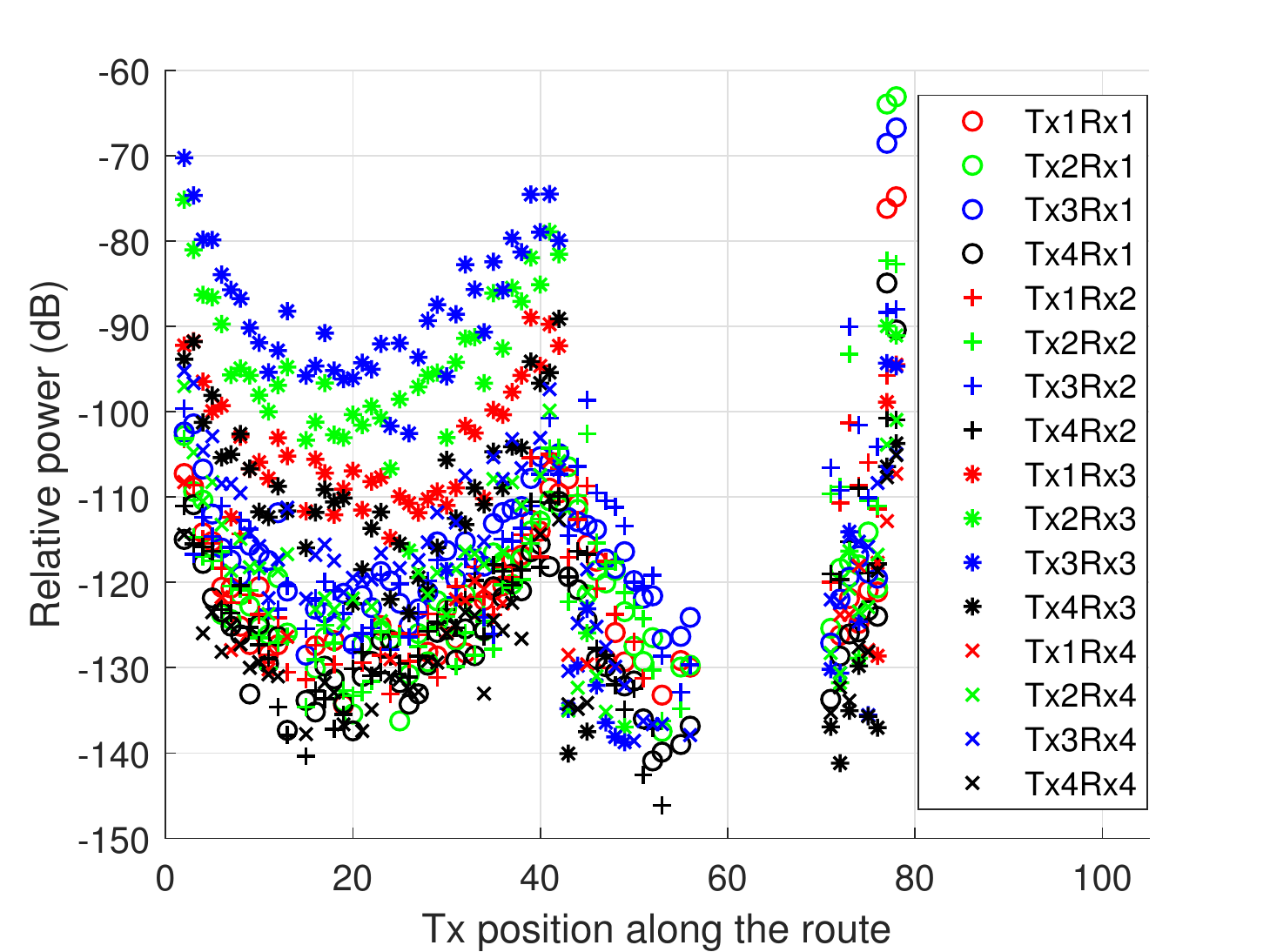} 
} 
\subfigure[32 GHz] { \label{fig:b} 
\includegraphics[width=3.5in,height=2.5in]{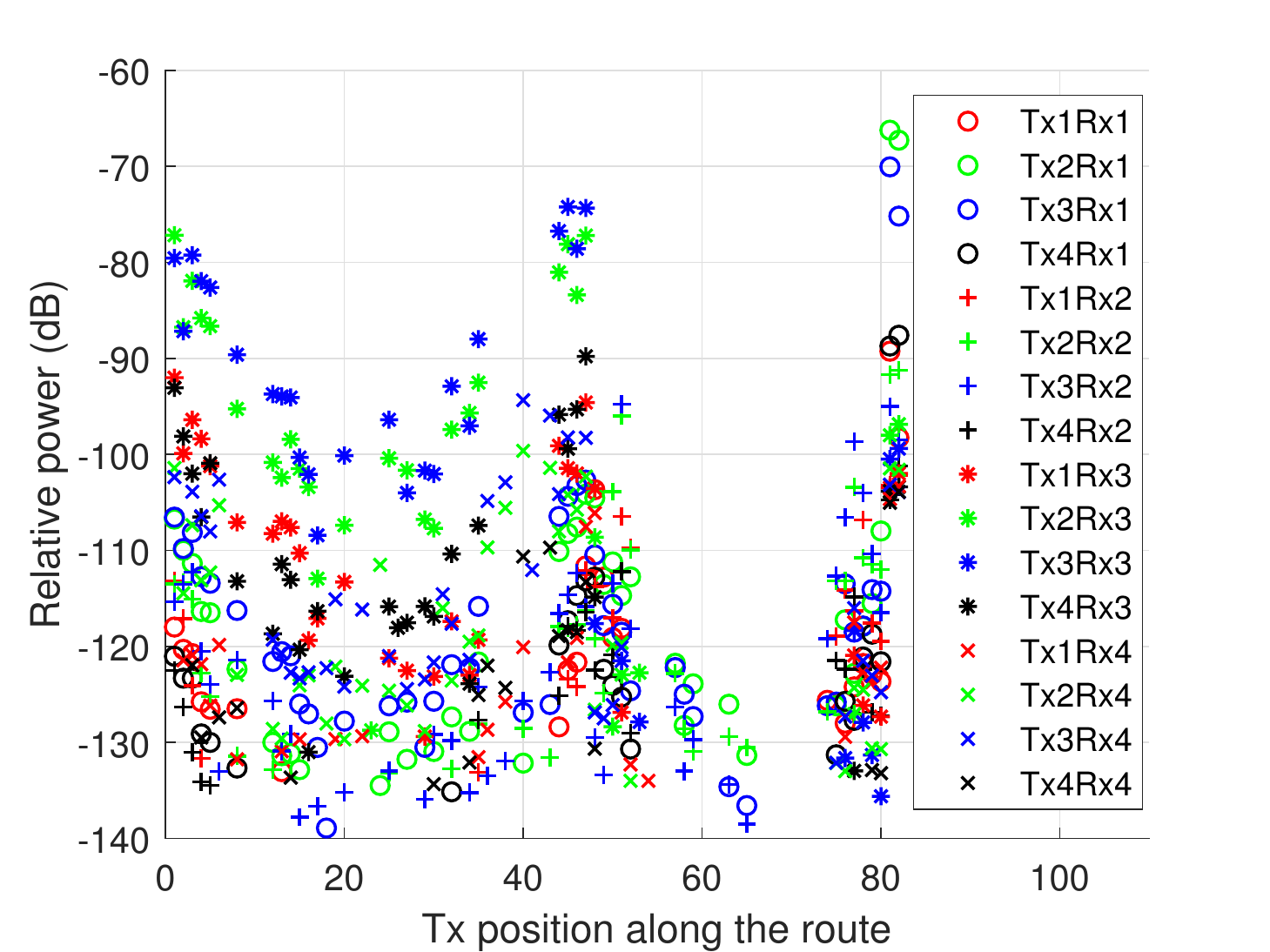} 
} 
\subfigure[39 GHz] { \label{fig:c} 
\includegraphics[width=3.5in,height=2.5in]{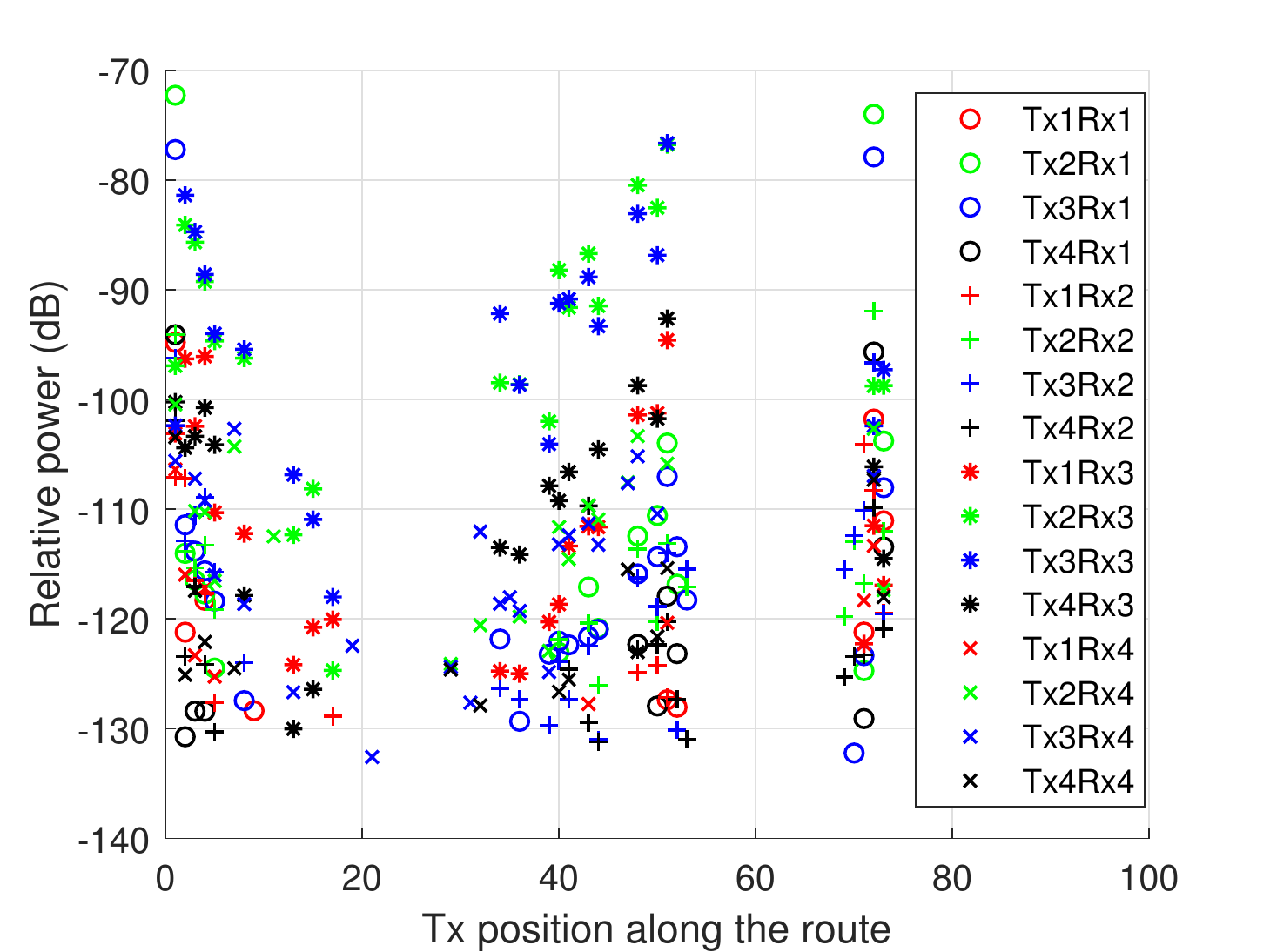} 
} 
\caption{Outdoor path loss measurement results at 28, 32, and 39 GHz bands.} 
\label{fig:pl} 
\end{figure}

Fig. \ref{fig:pl} shows the outdoor path loss measurement results at 28, 32, and 39 GHz bands with 4$\times$4 antenna configurations, respectively. The three frequency bands are measured on the same routes but not at exactly the same positions. From position 1 to about position 40, the Rx first goes away from the Tx and then goes back. Thus, the received power at first decreases and then increases. From about position 41 to the last position, the Rx moves away on the other road and then moves back.

For 28 GHz band, most of the positions can receive the signal. When the frequency goes up to 39 GHz band, it has a higher probability of outage. The difference of signal levels at the four directions can be up to 20 dB. For each channel, the received signal level variates continuously along the route, which validates the spatial consistency of mmWave propagations. When Rx moves to the other road, the Tx antenna from where the strongest signal comes also changes, as can be clearly seen from the comparison of the received signal level at position 1-40 and 41-60.  

An interesting phenomenon is that, all the four Rx antennas can receive the signals from all the four Tx antennas at most positions except the outage with a large distance, even when the Tx antennas are pointing to four directions. This indicates that signals can depart and arrive in a wide angle range in the space. It also proves that beamforming antennas can be used at BS to serve multiple users, and the users can achieve the multiplexing gain and make use of polarization.

\begin{figure} [tb!]
\centering 
\subfigure[28 GHz] { \label{fig:a} 
\includegraphics[width=3.5in,height=2.5in]{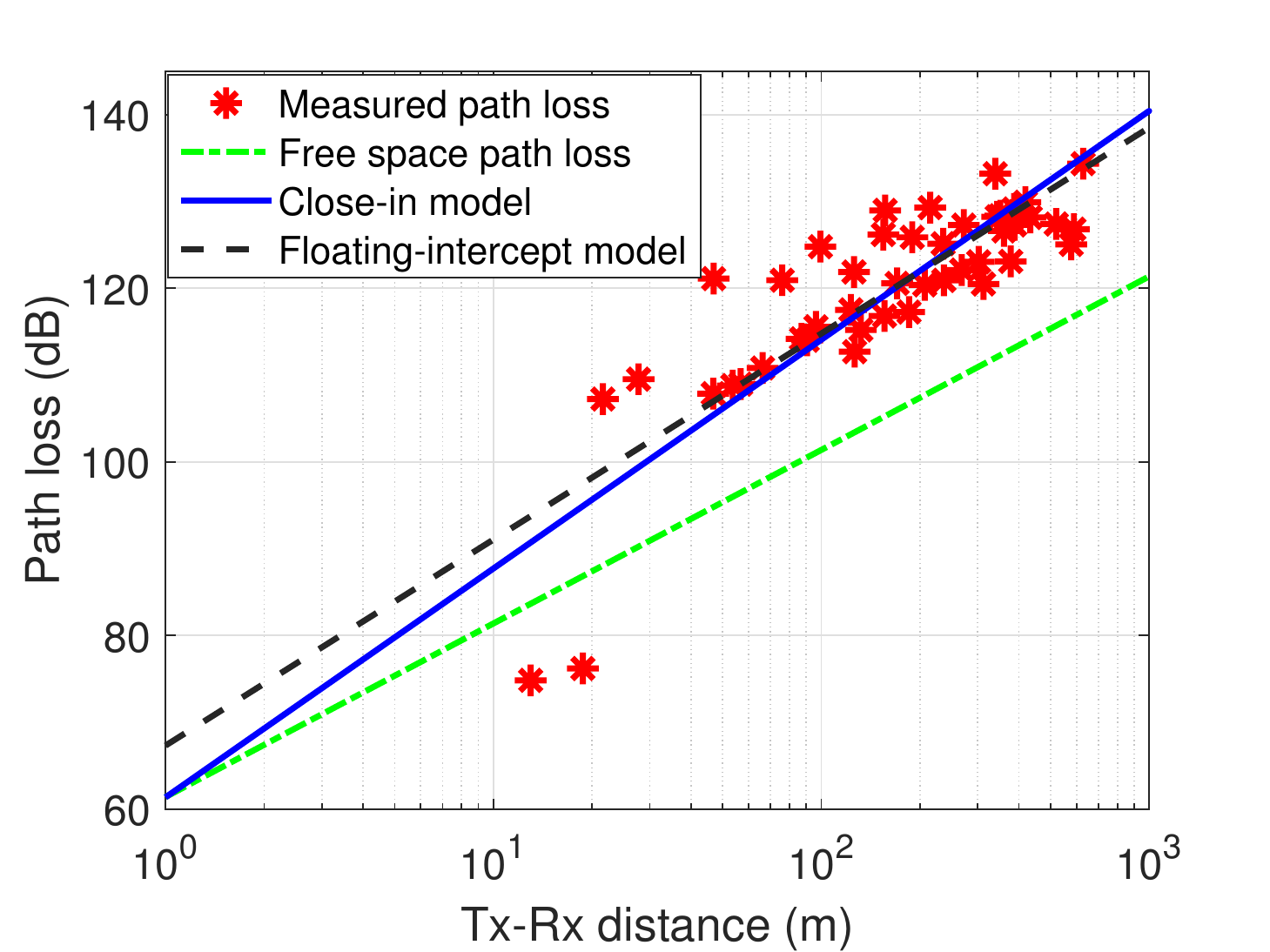} 
} 
\subfigure[32 GHz] { \label{fig:b} 
\includegraphics[width=3.5in,height=2.5in]{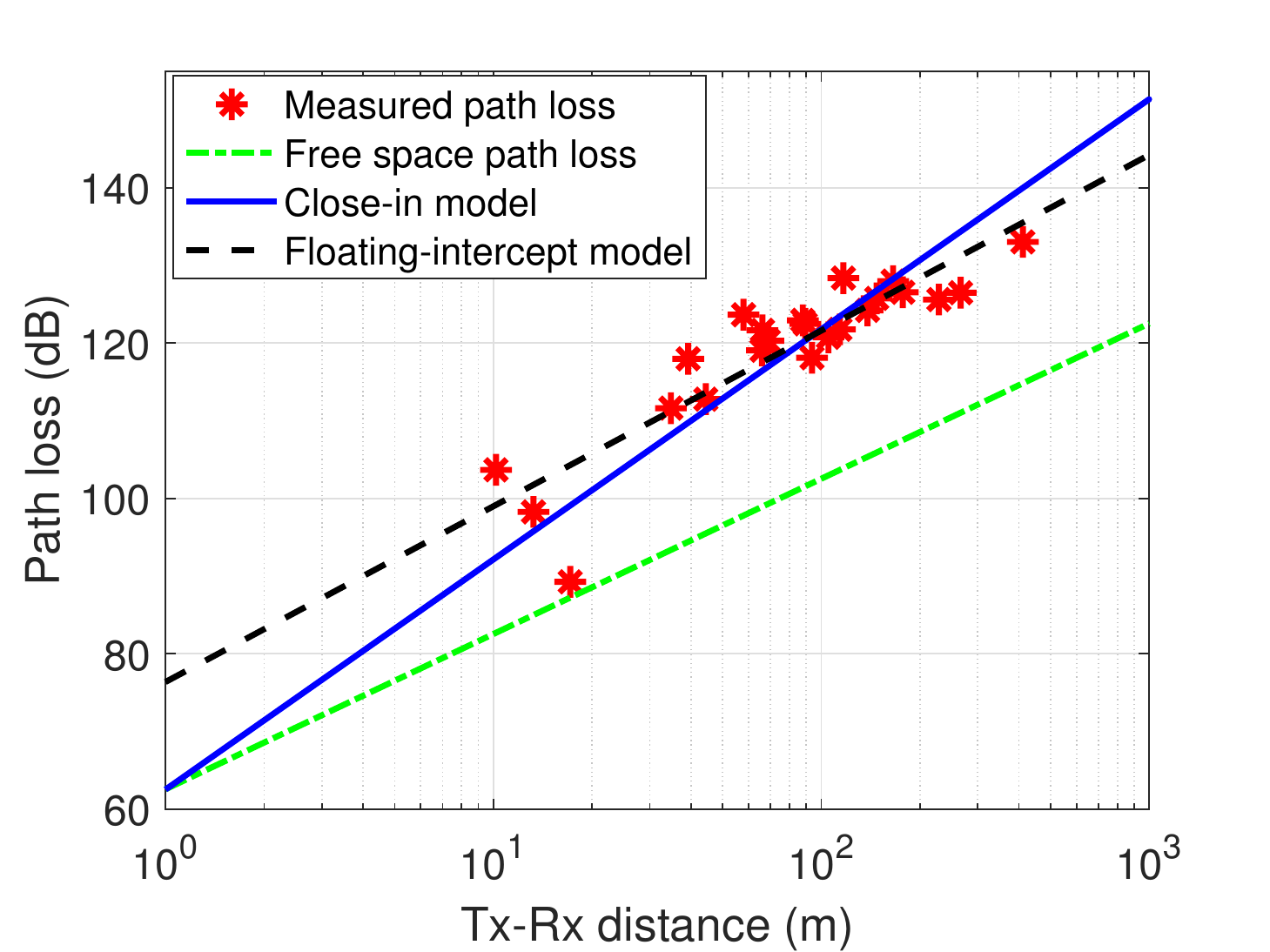} 
} 
\subfigure[39 GHz] { \label{fig:c} 
\includegraphics[width=3.5in,height=2.5in]{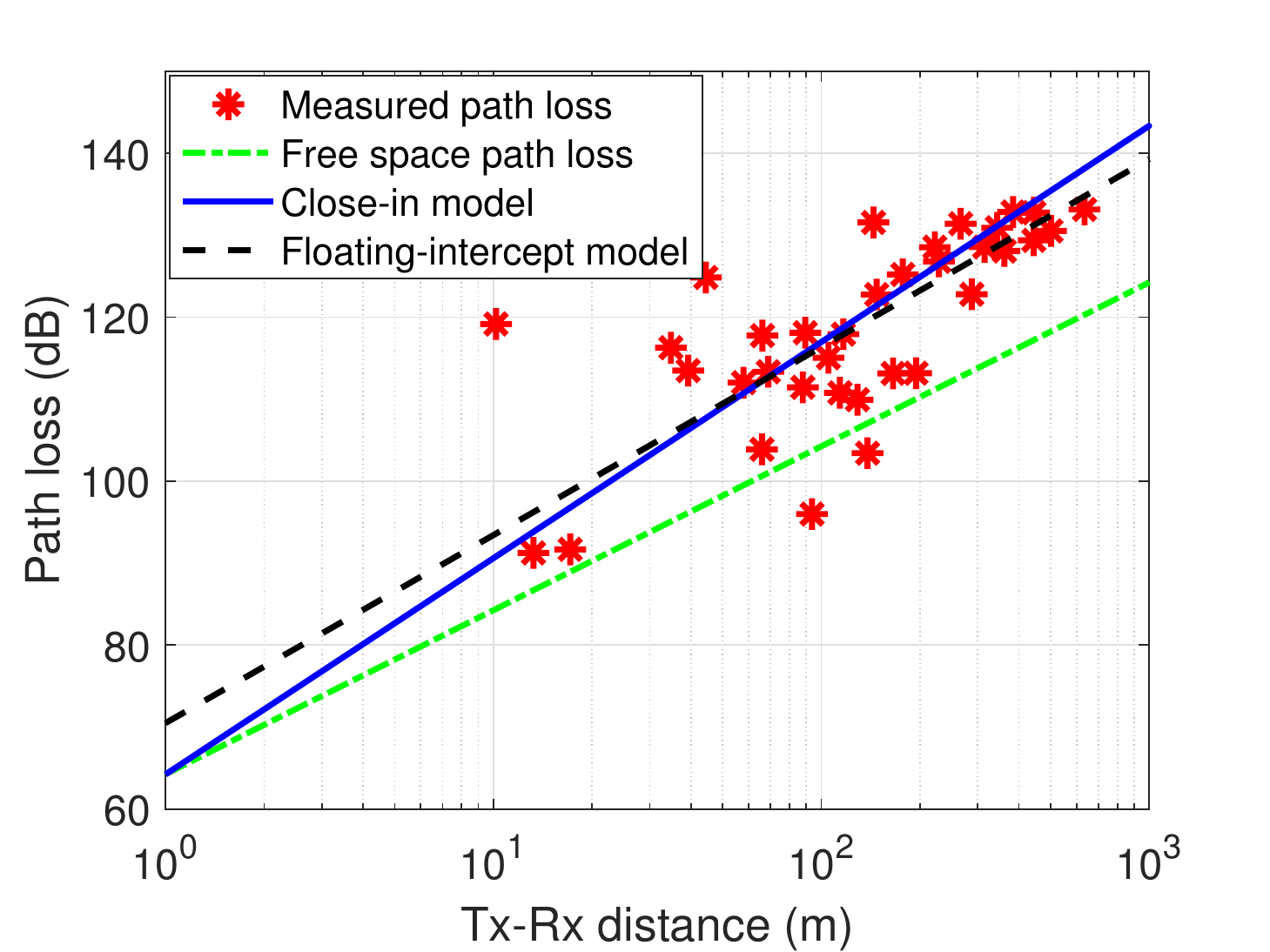} 
} 
\caption{Outdoor path loss measurement results at 28, 32, and 39 GHz bands.} 
\label{fig:pl_m} 
\end{figure}

\begin{figure*}[tb!]
\centering
\begin{minipage}[t]{0.48\linewidth}
\centerline{\includegraphics[width=3.3in]{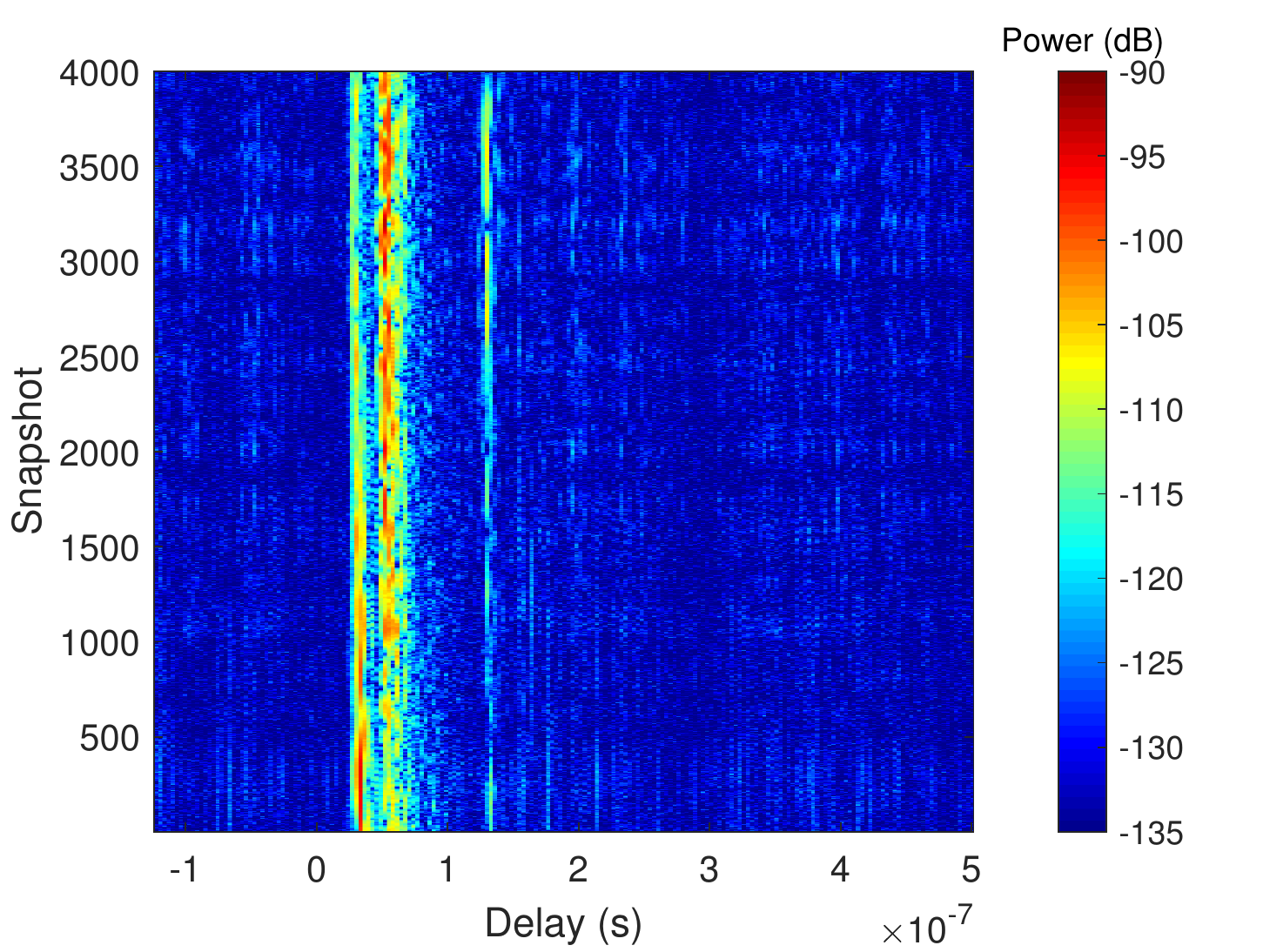}}
\footnotesize \centerline{(a) Tx1}
\end{minipage}
\begin{minipage}[t]{0.48\linewidth}
\centerline{\includegraphics[width=3.3in]{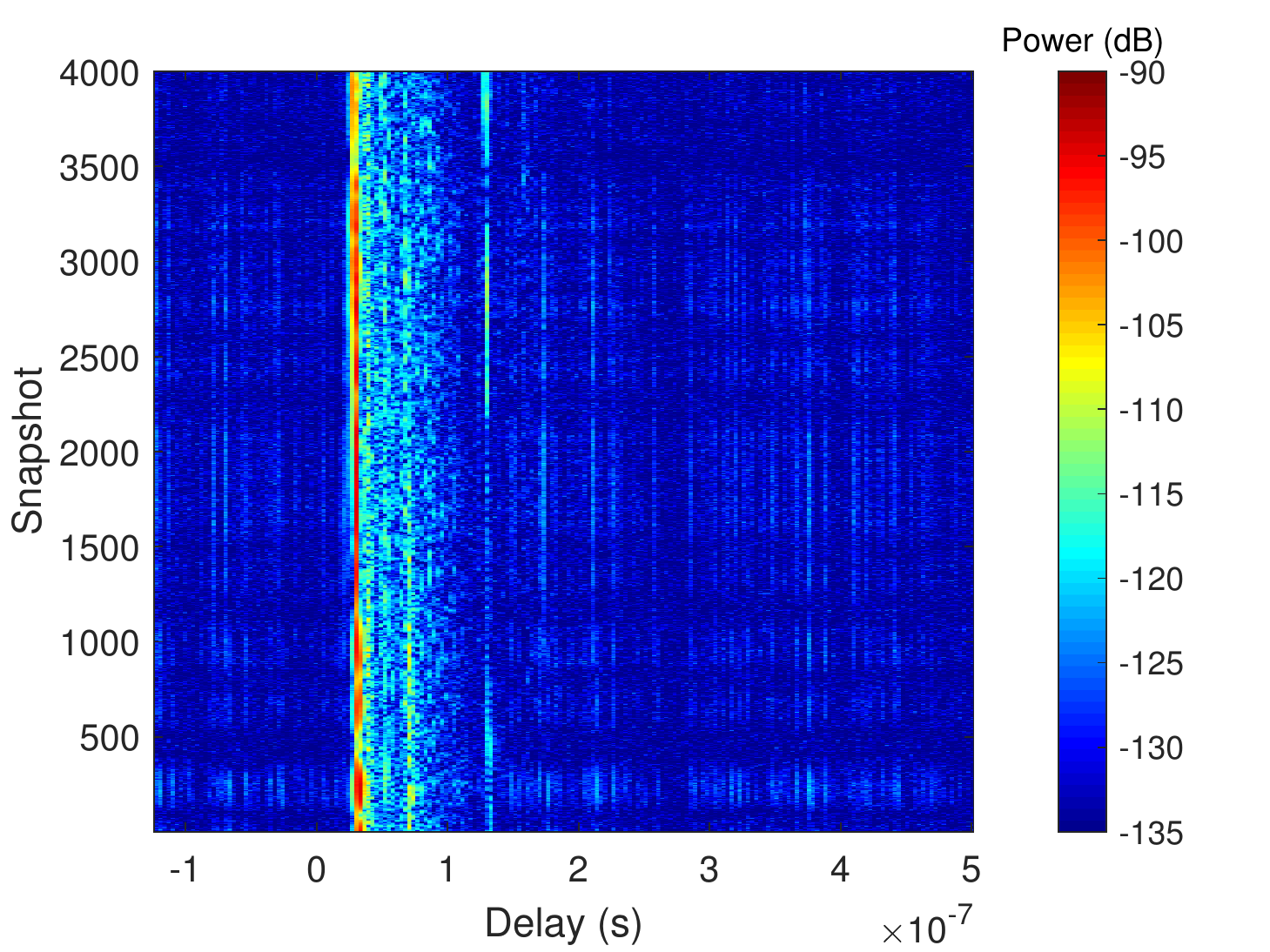}}
\footnotesize \centerline{(b) Tx2}
\end{minipage}

\begin{minipage}[t]{0.48\linewidth}
\centerline{\includegraphics[width=3.3in]{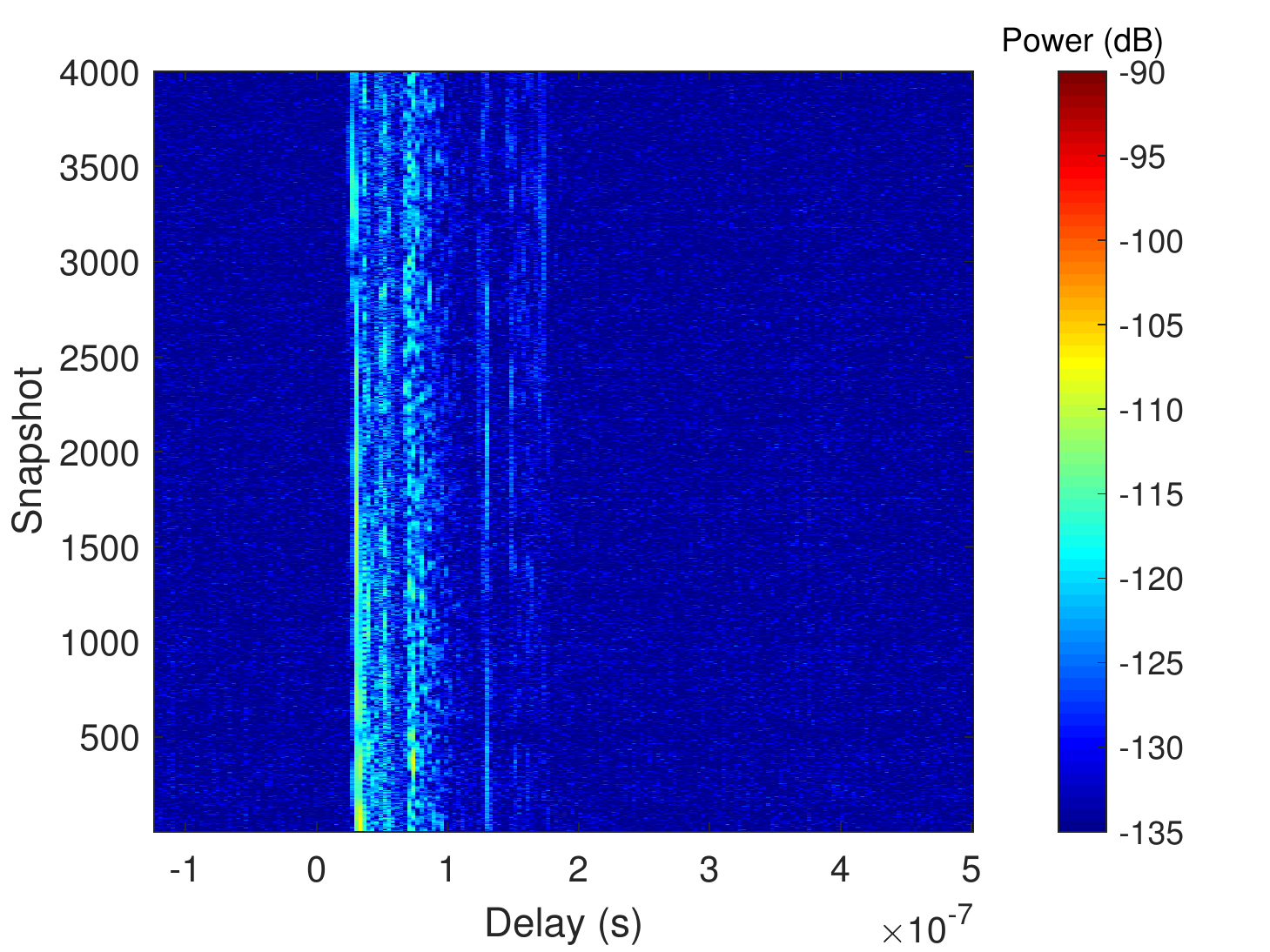}}
\footnotesize \centerline{(c) Tx3}
\end{minipage}
\begin{minipage}[t]{0.48\linewidth}
\centerline{\includegraphics[width=3.3in]{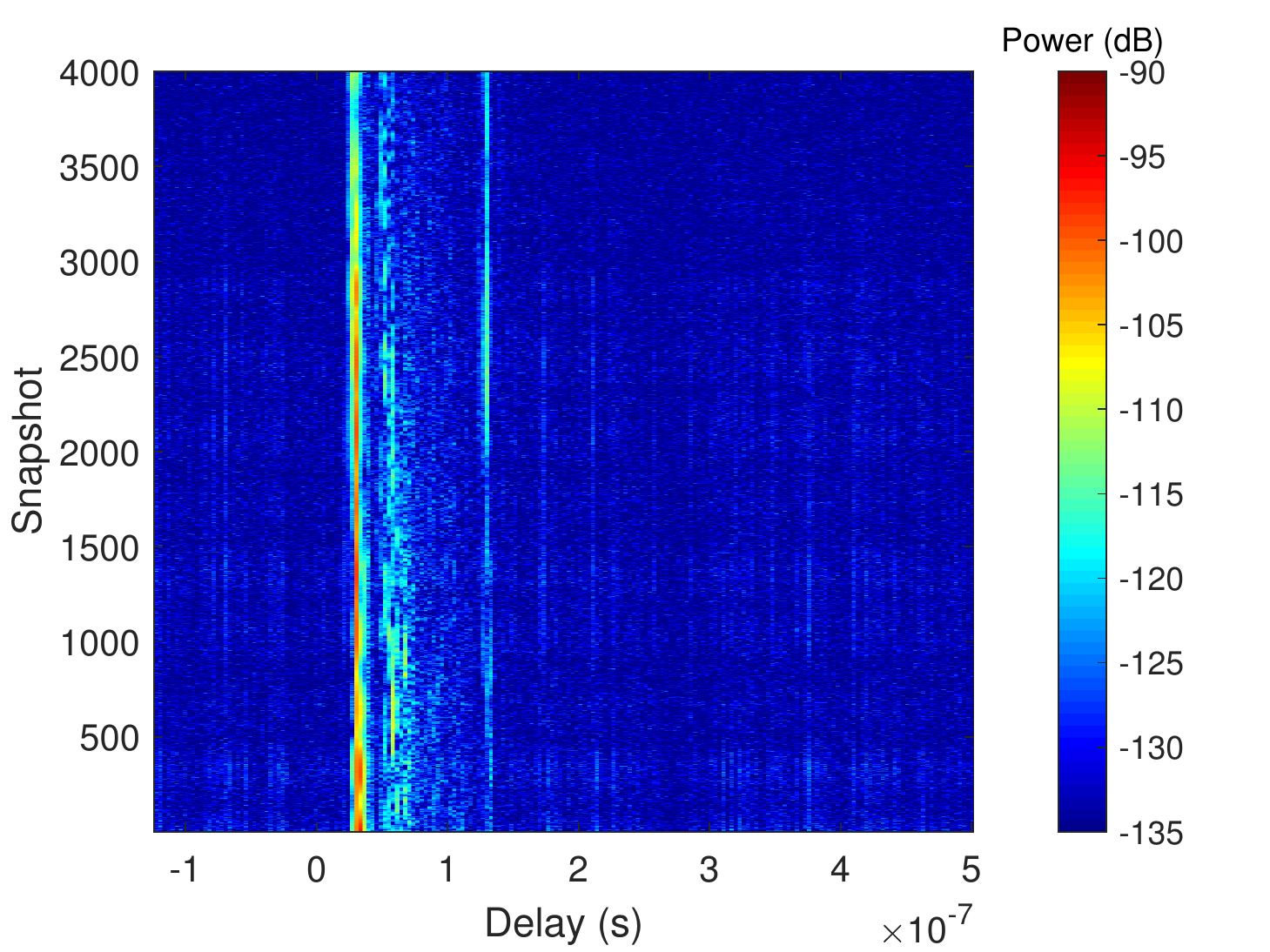}}
\footnotesize \centerline{(c) Tx4}
\end{minipage}
\caption{Time-varying PDPs for Tx1 to Tx4 at 28 GHz with Tx and Rx moving to the same direction.}
\label{fig:v2v_28}
\end{figure*}

The path loss modeling results for 28, 32, and 39 GHz bands are shown in Fig. \ref{fig:pl_m}. The CI model and FI model parameters for the three frequency bands are given in Table \ref{tab1}. As can be seen, the three frequency bands have similar PLEs, without a clear trend with frequency. The conclusion is the same with \cite{Sha18}, where it is said ``there is virtually no difference in the PLE beyond the first meter of free space, such that PLEs are of comparable values and independent of frequency''. However, the shadowing fading does show an increasing trend with the frequency. For 28 GHz band, the path loss at 1 km is about 140 dB, and it is less than 150 dB for 32 GHz and 39 GHz bands. This experiment indicates that mmWave is possible to be used with horn antennas for long-range communications up to 1 km. Note that Fig. \ref{fig:pl_m} is the path loss modeling results for one of the transmitting directions. Thus, the signal level may not be the strongest ones. If advanced beamforming and tracking technology is used, the Tx can always transmit signals toward the strongest directions, which will further reduce the path loss and extend the applicable distance of mmWave communications. 

\begin{table}[bt!]
\caption{CI model and FI model parameters for 28, 32, and 39 GHz bands.}
\setlength{\tabcolsep}{3pt}
\begin{tabular}{|p{80pt}|p{40pt}|p{40pt}|p{40pt}|}
\hline
Frequency band (GHz)& 28&32&39\\
\hline
CI model parameters, ($n$,$\sigma$)&	2.637, 5.47&2.964, 6.07&2.638, 9.72\\
\hline
FI model parameters, ($\alpha$,$\beta$,$\sigma$)&67.31, 2.374, 6.57&76.36, 2.263, 5.67&70.48, 2.294, 9.80\\
\hline
\end{tabular}
\label{tab1}
\end{table}

\subsection{Non-stationarity and spatial consistency}

\begin{figure*}[tb!]
\centering
\begin{minipage}[t]{0.48\linewidth}
\centerline{\includegraphics[width=3.3in]{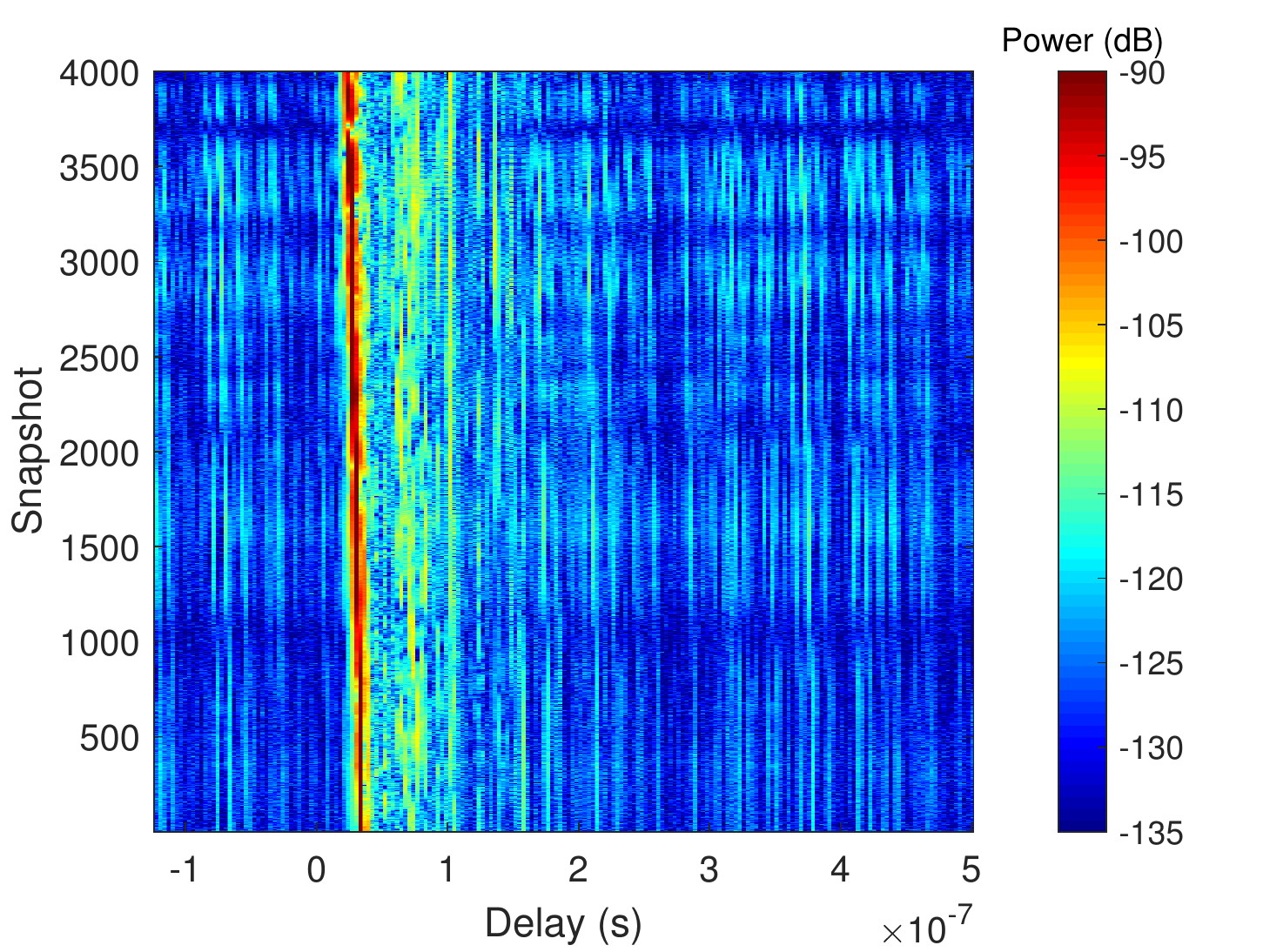}}
\footnotesize \centerline{(a) Tx1}
\end{minipage}
\begin{minipage}[t]{0.48\linewidth}
\centerline{\includegraphics[width=3.3in]{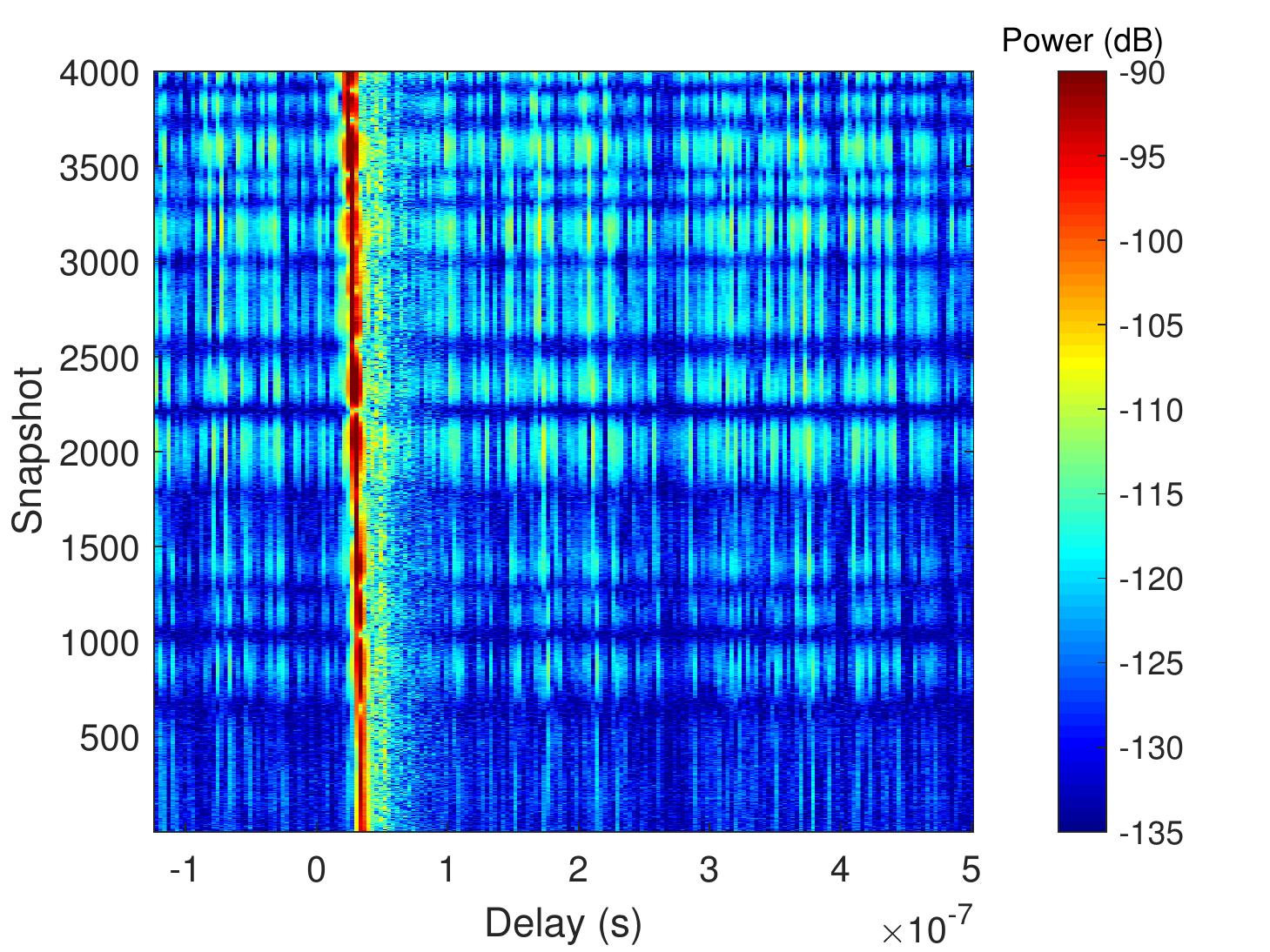}}
\footnotesize \centerline{(b) Tx2}
\end{minipage}

\begin{minipage}[t]{0.48\linewidth}
\centerline{\includegraphics[width=3.3in]{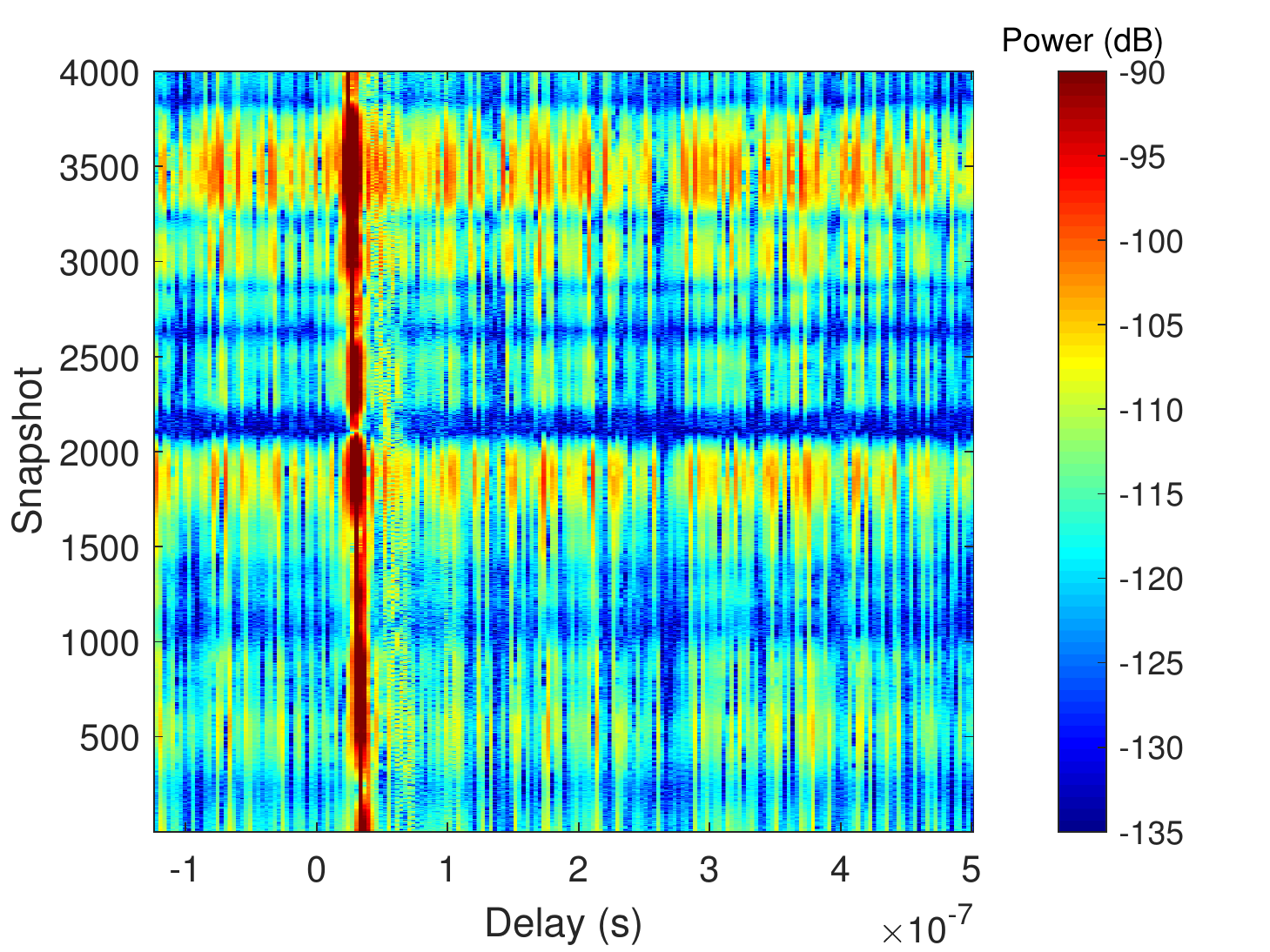}}
\footnotesize \centerline{(c) Tx3}
\end{minipage}
\begin{minipage}[t]{0.48\linewidth}
\centerline{\includegraphics[width=3.3in]{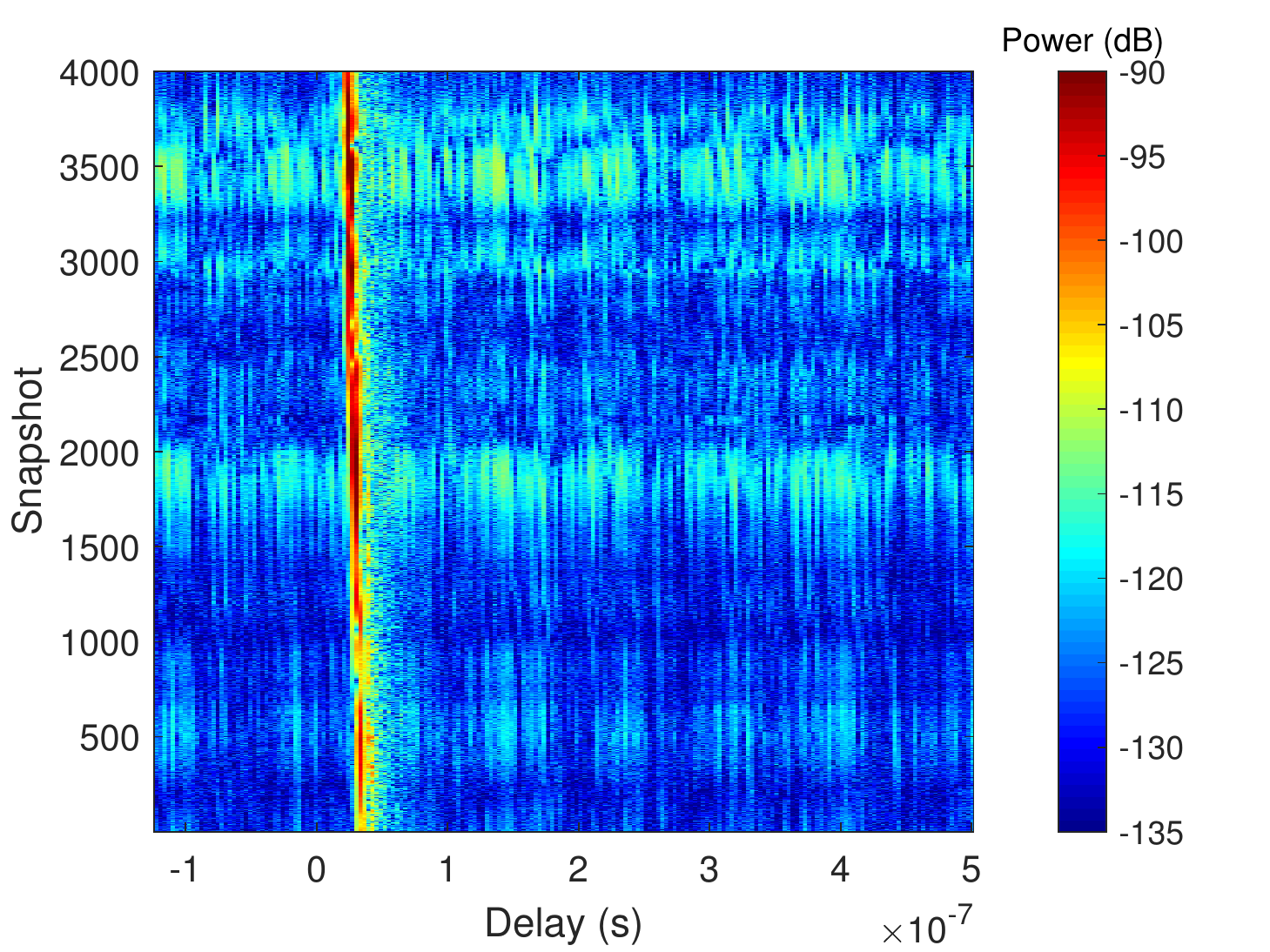}}
\footnotesize \centerline{(c) Tx4}
\end{minipage}
\caption{Time-varying PDPs for Tx1 to Tx4 at 28 GHz with Tx and Rx moving to the opposite directions.}
\label{fig:v2v_c_28}
\end{figure*}

\begin{figure*}[tb!]
\centering
\begin{minipage}[t]{0.32\linewidth}
\centerline{\includegraphics[width=2.4in]{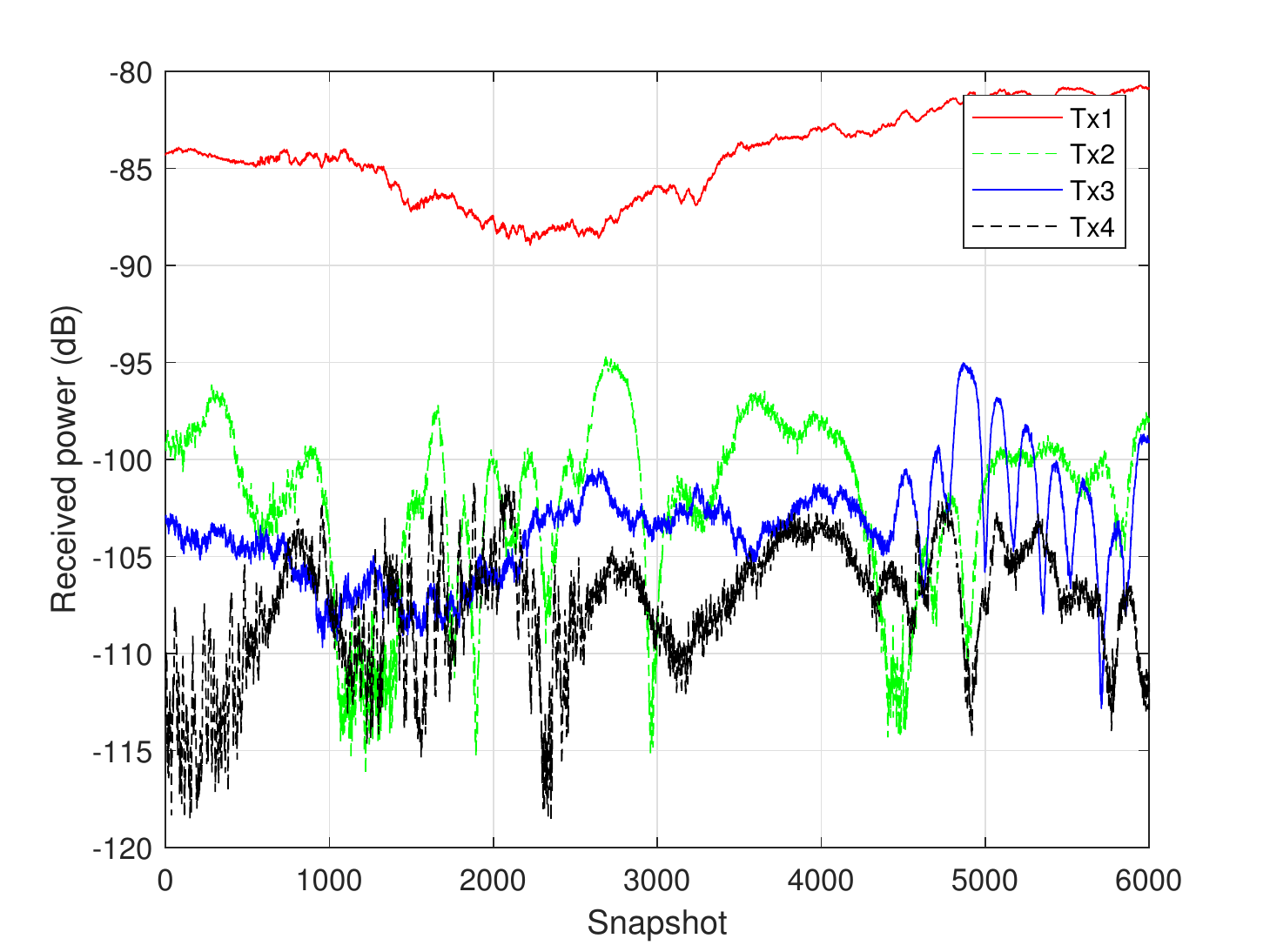}}
\footnotesize \centerline{(a) 28 GHz}
\end{minipage}
\begin{minipage}[t]{0.32\linewidth}
\centerline{\includegraphics[width=2.4in]{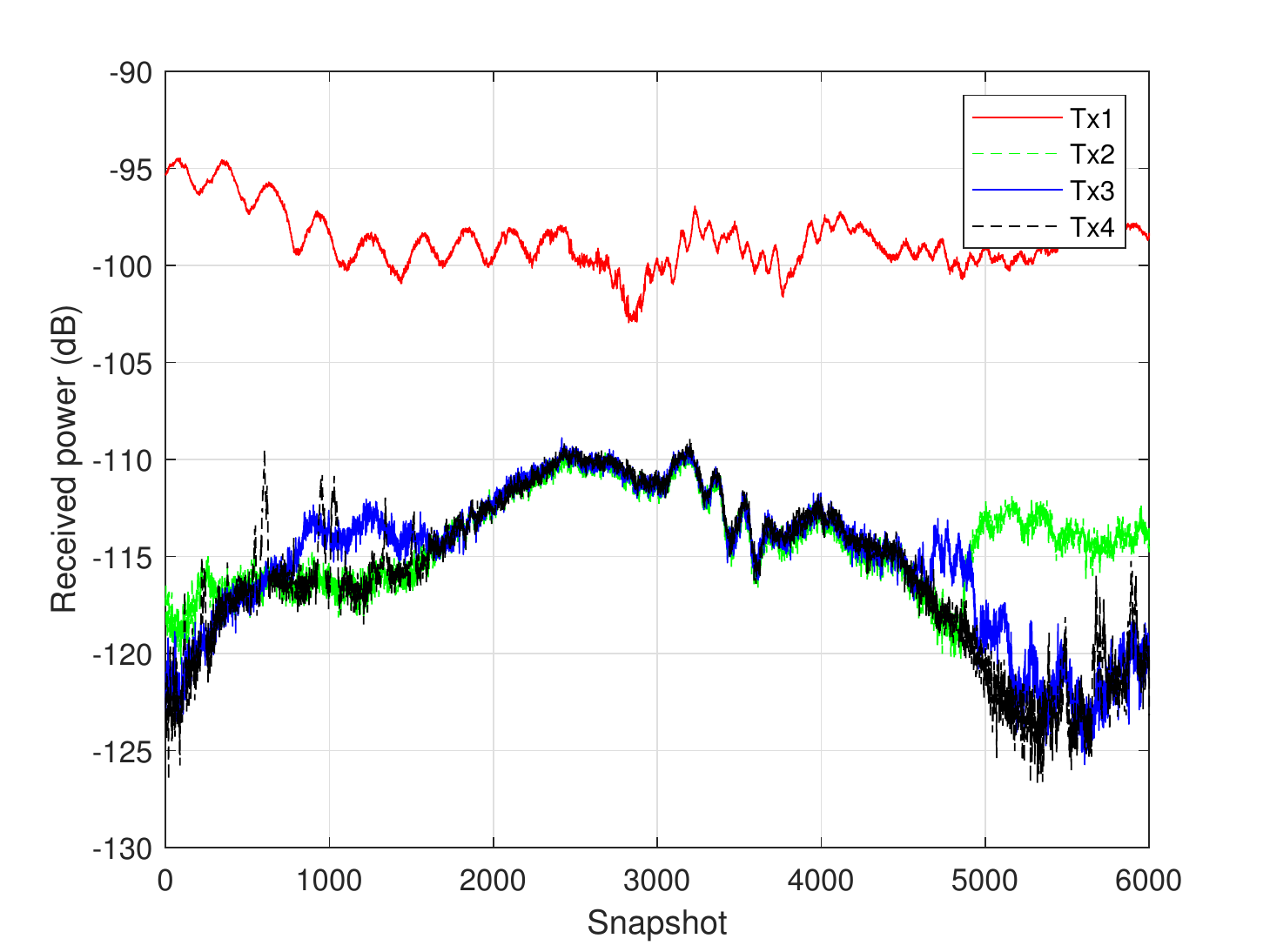}}
\footnotesize \centerline{(b) 32 GHz}
\end{minipage}
\begin{minipage}[t]{0.32\linewidth}
\centerline{\includegraphics[width=2.4in]{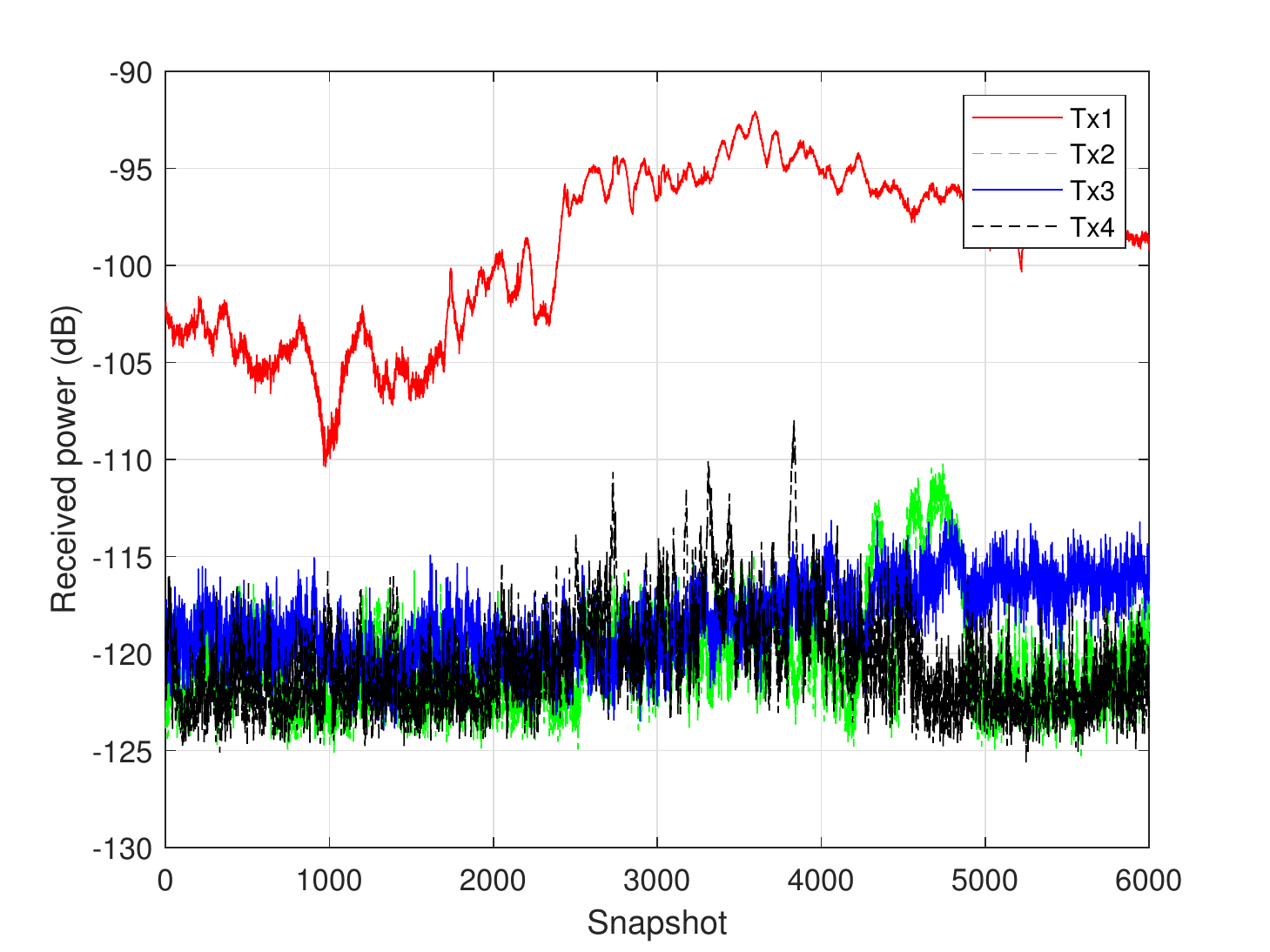}}
\footnotesize \centerline{(c) 39 GHz}
\end{minipage}
\caption{The LOS power variations at 28, 32, and 39 GHz bands with Tx and Rx moving to the same direction for the four antenna directions.}
\label{fig:v2v}
\end{figure*}

\begin{figure}[tb!]
\centering
\includegraphics[width=3.3in]{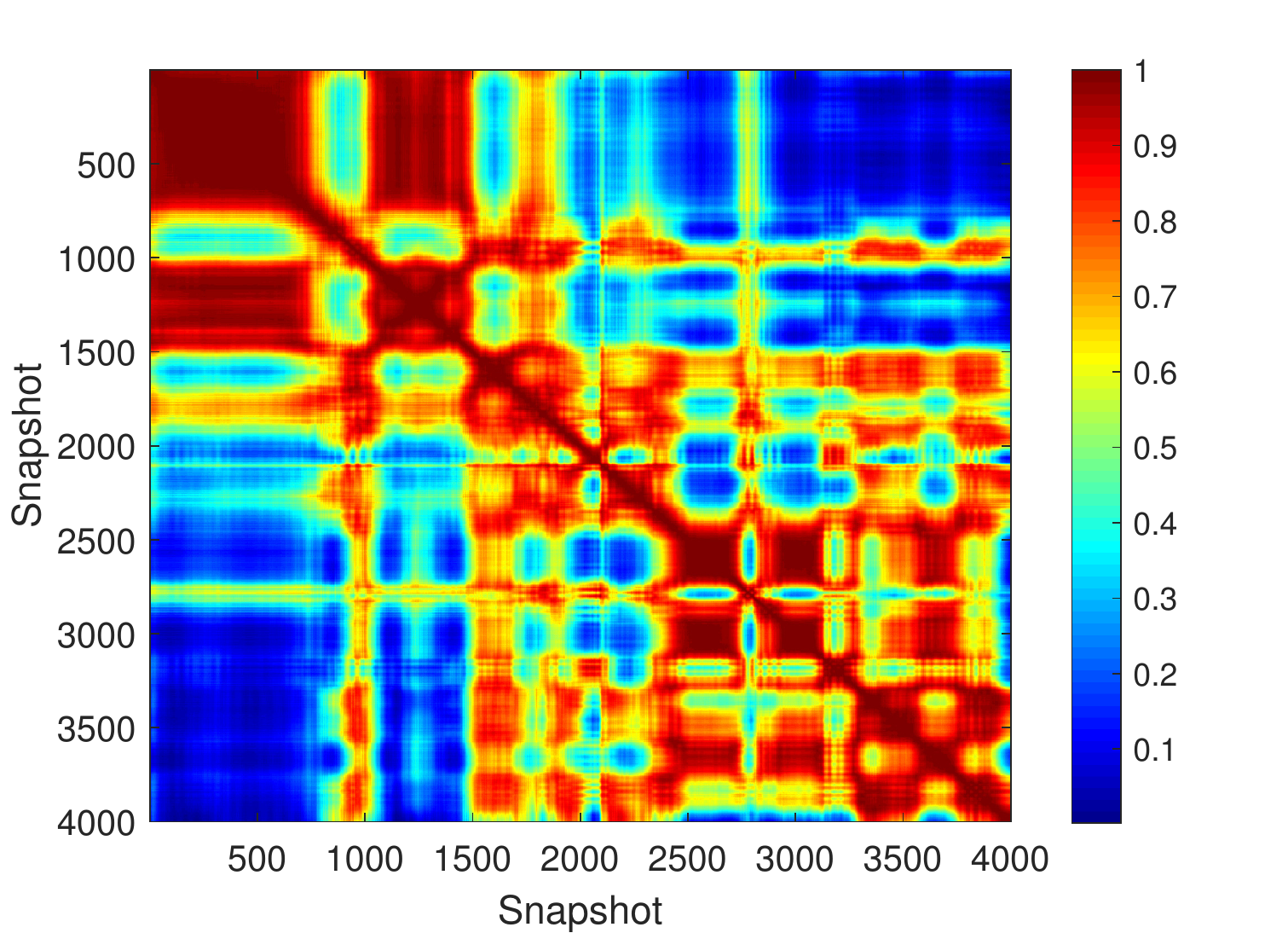}
\caption{The correlation matrix of different snapshots at 28 GHz band with Tx and Rx moving to the same direction.}
\label{fig:si}
\end{figure}

\begin{figure}[tb!]
\centering
\includegraphics[width=3.3in]{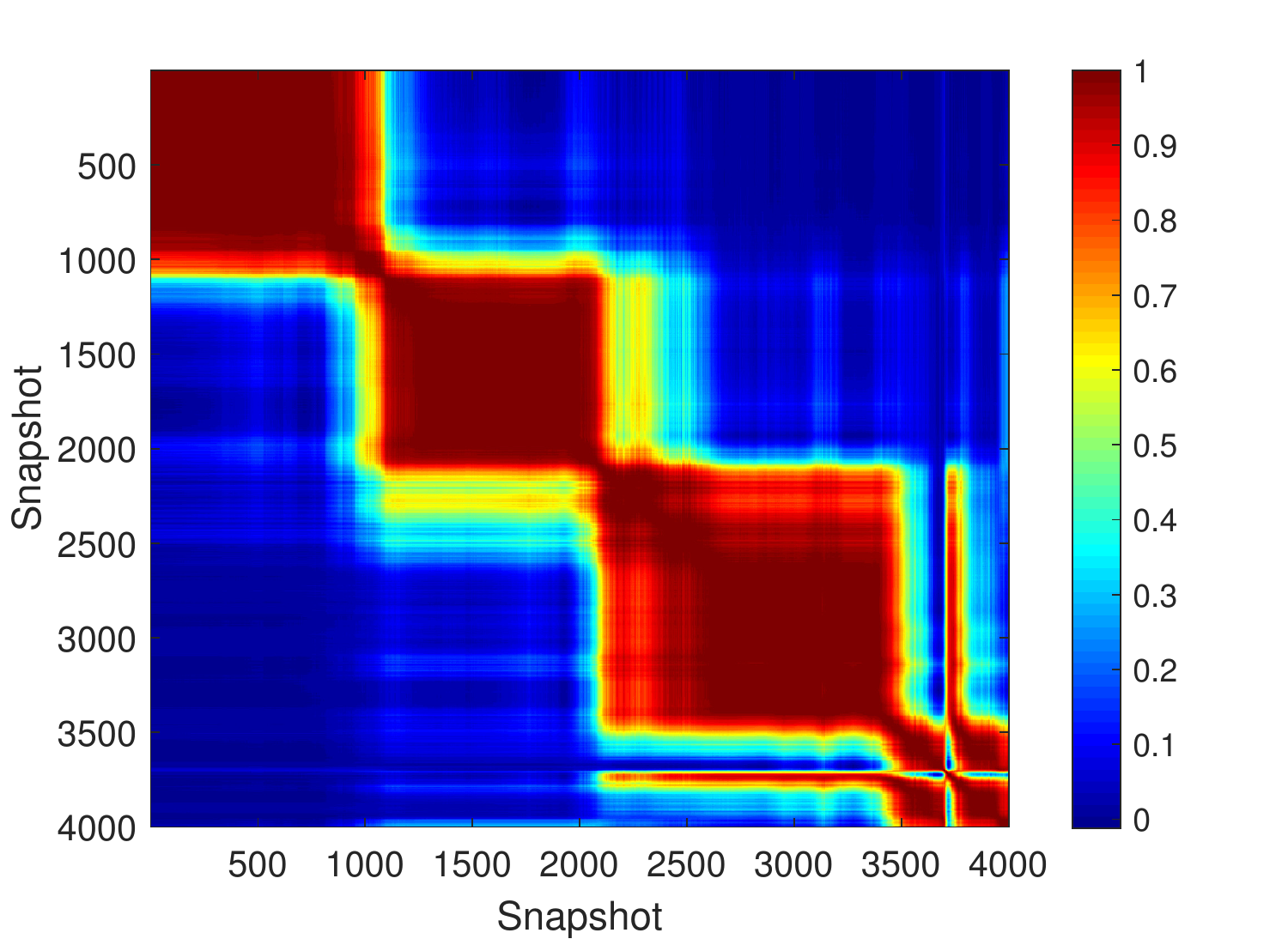}
\caption{The correlation matrix of different snapshots at 28 GHz band with Tx and Rx moving to the opposite directions.}
\label{fig:sic}
\end{figure}

\begin{figure}[tb!]
\centering
\includegraphics[width=3.3in]{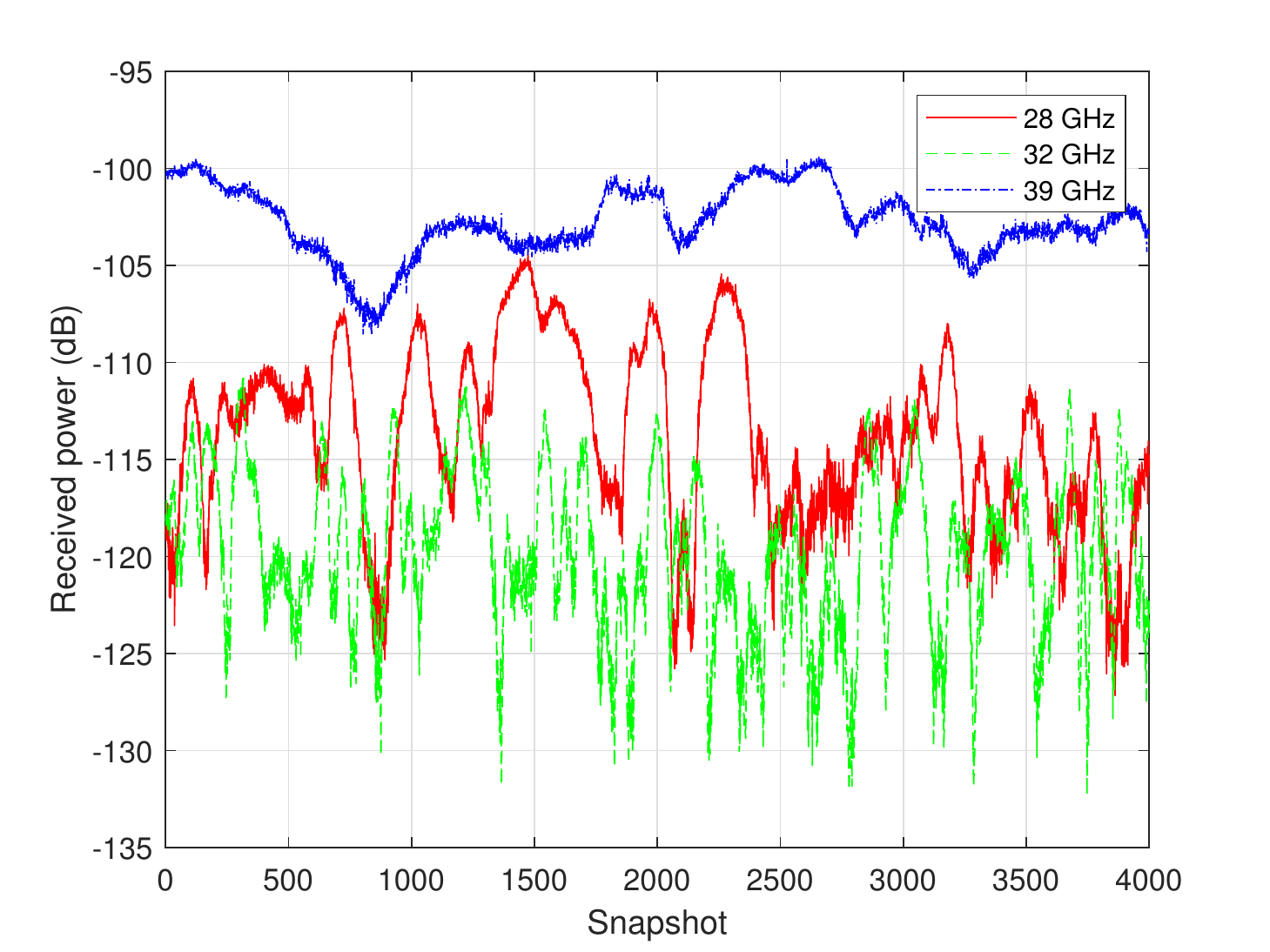}
\caption{The LOS power variations at 28, 32, and 39 GHz bands with Tx and Rx moving to the opposite directions.}
\label{fig:v2v_c}
\end{figure}

Mmwave can be used for vehicular networks to support high data transmission rate. MmWave V2V channel is challenging as it is highly dynamic. The type of vehicles, vehicle velocity, road situation, and surrounding environment will have effects on the channel characteristics. The mmWave V2V channel is essentially non-stationary and has Doppler frequency shift caused by the movements of vehicles.

Fig. \ref{fig:v2v_28} shows the measured PDP variations over time with Tx and Rx moving to the same direction at 28 GHz band. Note that a noise threshold of -135 dB is used, which can be seen from the lower limit of the colorbar. Up to 4000 snapshots are captured. There are several MPCs for the signals received from the four directions. The MPCs show non-stationarity and cluster birth-death over time. Fig.~\ref{fig:v2v_c_28} shows the measured PDP variations over time with Tx and Rx moving to the opposite directions. Compared to the results of the same direction, it has less MPCs as the Tx and Rx are moving away from each other.

Fig. \ref{fig:v2v} shows the LOS power variations at 28, 32, and 39 ~GHz bands. As Tx1 is toward the Rx, the received signal from Tx1 has the strongest power, which is about 20-30 dB higher than the other directions. The LOS power also variates over different snapshots, with a variation up to 10 dB. The results also show spatial consistency of mmWave V2V channel as the power is changing smoothly over different snapshots. 

Fig. \ref{fig:si} shows the correlation matrix of PDP in different snapshots at 28 GHz band with Tx and Rx moving to the same direction.  The correlation matrix is a measure of the stationary interval. The first 700 snapshots show high correlations, in which the channel can be seen as stationary. As a comparison, Fig. \ref{fig:sic} shows the correlation matrix of PDP in different snapshots at 28 GHz band with Tx and Rx moving to the opposite directions. The channel shows high correlations in three durations. 

As shown in Fig. \ref{fig:v2v_c}, for the case of Tx and Rx moving to the opposite directions, the LOS power variation is more severe than that of the same direction, as the relative moving speed is larger and the Doppler frequency shift is larger.

\section{Conclusions and Future Works}
In this paper, channel measurements have been conducted with 4$\times$4 MIMO antenna configurations at multi-frequency mmWave bands (28, 32, and 39 GHz) and in multi-scenarios for B5G wireless communication systems. The detailed measurement and modeling results have been analyzed.

The indoor human blockage, outdoor human blockage, and vehicle blockage have been investigated. For human blockage, the attenuation shows a small variation when a person walks along the LOS path. The attenuation is about 10-15 dB. When a person walks across the LOS path, a deep fading will happen with attenuation of 10-15 dB. The attenuation shows a slightly increasing trend with frequency.
The METIS KED model can act as the lower bound for the attenuation, while the Kirchhoff KED model can be the upper bound. By comparing the indoor and outdoor human blockage effects, we find that outdoor environment has less MPCs and the signal shows a larger variation when  it is blocked by human body. In addition, the vehicle blockage shows a much higher attenuation than human body and variates significantly for different parts of the vehicle, as the material and shape of a vehicle are very complex. These measurement and modeling results are meaningful for future indoor hotspot and vehicular network applications. 

The outdoor path loss measurements have been conducted in UMa scenario with Tx-Rx distance ranges up to 1 km. Though the signal has a probability of outage at some places, our measurement results prove that mmWave can propagate up to 1 km with a dynamic range of 150 dB. The PLE shows comparable values for different bands, which means the PLE is independent of the frequency. The shadowing fading, however, has an increasing trend with frequency. The non-stationary and spatial-consistency properties are also validated from the measurements. These measurement and modeling results are important for outdoor mmWave communications in UMa scenario.

In V2V measurements, we study the mmWave signal propagations with antennas toward four different directions. Two cases have been measured, i.e., vehicles move to the same direction and to the opposite directions. Surprisingly, we can receive the signals from the four directions with a strong path and some reflected paths, even the signal strength is not the same at the four directions. This implies that section antennas, massive MIMO, and advanced beamforming and tracking technologies can be used to achieve robust communication link. These beneficial results will provide useful guidelines for future multi-antenna technologies and vehicular networks.

In the future, we will conduct mmWave MIMO channel measurements for more frequency bands and more scenarios. Other channel characteristics such as atmosphere attenuation, vegetation attenuation, and O2I building entry loss will also be measured and modeled to have a deeper understanding of mmWave communication channels. 

Meanwhile, as the frequency range, signal bandwidth, antenna numbers, and application scenarios increase rapidly, advanced technologies such as artificial intelligence and machine learning will be fully applied to and integrated with B5G communication systems. Potential machine learning algorithms, such as artificial neural network, convolutional neural network, generative adversarial network, and clustering and classification algorithms may be used to channel modeling, such as channel characteristics prediction, MPCs clustering, and scenarios classification, based on the massive channel measurement datasets \cite{TBD18}. Other new massive MIMO technologies, such as cell-free massive MIMO \cite{Ngo17, Nay17, Ngo18}, ultra-massive MIMO \cite{Han18, Nie19}, and large intelligent surface \cite{Hu18, Hu18_2} should also be evaluated by real channel measurements.
 
%


%

\appendices



\ifCLASSOPTIONcaptionsoff
  \newpage
\fi

\begin{IEEEbiography}[{\includegraphics[width=1in,height=1.25in,clip,keepaspectratio]{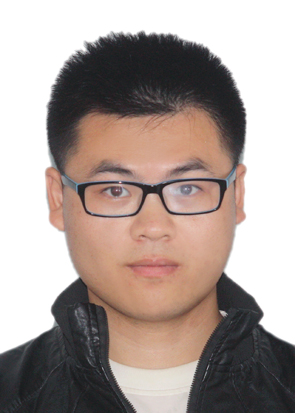}}]{Jie Huang} received the B.E. degree in Information Engineering from Xidian University, China, in 2013, and the Ph.D. degree in Communication and Information Systems from Shandong University, China, in 2018. 

From Jan. 2019 to Feb. 2020, he was a Postdoctoral Research Associate in Durham University, U.K. He is currently a Postdoctoral Research Associate in the National Mobile Communications Research Laboratory, Southeast University, China and also a researcher in Purple Mountain Laboratories, China. He received the Best Student Paper Award at WPMC'16. His research interests include millimeter wave and massive MIMO channel measurements and channel modeling, wireless big data, and B5G/6G wireless communications. 
\end{IEEEbiography}

\begin{IEEEbiography}
[{\includegraphics[width=1in,height=1.25in,clip,keepaspectratio]{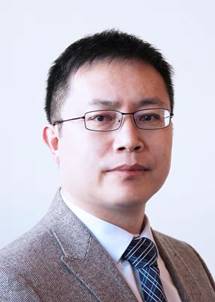}}]{Cheng-Xiang Wang} (S'01-M'05-SM'08-F'17) received the B.Sc. and M.Eng. degrees in Communication and Information Systems from Shandong University, China, in 1997 and 2000, respectively, and the Ph.D. degree in Wireless Communications from Aalborg University, Denmark, in 2004.

He was a Research Assistant with the Hamburg University of Technology, Hamburg, Germany, from 2000 to 2001, a Visiting Researcher with Siemens AG Mobile Phones, Munich, Germany, in 2004, and a Research Fellow with the University of Agder, Grimstad, Norway, from 2001 to 2005. He has been with Heriot-Watt University, Edinburgh, U.K., since 2005, where he was promoted to a Professor in 2011. In 2018, he joined Southeast University, China, as a Professor. He is also a part-time professor with the Purple Mountain Laboratories, Nanjing, China. He has authored three books, one book chapter, and more than 370 papers in refereed journals and conference proceedings, including 23 Highly Cited Papers. He has also delivered 18 Invited Keynote Speeches/Talks and 7 Tutorials in international conferences. His current research interests include wireless channel measurements and modeling, B5G wireless communication networks, and applying artificial intelligence to wireless communication networks.

Prof. Wang is a fellow of the IET, an IEEE Communications Society Distinguished Lecturer in 2019 and 2020, and a Highly-Cited Researcher recognized by Clarivate Analytics, in 2017-2019. He is currently an Executive Editorial Committee member for the IEEE TRANSACTIONS ON WIRELESS COMMUNICATIONS. He has served as an Editor for nine international journals, including the IEEE TRANSACTIONS ON WIRELESS COMMUNICATIONS from 2007 to 2009, the IEEE TRANSACTIONS ON VEHICULAR TECHNOLOGY from 2011 to 2017, and the IEEE TRANSACTIONS ON COMMUNICATIONS from 2015 to 2017. He was a Guest Editor for the IEEE JOURNAL ON SELECTED AREAS IN COMMUNICATIONS, Special Issue on Vehicular Communications and Networks (Lead Guest Editor), Special Issue on Spectrum and Energy Efficient Design of Wireless Communication Networks, and Special Issue on Airborne Communication Networks. He was also a Guest Editor for the IEEE TRANSACTIONS ON BIG DATA, Special Issue on Wireless Big Data, and is a Guest Editor for the IEEE TRANSACTIONS ON COGNITIVE COMMUNICATIONS AND NETWORKING, Special Issue on Intelligent Resource Management for 5G and Beyond. He has served as a TPC Member, TPC Chair, and General Chair for over 80 international conferences. He received ten Best Paper Awards from IEEE GLOBECOM 2010, IEEE ICCT 2011, ITST 2012, IEEE VTC 2013-Spring, IWCMC 2015, IWCMC 2016, IEEE/CIC ICCC 2016, WPMC 2016, and WOCC 2019. 
\end{IEEEbiography}

\begin{IEEEbiography}
[{\includegraphics[width=1in,height=1.25in,clip,keepaspectratio]{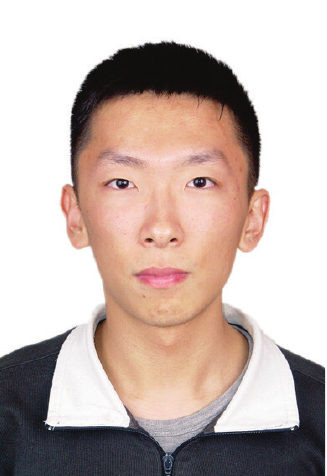}}]{Hengtai Chang} received the B.Sc. degree at School of Information Science and Engineering, Shandong University, Qingdao, China, in 2016. He is currently pursuing the Ph.D. degree at the School of Information Science and Engineering, Shandong University, Qingdao, China. His current research interests include UAV communications, channel measurements, wireless propagation channel modeling, and 5G channel modeling.
\end{IEEEbiography}

\begin{IEEEbiography}
[{\includegraphics[width=1in,height=1.25in,clip,keepaspectratio]{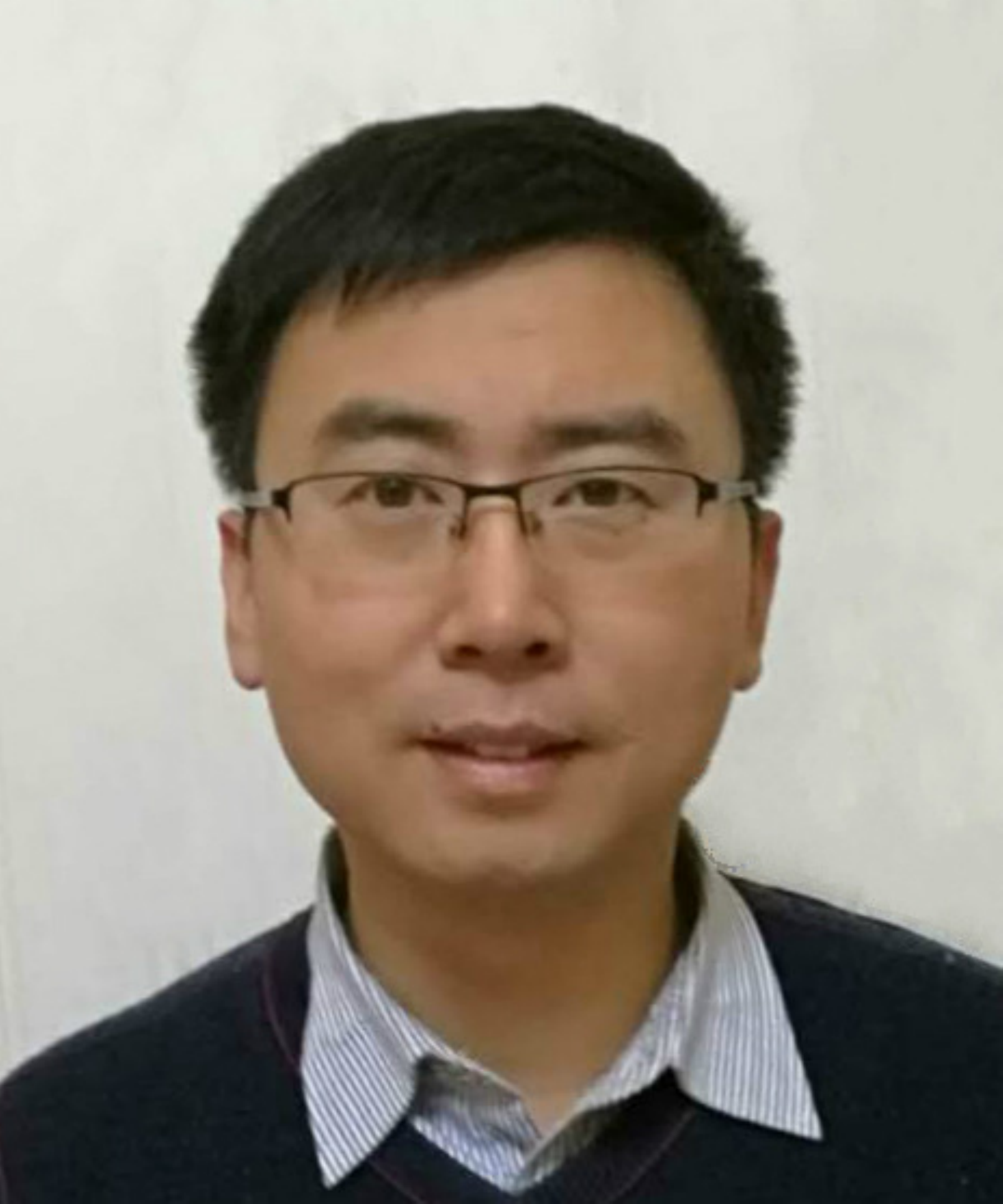}}]{Jian Sun} (M'08) received the B.Sc. degree in applied electronic technology, the M.Eng. degree in measuring and testing technologies and instruments, and the Ph.D. degree in communication
and information systems, all from Zhejiang University, Hangzhou, China, in 1996, 1999, and 2005, respectively.

From 2005 to 2018, he was a Lecturer with the School of Information Science and Engineering, Shandong University, China. Since 2018, he has been an Associate Professor. In 2008, he was a Visting Scholar with University of California San Diego (UCSD). In 2011, he was a Visiting Scholar with Heriot-Watt University, U.K., supported by U.K.–China Science Bridges: R\&D on (B)4G Wireless Mobile Communications project. His current research interests include signal processing for wireless communications, channel sounding and modeling, propagation measurement and parameter extraction, maritime communication, visible light communication, software defined radio, MIMO, multicarrier, wireless systems design and implementation.
\end{IEEEbiography}

\begin{IEEEbiography}
[{\includegraphics[width=1in,height=1.25in,clip,keepaspectratio]{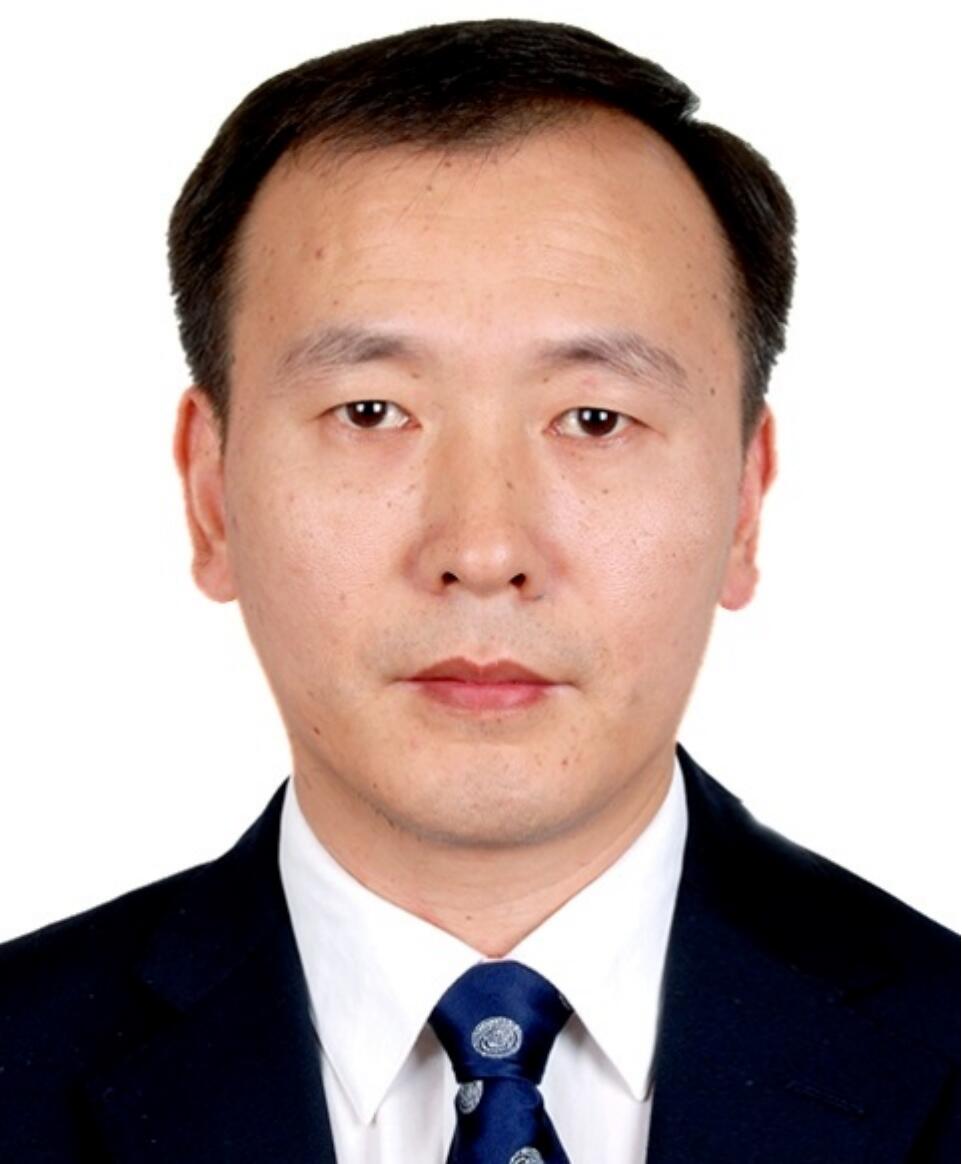}}]{Xiqi Gao} (S'92–AM'96–M'02–SM'07–F'15) received the Ph.D. degree in electrical engineering from Southeast University, Nanjing, China, in 1997.

He joined the Department of Radio Engineering, Southeast University, in April 1992. Since May 2001, he has been a professor of information systems and communications. From September 1999 to August 2000, he was a visiting scholar at Massachusetts Institute of Technology, Cambridge, MA, USA, and Boston University, Boston, MA. From August 2007 to July 2008, he visited the Darmstadt University of Technology, Darmstadt, Germany, as a Humboldt scholar. His current research interests include broadband multicarrier communications, MIMO
wireless communications, channel estimation, and turbo equalization, and multirate signal processing for wireless communications. From 2007 to 2012, he served as an Editor for the IEEE TRANSACTIONS ON WIRELESS
COMMUNICATIONS. From 2009 to 2013, he served as an Editor for the IEEE
TRANSACTIONS ON SIGNAL PROCESSING. From 2015 to 2017, he served as
an Editor for the IEEE TRANSACTIONS ON COMMUNICATIONS. 

Dr. Gao received the Science and Technology Awards of the State Education Ministry of China in 1998, 2006 and 2009, the National Technological Invention Award of China in 2011, and the 2011 IEEE Communications Society Stephen O. Rice Prize Paper Award in the Field of Communications Theory.
\end{IEEEbiography}


\begin{thebibliography}{1}

\bibitem{Huang17} J. Huang, C.-X. Wang, R. Feng, J. Sun, W. Zhang, and Y. Yang, ``Multi-frequency mmWave massive MIMO channel measurements and characterization for 5G wireless communication systems,'' \emph{IEEE J. Sel. Areas Commun.}, vol. 35, no.~7, pp. 1591--1605, Jul. 2017.

\bibitem{Huang19} J. Huang, Y. Liu, C.-X. Wang, J. Sun, and H. Xiao, ``5G millimeter wave channel sounders, measurements, and models: Recent developments and future challenges,'' \emph{IEEE Commun. Mag.}, vol. 57, no.~1, pp. 138--145, Jan. 2019.

\bibitem{COMST19} C.-X. Wang, J. Bian, J. Sun, W. Zhang, and M. Zhang, ``A survey of 5G channel measurements and models,'' \emph{IEEE Commun. Surveys Tuts.}, vol.~20, no. 4, pp. 3142--3168, 4th Quart. 2018.

\bibitem{Rap13} T. S. Rappaport, S. Sun, R. Mayzus, H. Zhao, Y. Azar, K. Wang, G. N. Wong, J. K. Schulz, M. Samimi, and F. Gutierrez, ``Millimeter wave mobile communications for 5G cellular: It will work!'' \emph{IEEE Access}, vol.~1, pp. 335--349, May 2013.

\bibitem{Ben11} E. Ben-Dor, T. S. Rappaport, Y. Qiao, and S. J. Lauffenburger, ``Millimeter-wave 60 GHz outdoor and vehicle AOA propagation measurements using a broadband channel sounder,'' in \emph{Proc. IEEE Globecom'11}, Houston, Texas, Dec. 2011, pp. 1--6.

\bibitem{Rap13_2} T. S. Rappaport, F. Gutierrez, E. Ben-Dor, J. N. Murdock, Y. Qiao, and J. I. Tamir, ``Broadband millimeter-wave propagation measurements and models using adaptive-beam antennas for outdoor urban cellular communications,'' \emph{IEEE Trans. Antennas Propag.}, vol. 61, no.~4, pp.~1850--1859, Apr. 2013.

\bibitem{Sam13} M. Samimi, K. Wang, Y. Azar, G. N. Wong, R. Mayzus, H. Zhao, J. K. Schulz, S. Sun, F. Gutierrez, and T. S. Rappaport, ``28 GHz angle of arrival and angle of departure analysis for outdoor cellular communications using steerable beam antennas in New York city,'' in \emph{Proc. IEEE VTC'13-Spring}, Dresden, Germany, Jun. 2013, pp. 1--6.

\bibitem{Aza13} Y. Azar, G. N. Wong, K. Wang, R. Mayzus, J. K. Schulz, H. Zhao, F. Gutierrez, D. Hwang, and T. S. Rappaport, ``28 GHz propagation measurements for outdoor cellular communications using steerable beam antennas in New York city,'' in \emph{Proc. IEEE ICC'13}, Budapest, Hungary, Jun. 2013, pp. 5143--5147.

\bibitem{Mac14} G. R. MacCartney and T. S. Rappaport, ``73 GHz millimeter wave propagation measurements for outdoor urban mobile and backhaul communications in New York city,'' in \emph{Proc. IEEE ICC'14}, Sydney, NSW, Jun. 2014, pp. 4862--4867.

\bibitem{Sam15} M. K. Samimi and T. S. Rappaport, ``3-D statistical channel model for millimeter-wave outdoor mobile broadband communications,'' in \emph{Proc. IEEE ICC'15}, London, UK, Jun. 2015, pp. 2430--2436.

\bibitem{Sun16} S. Sun, G. R. MacCartney, and T. S. Rappaport, ``Millimeter-wave distance-dependent large-scale propagation measurements and path loss models for outdoor and indoor 5G systems,'' in \emph{Proc. EuCAP'16}, Davos, Switzerland, Apr. 2016, pp. 1--5.

\bibitem{NYUSIM} NYUSIM, https://wireless.engineering.nyu.edu/nyusim/

\bibitem{ITU2406} ITU-R P.2406-0, \emph{Studies for short-path propagation data and models for terrestrial radio communication systems in the frequency range 6 GHz to 100 GHz}, ITU, Sept. 2017.

\bibitem{Wei14} R. J. Weiler, M. Peter, W. Keusgen, and M. Wisotzki, ``Measuring the busy urban 60 GHz outdoor access radio channel,'' in \emph{Proc. ICUWB'14}, Paris, France, Sept. 2014, pp. 166--170.

\bibitem{Zhao17} X. Zhao, S. Li, Q. Wang, M. Wang, S. Sun, and W. Hong, ``Channel measurements, modeling, simulation and validation at 32 GHz in outdoor microcells for 5G radio systems,'' \emph{IEEE Access}, vol. 5, pp. 1062--1072, Jan. 2017.

\bibitem{Zhou17} L. Zhou, L. Xiao, Z. Yang, J. Li, J. Lian, and S. Zhou, ``Path loss model based on cluster at 28 GHz in the indoor and outdoor environments,'' \emph{Sci. China Inf. Sci.}, vol. 60, no. 8, pp. 1--11, Aug. 2017.

\bibitem{Wang19_4} H. Wang, P. Zhang, J. Li, and X. You, ``Radio propagation and wireless coverage of LSAA-based 5G millimeter-wave mobile communication systems,'' \emph{China Commun.}, vol. 16, no. 5, pp. 1--18, May 2019.

\bibitem{Zhang19} P. Zhang, J. Li, H. Wang, and X. You, ``Millimeter-wave space-time propagation characteristics in urban macrocell scenarios,'' in \emph{Proc. IEEE ICC'19}, Shanghai, China, May 2019, pp. 1--6.

\bibitem{Sana16} S. Salous, S. M. Feeney, X. Raimundo, and A. A. Cheema, ``Wideband MIMO channel sounder for radio measurements in the 60 GHz band,'' \emph{IEEE Trans. Wireless Commun.}, vol. 15, no. 4, pp. 2825--2832, Apr. 2016.

\bibitem{Pap16} P. B. Papazian, C. Gentile, K. A. Remley, J. Senic, and N. Golmie, ``A radio channel sounder for mobile millimeter-wave communications: System implementation and measurement assessment,'' \emph{IEEE Trans. Microwave Theory Tech.}, vol. 64, no. 9, pp. 2924--2932, Sept. 2016.

\bibitem{Cau19} D. Caudill, P. B. Papazian, C. Gentile, J. Chuang, and N. Golmie, ``Omnidirectional channel sounder with phased-array antennas for 5G mobile communications,'' \emph{IEEE Trans. Microw. Theory Techn.}, vol. 67, no. 7, pp. 2936--2945, Jul. 2019.

\bibitem{Wang18} R. Wang, O. Renaudin, C. U. Bas, S. Sangodyin, and A. F. Molisch, ``Antenna switching sequence design for channel sounding in a fast time-varying channel,'' in \emph{Proc. IEEE ICC'18}, Kansas City, MO, May 2018, pp. 1--6.

\bibitem{Bas19} C. U. Bas, R. Wang, S. Sangodoyin, D. Psychoudakis, T. Henige, R. Monroe, J. Park, C. J. Zhang, and A. F. Molisch, ``Real-time millimeter-wave MIMO channel sounder for dynamic directional measurements,'' \emph{IEEE Trans. Veh. Technol.}, vol. 68, no. 9, pp. 8775--8789, Sept. 2019.

\bibitem{Wang19} R. Wang, O. Renaudin, C. U. Bas, S. Sangodoyin, and A. F. Molisch, ``On channel sounding with switched arrays in fast time-varying channels,'' \emph{IEEE Trans. Wireless Commun.}, vol. 18, no. 8, pp. 3843--3855, Aug. 2019.

\bibitem{Blu17_2} J. Blumenstein, A. Prokes, A. Chandra, T. Mikulasek, R. Marsalek, T. Zemen, and C. Mecklenbräuker, ``In-vehicle channel measurement, characterization, and spatial consistency comparison of 3-11 GHz and 55-65 GHz frequency bands,'' \emph{IEEE Trans. Veh. Technol.}, vol. 66, no. 5, pp. 3526--3537, May 2017.

\bibitem{Blu18} J. Blumenstein, A. Prokes, J. Vychodil, T. Mikulasek,  J. Milos, E. Zöchmann, H. Groll, C. F. Mecklenbräuker, M. Hofer, D. Löschenbrand, L. Bernadó, T. Zemen, S. Sangodoyin, and A. Molisch, ``Measured high-resolution power-delay profiles of nonstationary vehicular millimeter wave channels,'' in \emph{Proc. IEEE PIMRC'18}, Bologna, Italy, Sept. 2018, pp.~1--5.

\bibitem{Rah19} A. U. Rahman, A. Chandra, A. Prokes, J. Blumenstein, T. Mikulasek, and J. Vychodil, ``Doppler characteristics of 60 GHz mmWave I2I channels,'' in \emph{Proc. IEEE ICC'19}, Shanghai, China, May 2019, pp.~1--6.

\bibitem{Cha19} A. Chandra, A. Ur Rahman, U. Ghosh, J. A. García-Naya, A. Prokeš, J. Blumenstein, and C. F. Mecklenbräuker, ``60-GHz millimeter-wave propagation inside bus: Measurement, modeling, simulation, and performance analysis,'' \emph{IEEE Access}, vol. 7, pp. 97815--97826, Jun. 2019.

\bibitem{Zoc19} E. Zöchmann, M. Hofer, M. Lerch, S. Pratschner, L. Bernadó, J. Blumenstein, S. Caban, S. Sangodoyin, H. Groll, T. Zemen, A. Prokeš, M. Rupp, A. F. Molisch, and C. F. Mecklenbräuker, ``Position-specific statistics of 60 GHz vehicular channels during overtaking,'' \emph{IEEE Access}, vol. 7, pp. 14216--14232, Jan. 2019.

\bibitem{He19} R. He, C. Schneider, B. Ai, G. Wang, D. Dupleich, R. Thomae, M. Boban, J. Luo, Z. Zhong, and Y. Zhang, ``Propagation channels of 5G millimeter wave vehicle-to-vehicle communications: Recent advances and future challenges,'' \emph{IEEE Veh. Technol. Mag.}, 2020, in press.

\bibitem{Wang17} R. Wang, C. U. Bas, S. Sangodoyin, S. Hur, J. Park, J. Zhang, and A. F. Molisch, ``Stationarity region of mm-Wave channel based on outdoor microcellular measurements at 28 GHz,'' in \emph{Proc. IEEE MILCOM'17}, Baltimore, MD, Oct. 2017, pp. 782--787.
 
\bibitem{Bas19_2} C. U. Bas, R. Wang, S. Sangodoyin, T. Choi, S. Hur, K. Whang, J. Park, C. J. Zhang, and A. F. Molisch, ``Outdoor to indoor propagation channel measurements at 28 GHz,'' \emph{IEEE Trans. Wireless Commun.}, vol.~18, no.~3, pp. 1477--1489, Mar. 2019.

\bibitem{Li19} G. Li, B. Ai, G. L. Stüber, K. Guan, and G. Shi, ``On the modeling of near-field scattering of vehicles in vehicle-to-X wireless channels based on scattering centers,'' \emph{IEEE Access}, vol. 7, pp. 3264--3274, Jan. 2019.

\bibitem{Qi17} W. Qi, J. Huang, J. Sun, Y. Tan, C. Wang, and X. Ge, ``Measurements and modeling of human blockage effects for multiple millimeter wave bands,'' in \emph{Proc. IWCMC'17}, Valencia, Spain, Jun. 2017, pp. 1604--1609.


\bibitem{KEDMETIS} V. Nurmela etc., METIS, ICT-317669-METIS/D1.4, ``METIS channel models,'' Jul. 2015.

\bibitem{mmMAGIC} M. Peter etc., mmMAGIC, H2020-ICT-671650-mmMAGIC/D2.1, ``Measurement campaigns and initial channel models for preferred suitable frequency ranges,'' Mar. 2016.

\bibitem{Hri00} H. D. Hristov, \emph{Fresnel zones in wireless links, Zone Plate Lenses and Antennas}, London: Artech House, 2000.

\bibitem{Fon07} F. P. Fontan, A. Abele, B. Montenegro, F. Lacoste, V. Fabbro, L. Castanet, B. Sanmartin, and P. Valtr, ``Modelling of the land mobile satellite channel using a virtual city approach,'' in \emph{Proc. EuCAP'07}, Edinburgh, UK, Nov. 2007, pp. 1--7.

\bibitem{GTDbook} G. L. James, \emph{Geometrical theory of diffraction for electromagnetic diffraction}, 3rd edition, London: The Institution of Engineering and Technology, 1979.
 
\bibitem{Sun16_2} S. Sun, T. S. Rappaport, T. A. Thomas, A. Ghosh, H. C. Nguyen, I. Z. Kovács, I. Rodriguez, O. Koymen, and A. Partyka, ``Investigation of prediction accuracy, sensitivity, and parameter stability of large-scale propagation path loss models for 5G wireless communications,''  \emph{IEEE Trans. Veh. Technol.}, vol. 65, no. 5, pp.~2843--2860, May 2016.
 
\bibitem{Wang18_5G} S. Wu, C.-X. Wang, e. M. Aggoune, M. M. Alwakeel, and X. You, ``A general 3-D non-stationary 5G wireless channel model,''  \emph{IEEE Trans. Commun.}, vol. 66, no. 7, pp. 3065--3078, Jul. 2018.

\bibitem{Wu14} S. Wu, C.-X. Wang, H. Aggoune, M. M. Alwakeel, and Y. He, ``A non-stationary 3D wideband twin-cluster model for 5G massive MIMO channels,''  \emph{IEEE J. Sel. Areas Commun.}, vol. 32, no. 6, pp. 1207--1218, Jun. 2014.

\bibitem{Wu15} S. Wu, C.-X. Wang, H. Haas, H. Aggoune, M. M. Alwakeel, and B. Ai, ``A non-stationary wideband channel model for massive MIMO communication systems,''  \emph{IEEE Trans. Wireless Commun.}, vol. 14, no. 3, pp. 1434--1446, Mar. 2015.

\bibitem{Gha15} A. Ghazal, C.-X. Wang, B. Ai, D. Yuan, and H. Haas, ``A non-stationary wideband MIMO channel model for high-mobility intelligent transportation systems,''  \emph{IEEE Trans. Intell. Transp. Syst.}, vol. 16, no. 2, pp. 885--897, Apr. 2015.

\bibitem{Gha17} A. Ghazal, Y. Yuan, C.-X. Wang, Y. Zhang, Q. Yao, Y. Yuan, H. Zhou, and W. Duan, ``A non-stationary IMT-A MIMO channel model for high-mobility wireless communication systems,''  \emph{IEEE Trans. Wireless Commun.}, vol. 16, no. 4, pp. 2057--2068, Apr. 2017.

\bibitem{Bian18} J. Bian, J. Sun, C.-X. Wang, R. Feng, J. Huang, Y. Yang, and M. Zhang, ``A WINNER+ based 3D non-stationary wideband MIMO channel model,''  \emph{IEEE Trans. Wireless Commun.}, vol. 17, no. 3, pp. 1755--1767, Mar. 2018.

\bibitem{Liu19} Y. Liu, C.-X. Wang, J. Huang, J. Sun, and W. Zhang, ``Novel 3D non-stationary mmWave massive MIMO channel models for 5G high-speed train wireless communications,'' \emph{IEEE Trans. Veh. Technol.}, vol. 68, no.~3, pp. 2077--2086, Mar. 2019.

\bibitem{Yuan14} Y. Yuan, C.-X. Wang, X. Cheng, B. Ai, and D. I. Laurenson, ``Novel 3D geometry-based stochastic models for non-isotropic MIMO vehicle-to-vehicle channels,'' \emph{IEEE Trans. Wireless Commun.}, vol. 13, no. 1, pp.~298--309, Jan. 2014.
 
\bibitem{Yuan15} Y. Yuan, C.-X. Wang, Y. He, M. M. Alwakeel, and H. Aggoune, ``3D wideband non-stationary geometry-based stochastic models for non-isotropic MIMO vehicle-to-vehicle channels,'' \emph{IEEE Trans. Wireless Commun.}, vol. 14, no. 12, pp. 6883--6895, Dec. 2015.

\bibitem{Bian19} J. Bian, C.-X. Wang, J. Huang, Y. Liu, J. Sun, M. Zhang, and H. Aggoune, ``A 3D wideband non-stationary multi-mobility model for vehicle-to-vehicle MIMO channels,''  \emph{IEEE Access}, vol. 7, no. 1, pp.~32562--32577, Mar. 2019.

\bibitem{Wu17} X. Wu, C.-X. Wang, J. Sun, J. Huang, R. Feng, Y. Yang, and X. Ge, ``60-GHz millimeter-wave indoor channel measurements and modeling for 5G systems,''  \emph{IEEE Trans. Antennas Propag.}, vol. 65, no. 4, pp. 1912--1924, Apr. 2017.

\bibitem{Huang18} J. Huang, C.-X. Wang, Y. Liu, J. Sun, and W. Zhang, ``A novel 3D GBSM for mmWave channels,''  \emph{Sci. China Inf. Sci.}, vol. 61, no. 10, pp.~1--15, Oct. 2018. 

\bibitem{Sha18} M. Shafi, J. Zhang, H. Tataria, A. F. Molisch, S. Sun, T. S. Rappaport, F. Tufvesson, S. Wu, and K. Kitao, ``Microwave vs. millimeter-wave propagation channels: Key differences and impact on 5G cellular systems,''  \emph{IEEE Commun. Mag.}, vol. 56, no. 12, pp. 14--20, Dec. 2018.

\bibitem{TBD18} J. Huang, C.-X. Wang, L. Bai, J. Sun, Y. Yang, J. Li, O. Tirkkonen, and M.-T. Zhou, ``A big data enabled channel model for 5G wireless communication systems,''  \emph{IEEE Trans. Big Data}, vol. 6, no. 1, Mar. 2020.


\bibitem{Ngo17} H. Q. Ngo, A. Ashikhmin, H. Yang, E. G. Larsson, and T. L. Marzetta, ``Cell-free Massive MIMO versus small cells,'' \emph{IEEE Trans. Wireless Commun.}, vol. 16, no. 3, pp. 1834--1850, Mar. 2017.

\bibitem{Nay17} E. Nayebi, A. Ashikhmin, T. L. Marzetta, H. Yang, and B. D. Rao, ``Precoding and power optimization in cell-free massive MIMO systems,'' \emph{IEEE Trans. Wireless Commun.}, vol. 16, no. 7, pp. 4445--4459, Jul. 2017.

\bibitem{Ngo18} H. Q. Ngo, L. Tran, T. Q. Duong, M. Matthaiou, and E. G. Larsson, ``On the total energy efficiency of cell-free massive MIMO,'' \emph{IEEE Trans. Green Commun. Netw.}, vol. 2, no. 1, pp. 25--39, Mar. 2018.

\bibitem{Han18} C. Han, J. M. Jornet, and I. Akyildiz, ``Ultra-massive MIMO channel modeling for graphene-enabled terahertz-band communications,'' in \emph{Proc. IEEE VTC'18-Spring}, Porto, Portugal, Jun. 2018, pp. 1--5.

\bibitem{Nie19} S. Nie, J. M. Jornet, and I. F. Akyildiz, ``Intelligent environments based on ultra-massive MIMO platforms for wireless communication in millimeter wave and terahertz bands,'' in \emph{Proc. IEEE ICASSP'19}, Brighton, UK, May 2019, pp. 7849--7853.

\bibitem{Hu18} S. Hu, F. Rusek, and O. Edfors, ``Beyond massive MIMO: The potential of data transmission with large intelligent surfaces,'' \emph{IEEE Trans. Signal Process.}, vol. 66, no. 10, pp. 2746--2758, May 2018.

\bibitem{Hu18_2} S. Hu, F. Rusek, and O. Edfors, ``Beyond massive MIMO: The potential of positioning with large intelligent surfaces,'' \emph{IEEE Trans. Signal Process.}, vol. 66, no. 7, pp. 1761--1774, Apr. 2018.




\end{thebibliography}
\end{document}